\documentclass[apj]{emulateapj}

\shorttitle{REMOVING BIASES IN STELLAR MASS-MAPS}
\shortauthors{Mart\'{\i}nez-Garc\'ia et al.}

\usepackage{amsmath} 
\usepackage{hyperref}
\usepackage{graphicx}
\usepackage{color}

\newcommand{\samename}{\vrule height0.4pt depth0.0pt width1.0in \thinspace.}
\newcommand{\tn}{\tablenotemark}

\begin{document}

\title{REMOVING BIASES IN RESOLVED STELLAR MASS-MAPS OF GALAXY DISKS THROUGH SUCCESSIVE BAYESIAN MARGINALIZATION} 

\author{Eric E. Mart\'inez-Garc\'ia\altaffilmark{1}, 
Rosa A. Gonz\'alez-L\'opezlira\altaffilmark{2,3}, 
Gladis Magris C.\altaffilmark{4},
and Gustavo Bruzual A.\altaffilmark{2}}
\affil{1 Cerrada del Rey 40-A, Chimalcoyoc Tlalpan, Ciudad de M\'exico, C.P. 14630; 
{\color{blue}martinezgarciaeric@gmail.com}} 

\affil{2 Instituto de Radioastronom\'ia y Astrof\'isica, UNAM, Campus Morelia,
     Michoac\'an, M\'exico, C.P. 58089}

\affil{3 Argelander Institut f\"ur Astronomie, Universit\"at Bonn, Auf dem H\"ugel 71, D-53121 Bonn, Germany}

\affil{4 Centro de Investigaciones de Astronom\'ia, Apartado Postal 264, M\'erida 5101-A, Venezuela}

\begin{abstract}
Stellar masses of galaxies are frequently obtained by fitting stellar population synthesis models
to galaxy photometry or spectra. The state of the art method resolves spatial
structures within a galaxy to assess the total stellar mass content. In comparison
to unresolved studies, resolved methods yield, on average, higher fractions of stellar mass for galaxies.
In this work we improve the current method in order to mitigate a
bias related to the resolved spatial distribution derived for the mass.
The bias consists in an apparent filamentary mass distribution, and a spatial coincidence
between mass structures and dust lanes near spiral arms.
The improved method is based on iterative Bayesian marginalization,
through a new algorithm we have named Bayesian Successive Priors (BSP).
We have applied BSP to M~51, and to a pilot sample of 90 spiral galaxies from the
Ohio State University Bright Spiral Galaxy Survey.
By comparing quantitatively both methods, we find that the average fraction of stellar mass missed by 
unresolved studies is only half than previously thought. 
In contrast with the previous method, the output BSP mass-maps
bear a better resemblance to near infrared images.

\end{abstract}


\section{Introduction}

How galaxies form and assemble their mass is a primordial
question in modern astrophysics. Galaxy masses are crucial for 
their evolution, and for the evolution of cosmic structures at all scales.
The determination of the {\it stellar} mass content of galaxies can help
constrain, e.g., the dark matter fraction,
the specific star formation rate ($\Psi_{\rm S}$, the star formation rate, $\Psi$, per unit stellar mass),
the stellar mass function, and the universe's stellar mass density and
star formation history (SFH).

There are different methods to estimate the mass of a galaxy,
e.g., dynamical or through gravitational lensing ~\citep[see][for a review]{cou14}.
Regarding the stellar mass component,
the use of stellar population synthesis (SPS) models
to estimate mass through the
stellar mass-to-light ratio, $\Upsilon_{*}$\footnote{
Throughout this work $\Upsilon_{*}$ refers to the stellar (including remnants) mass-to-light 
ratio in units of $M_{\sun}/L_{\sun}$, i.e., we do not include dark matter,
nor gas mass in $\Upsilon_{*}$.},
has been frequently advocated~\citep[e.g.][]{bel01,bel03}.
Notwithstanding their common degeneracies, SPS models can in general yield 
reliable mass estimates. One novel technique 
is the {\it resolved stellar mass-map} method~\citep[][ZCR hereafter]{zcr09},
that delivers a map of the stellar mass surface density by photometric means.
Galaxy masses determined by treating the galaxies as point sources are 
often underestimated~\citep[and sometimes overestimated, see][]{rod15},
thus the need to resolve structures~(ZCR; Sorba \& Sawicki 2015). 
Even more, if the stellar mass of each galaxy in a cluster is estimated
separately, the total stellar mass fraction is lower 
than when a constant $\Upsilon_{*}$ is assumed~\citep{lea12}.

The resolved stellar mass method is truly powerful, since it can solve
not only for the mass, but for other physical parameters
of the SPS models, based solely on photometry.
Resolved maps of stellar mass are also important for studies 
aimed at understanding the dynamics of bars and/or spirals (since gravity is the main driver),
and their secular evolution~\citep[e.g.,][]{foy10,mart13,egu16}.
Additionally, they can be used to determine the baryonic contribution to rotation
curves~\citep[e.g.,][]{rep13,rep15,mcg16}. The method can also be extended to
higher-redshift studies~\citep[e.g.,][]{lan07,wuy12}.

Despite their potential, the resulting mass-maps may be biased, in the sense that the stellar 
mass shows a filamentary structure and is concentrated in dust lanes. 
In this paper we aim to understand the origin of this shortcoming and improve 
the method to derive resolved stellar mass-maps.
We must also mention that in this research we use SPS models that assume a constant metallicity
along the SFH.~\citet{gal09} studied the effects of using a variable metallicity SPS library
and found no significant biases when estimating $\Upsilon_{*}$. Nevertheless,~\citet{int13}
indicate that the color-mass-to-light ratio relations~\citep[CMLR, see e.g.][]{mcg14}
resulting from an evolving metallicity along a coherent SFH within an individual galaxy
are probably different from the CMLR established for the general galaxy population.
Furthermore, biases in mass determinations from CMLR can be even more significant at high redshifts than for 
local studies~\citep[see e.g.,][]{mit13}. In this work we do not use CMLR to recover $\Upsilon_{*}$;
instead, we use a statistically robust Bayesian technique to infer
the predicted $\Upsilon_{*}$ via the comparison of observed colors
with a comprehensive library of SPS models.

The paper is organized as follows. In section~\ref{mass_maps} we describe the 
resolved stellar mass-map method in its present form and explain/investigate the source of the bias.
We introduce a new method (based on the former) in section~\ref{bay_object}.
In section~\ref{M51_BSP} we apply the new method to the spiral galaxy M~51 (NGC~5194);
comparisons with other methods are also briefly described.
In section~\ref{pilot_sample} we apply the new method to a pilot sample of spiral galaxies,
and discuss and analyze the results.
The uncertainties in the stellar mass estimates are discussed in section~\ref{mass_errors}.
Finally, we give our conclusions in section~\ref{conclu}.

\section{Resolved maps of stellar mass}~\label{mass_maps}

The ZCR method uses a Monte Carlo library of SPS models obtained from the
2007 version of~Bruzual \& Charlot (2003; CB07) models with the~\citet{cha03}
stellar initial mass function (IMF).
The library was built by adopting prior probability
distributions for parameters such as the 
SFH, the dust attenuation~\citep[treated as in the two-component model of][]{cha00},
and a non-evolving metallicity.
By randomly drawing the parameters from the prior distributions~\citep[cf.][]{daC08},
the resulting library consists of $\approx5\times10^{4}$ templates (or models).

\begin{figure*}
\centering
\epsscale{1.0}
\plotone{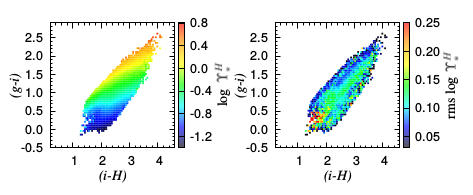}
\caption[f1]{
{\it{Left}}:
decimal logarithm of the {\it effective}, i.e., as seen by the observer~(cf. ZCR),
mass-to-light ratio at the $H$ band,  $\Upsilon_{*}^{H}$,
derived from the $(g-i)$ vs.\ $(i-H)$ color-color diagram. The data are taken
from the MAGPHYS-CB07 Monte Carlo SPS library, and grouped in bins $0.05\times0.05$ mag$^2$.
SDSS $g$ and $i$ magnitudes are in the AB magnitude system, $H$ magnitudes are Vega.
{\it{Right}}:
root mean square (rms) errors of log $\Upsilon_{*}^{H}$ 
in the left panel. The global median rms error is 0.1 dex.
~\label{fig1}}
\end{figure*}

The ZCR fiducial method is based on surface brightness photometry
at the $g$ and $i$ Sloan Digital Sky Survey (SDSS) optical bands,
and one near-infrared (NIR) filter such as $J$, $H$,
or $K$. The method was extended to include the
Spitzer Space Telescope Infrared Array Camera (IRAC) $3.6\micron$-band by~\citet{rep15}.
Other optical color combinations are possible, with the disadvantage of having
more degeneracy in $\Upsilon_{*}$, and thus more uncertain
results~\citep[see e.g.,][their Figures 1 and 2, respectively]{rep15,bel01}.
The templates from the SPS library are binned in colors $(g-i)$ and $(i-H)$,
using a bin width of 0.05 magnitude (see Figure~\ref{fig1}).
The median mass-to-light ratio at the $H$-band, $\Upsilon^{H}_{*}$,
is estimated for each bin. A look-up table can thus be constructed to compare
with observed photometry on a pixel-by-pixel basis.
The $\Upsilon^{H}_{*}$ is the {\it effective} mass-to-light ratio, i.e., refers to the light that
reaches the observer, as opposed to the light that is emitted.
The effective $\Upsilon^{H}_{*}$ may be affected by extinction~(ZCR).

Earlier studies concerning pixel-by-pixel spatially resolved properties of galaxies can be found
in~\citet{bot86},~\citet{abr99},~\citet{cont03},~\citet{esk03},~\citet{kas03},
\citet{lan07}, and~\citet{wel08}.

\subsection{Application to M~51. A filamentary mass structure?}~\label{m51_maxL}

\begin{figure*}
\centering
\epsscale{1.0}
\plotone{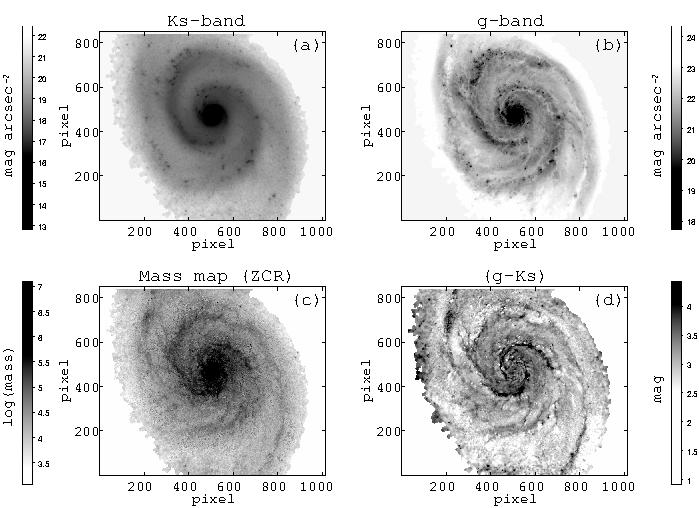}
\caption[f2]{
{\it{Top left}}: $K_{s}$-band mosaic of M~51; grayscale in Vega mag.
{\it{Top right}}: $g$-band mosaic of M~51; grayscale in AB mag.
{\it{Bottom left}}: M~51 stellar mass-map derived with the ZCR method, based
on $(g-i)$ and $(i-K_{s})$ colors,
and $K_{s}$ mass-to-light ratio, $\Upsilon_{*}^{K_{s}}$; mass in $M_{\sun}$.
{\it{Bottom right}}: $(g-K_{s})$ color map of M~51. Notice the similarities of the 
features in this extinction map and in the stellar mass-map in the bottom left panel (c).
Higher extinction is indicated by darker features. North is up, East is to the left.
~\label{fig2}}
\end{figure*}

Now we present results obtained by applying the ZCR method to the spiral
galaxy M~51. We use $g$ and $i$-band imaging from the twelfth
data release (DR12) of the SDSS~\citep{ala15}, as well as 
the $K_{s}$-band mosaic from~\citet{gon96}.
The NIR images were obtained at Kitt Peak National Observatory (KPNO),
with the IR Imager (IRIM) camera on the 
1.3 meter telescope;
the IRIM had a $256^2$ NICMOS3 array with
a $2\arcsec$ pixel$^{-1}$ plate scale.
The observations were performed during March 1994, in
non-photometric conditions, and 
the exposures were resampled with sub-pixel accuracy before
combining. The final $K_{s}$-band
mosaic has $0.5\times 0.5$ arcsec$^2$ pixels, and a 
total exposure time of 22 minutes; 
it was photometrically calibrated\footnote{
Throughout this work NIR magnitudes are Vega, SDSS magnitudes are in the AB magnitude system.}
with the Two Micron All Sky Survey~\citep[2MASS,][]{skr06}.
The SDSS frames were re-sampled to the resolution of the NIR data, 
and registered with the $K_{s}$-band image. The registration
was done with the IRAF\footnote{IRAF is distributed by the National Optical Astronomy Observatories,
which are operated by the Association of Universities for Research in
Astronomy, Inc., under cooperative agreement with the National Science
Foundation.}~\citep{tod93} tasks {\tt{GEOMAP}} and {\tt{GREGISTER}}.
No point spread function (PSF) match was done to the images, since
the data have similar PSFs and the process can corrupt the noise properties~\citep{zcr09}.
In Figures~\ref{fig2}a (top left panel), and~\ref{fig2}b (top right panel),
we show the $K_{s}$-band and $g$-band final images, respectively.
The foreground stars and background galaxies were removed and
their pixels replaced with values from the background-subtracted ``sky''.
With the purpose of isolating the disk from the
lower signal-to-noise (S/N) background, the final mosaics were treated with the
{\tt{Adaptsmooth}} code of~\citet{zbt09}, as follows.
A first run of {\tt{Adaptsmooth}} was performed on the the $K_{s}$-band data
(which have a lower S/N ratio than the SDSS images),
with the requirement of a minimum S/N ratio per pixel of 20,
a maximum smoothing radius of 10, and the assumption of background-dominated noise.
In order to homogenize the lower limit of the S/N ratio per pixel,
the output smoothing $K_{s}$-band mask was then used as an input, in subsequent runs
of {\tt{Adaptsmooth}}, for the SDSS $g$ and $i$ bands.

The SPS library was obtained from the Multi-wavelength Analysis of Galaxy
Physical Properties package (MAGPHYS-CB07 library, hereafter) by~\citet{daC08}.\footnote{
\url{http://www.iap.fr/magphys/magphys/MAGPHYS.html}}
The absolute magnitudes of the Sun were taken from~\citet{bla07}.
We assume a distance to M~51 of $9.9\pm0.7$~Mpc~\citep{tik09},
and correct the models for Galactic extinction~\citep{schl11}.

\begin{figure}
\centering
\epsscale{1.0}
\plotone{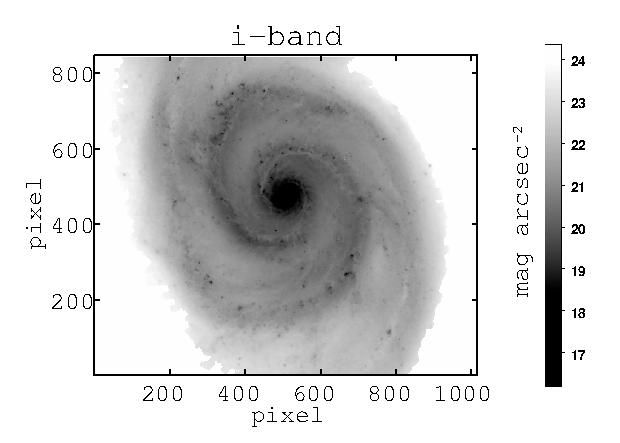}
\caption[f3]{
SDSS $i$-band mosaic of M~51. Grayscale in AB mag. North is up, East is to the left.
~\label{fig3}}
\end{figure}

\begin{figure*}
\centering
\epsscale{1.0}
\plotone{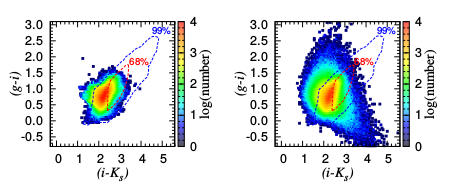}
\caption[f4]{
2-D histograms of the observed $(g-i)$ and $(i-K_{s})$ colors of M~51's pixels.
The areas inside the blue and red dashed lines contain 
$99\%$ and $68\%$, respectively, of the templates in the MAGPHYS-CB07 SPS library
corrected for Galactic extinction.
{\it{Left}}: after applying the {\tt{Adaptsmooth}} procedure as described in the text.
The maximum of $\log(\rm number)$ occurs near $(i-K_{s})\sim 2.29$ and $(g-i)\sim 0.85$.
{\it{Right}}: without applying the {\tt{Adaptsmooth}} procedure.
~\label{fig4}}
\end{figure*}

The resulting mass-map is presented in Figure~\ref{fig2}c (bottom left panel).
For comparison purposes we show in Figure~\ref{fig3} the $i$-band image.
The color range covered by the observed photometry of M~51 is shown,
as a 2-D histogram, in Figure~\ref{fig4}. In the left panel
we show the observed colors of the pixels after applying the {\tt{Adaptsmooth}} procedure
as described earlier. The right panel shows the observed colors of the same pixels
without using the {\tt{Adaptsmooth}} procedure. 
From the comparison of these plots we appreciate the advantage of
increasing the S/N ratio in the outskirts of the disk, otherwise the
uncertainties in the fits would be quite large.
In these figures we also demarcate the color range covered by $99\%$
and $68\%$ of the total templates in the MAGPHYS-CB07 library,
with a blue and a red contour, respectively.
Most of the observed colors fall within the span of the SPS library.
The plots are illustrative and do not reflect the observational
uncertainties of the data. 

\begin{figure}
\centering
\epsscale{1.0}
\plotone{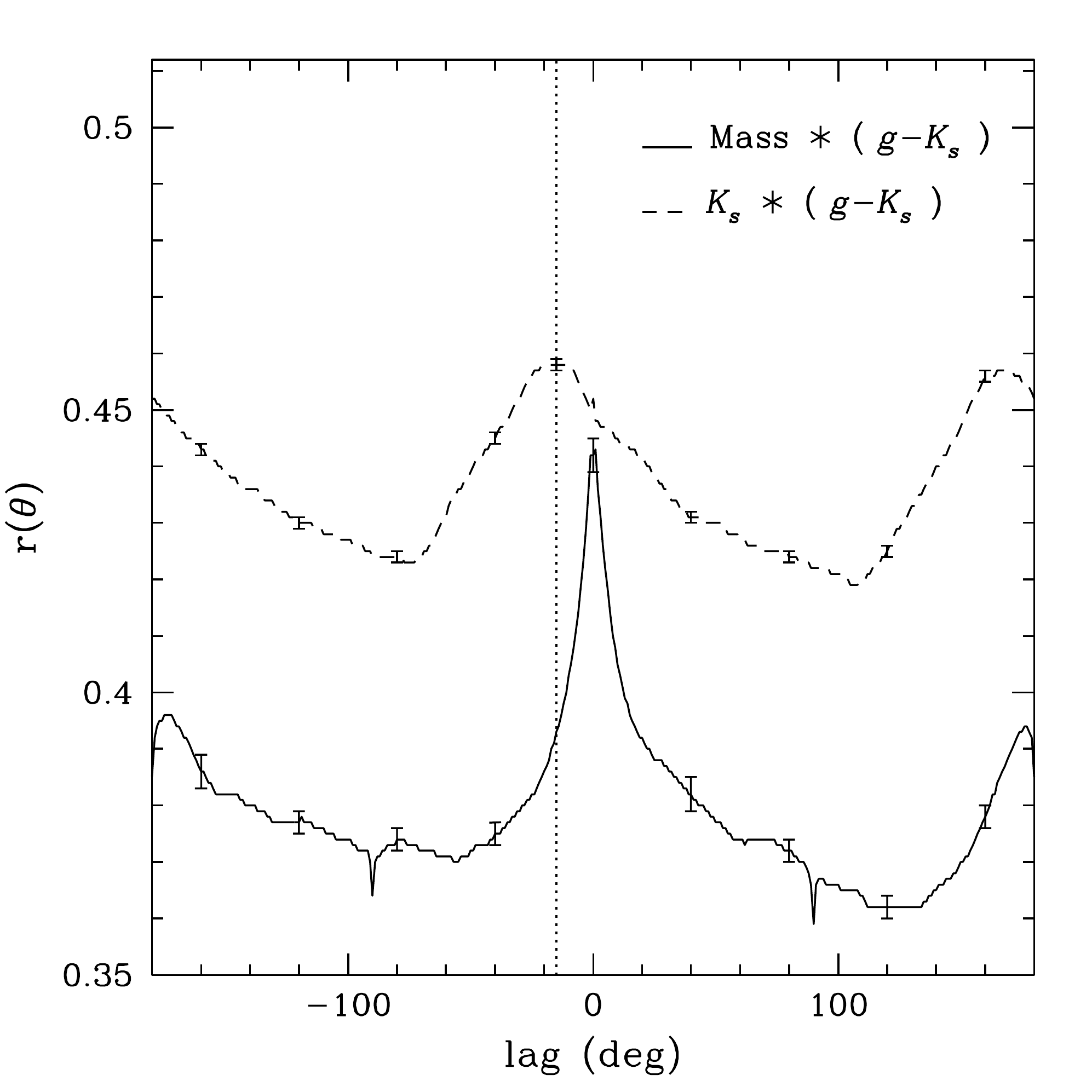}
\caption[f5] {
Cross correlation functions $r(\theta)$ (see text). {\it Solid line:} between the mass structure resulting from 
the ZCR method and the intensity ratio in the $(g-K_{s})$ image.
The absolute maximum is at $\theta=0\degr$, indicating similarity. {\it Dashed line:} 
between the intensity in the $K_{s}$-band image and the intensity ratio in the $(g-K_{s})$ extinction map.
The maximum occurs at $\theta\sim-15\degr$ (marked by the vertical dotted line), and 
corresponds to the angular lag between the dust lanes and the stellar arms.
The total height of each error bar is $2\sigma_{cc}$.
~\label{fig5}}
\end{figure}

One striking thing to notice about the mass-map (Figure~\ref{fig2}c)
is that it does not present a smooth spiral arm structure.
There is a well defined two-arm spiral pattern, but 
many filamentary structures are also observed.  
In addition, a visual comparison of the mass structure with the optical $g$-band indicates that, presumably,
most of the structure is coincident with the dust lanes, as inferred from optical extinction.
This can be seen more easily in Figure~\ref{fig2}d (bottom right panel), where we show the $(g-K_{s})$ image.
To test the similarities between the mass-map and the $(g-K_{s})$ image quantitatively, we use cross-correlation
techniques. The Pearson correlation coefficient is defined as

\begin{equation}~\label{crossco}
  r = \frac{\sum\limits_{j}\sum\limits_{i}(f_{ij}-\bar{f})(g_{ij}-\bar{g})}
              {\sqrt{\sum\limits_{j}\sum\limits_{i}(f_{ij}-\bar{f})^2}
               \sqrt{\sum\limits_{j}\sum\limits_{i}(g_{ij}-\bar{g})^2}},
\end{equation}

\noindent where $f_{ij}$ is the intensity of the $i^{\rm th},j^{\rm th}$ pixel in the first image,
$g_{ij}$ is the intensity of the $i^{\rm th},j^{\rm th}$ pixel in the second image,
$\bar{f}$ is the mean intensity of the first image,
and $\bar{g}$ is the mean intensity of the second image.
The cross-correlation function, $(f\star g)(\theta)$, is then obtained by rotating the first image
with respect to the second one, while fixing the center of rotation at the
center of the object (the nuclei of M~51 in this case).
We obtain $r(\theta)$ from equation~\ref{crossco} by varying $\theta$
from $-180\degr$ to $180\degr$ in increments of $1\degr$;
we assume that the angle $\theta$ increases counterclockwise.
All the M~51 data were deprojected assuming an inclination angle of $20\degr$, 
and a position angle of $172\degr$~\citep{ler08}.
The result of the cross-correlation between the output
mass-map of the ZCR method and the intensity ratio
in the $(g-K_{s})$ image is shown in Figure~\ref{fig5}.
By ``intensity ratio'', we mean the ratio between the intensity in the $g$-band
image and the intensity in the $K_{s}$-band image. We use this
ratio instead of the $(g-K_{s})$ color because the latter scales logarithmically 
and cannot be compared with the mass distribution, that scales linearly.
Note that we actually take the intensity ratio in the minus $(g-K_{s})$ image;
this is done with the purpose of getting positive values of $r$ (when using equation~\ref{crossco}).
Error bars were estimated with bootstrap methods~\citep{bha90,lep92}.
We replace each pixel separately with a random value, drawn from a Gaussian probability distribution,
and for each $\theta$ recalculate equation~\ref{crossco}. We repeat this process a total of 30 times
and calculate~$\sigma_{cc}$, the standard deviation of the resulting distribution.

There is clearly a peak in the cross-correlation function near $\theta=0\degr$,
indicating a similarity between the structures. For comparison, we also show the
cross-correlation between the intensity in the $K_{s}$-band image and the intensity
ratio in the $(g-K_{s})$ image.
The absolute maximum in this case occurs around $\theta=-15.5\degr\pm0.8$,
and marks the angular offset between the spiral arms in the $K_{s}$-band and the dust
lanes in the $(g-K_{s})$ image. This means that if we rotate the spiral arms in
the $K_{s}$-band by $15\degr$, clockwise, they will match the spatial location of the dust lanes.

As is well known, disk galaxies, when studied at
different wavelengths, often show significant differences~\citep[e.g.,][]{blo91,blo94}.
Even if at NIR wavelengths young stars and clusters can contribute $20\%$ -- $30\%$
to the total radiation in spiral arm regions~\citep[e.g.,][]{rix93,gon96,rho98,jam99,pat01,gros06,gros08},
most of the light in the disk comes from evolved giant stars,
and most of the mass is concentrated in low mass main sequence stars.
Hence, any structures present in resolved stellar mass-maps
should resemble the NIR surface brightness morphology to a significant degree.
This is not the case of the stellar mass-map shown in Figure~\ref{fig2}c (bottom left panel), 
where we see filamentary structure not present in the $K_{s}$ light distribution,
Figure~\ref{fig2}a (top left panel).

\begin{figure*}
\centering
\epsscale{1.0}
\plotone{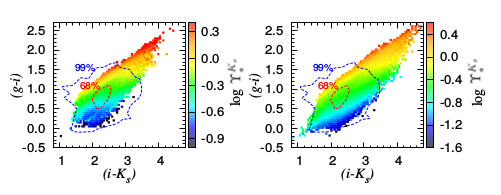}
\caption[f6] {
{\it{Left}}:
decimal logarithm of the {\it effective} mass-to-light ratio at the $K_{s}$-band, 
$\Upsilon_{*}^{K_{s}}$, derived from the 
$(g-i)$ vs.\ $(i-K_{s})$ color-color diagram.
The data are taken from the SSAG-BC03 Monte Carlo SPS library~\citep[][]{mag15},
corrected for Galactic extinction towards M~51.
SDSS $g$ and $i$ magnitudes are in the AB magnitude system, $K_{s}$ magnitudes are Vega.
The blue/red dashed contour delimits $99\%$/68\% of the observed colors for M~51 (see Figure~\ref{fig4}, left panel).
{\it{Right}}: analogous to left panel, but for the MAGPHYS-BC03 Monte Carlo SPS library.
~\label{fig6}}
\end{figure*}

\begin{figure}
\centering
\epsscale{1.0}
\plotone{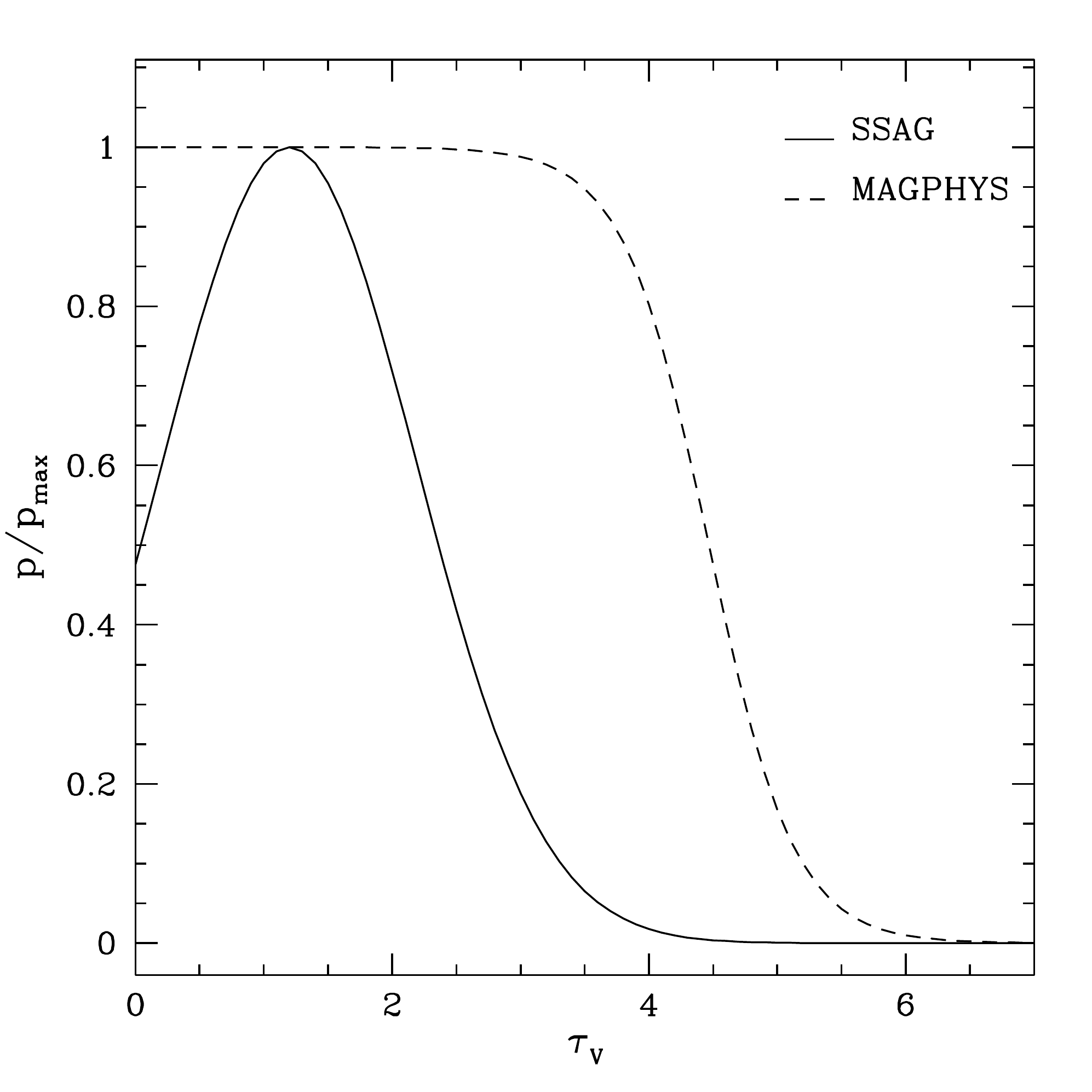}
\caption[f7] {
Probability distribution functions of the $V$-band optical depth of the dust seen
by young stars, $\tau_{V}$, used by the SSAG-BC03 (solid line),
and the MAGPHYS (dashed line) Monte Carlo SPS libraries, respectively.
~\label{fig7} }
\end{figure}

We perform three other different and independent tests,
and compare the resulting stellar mass-maps as described below.
\begin{enumerate}

\item We do not use the NIR band, and rely only on the optical SDSS colors, e.g.,
$(u-i)$ and $(g-i)$,
and on the mass-to-light ratio estimated in the $i$-band, $\Upsilon^{i}_{*}$.

\item We remove the binning of the models and use the full 
$5\times 10^{4}$ templates of the MAGPHYS-CB07 library in the computations.

\item We use a new Monte Carlo SPS (optical-NIR) library taken from the
Synthetic Spectral Atlas of Galaxies~\citep[SSAG;][]{mag15}.
SSAG\footnote{\url{http://www.astro.ljmu.ac.uk/~asticabr/SSAG.html}}
assumes random SFHs according to the~\citet{che12} prescription,
that includes a burst and a truncation event.
Dust is treated as in~\citet{cha00}, and metallicity is distributed between 0.02 $Z_{\sun}$ and 2.5 $Z_{\sun}$,
with 95\% galaxy templates having $Z > 0.2 Z_{\sun}$. The adopted IMF is Chabrier.
The library contains $6.7\times~10^{4}$ templates (SSAG-BC03 library henceforth).
The range in these models of the {\it effective}
mass-to-light ratio in the $K_s$-band, $\Upsilon^{K_{s}}_{*}$,
as determined by a $(g-i)$ vs.\ $(i-K_{s})$ color-color diagram, is shown in Figure~\ref{fig6}, left panel.
For comparison purposes we show the same diagram
for the BC03 version of the MAGPHYS library (MAGPHYS-BC03) in the right panel.
The MAGPHYS library extends to redder colors due to the different
probability distribution functions used to model the optical depth in the $V$-band,
$\tau_{V}$ (see Figure~\ref{fig7}).

\end{enumerate}

The filamentary structure, and the spatial coincidence between mass
and dust lanes prevail in all the tests.
A similar result is obtained for other spiral galaxies as well,
and was already noticed although not sufficiently discussed
in ZCR (their Figure 8).
It is noteworthy that this is not a problem of resolution in the SPS libraries,
since the mean sampling is $\sim0.005\%$ or less, both for colors and for $\Upsilon^{K_{s}}_{*}$;
hence, the template set is densely populated.

In this work we will focus on the structural properties
of the mass-maps. We will only mention here that both the local and the
integrated stellar masses derived from SPS models may vary on account of 
different treatments of the thermally-pulsating asymptotic
giant branch~\citep[TP-AGB; see, e.g.,][]{mara06,bru07,conr09},
and the choice of IMF in the libraries. The mass determinations may also differ if obtained
from different bands, even when using the same models~\citep{mcg14}.

\subsection{The level of accuracy in mass-to-light ratio estimates}

~\citet{gal09} discuss thoroughly the $\Upsilon_{*}$ accuracy that can be achieved 
by comparing colors with predictions from a large library of SFHs.
Typical accuracies are of the order of 0.1-0.15 dex.
A similar result is deduced by other authors~(e.g., Bell \& de Jong 2001; ZCR; Taylor et al. 2011).
This level of accuracy is barely improved with spectroscopic data~\citep{gal09}.

\begin{figure}
\centering
\epsscale{1.0}
\plotone{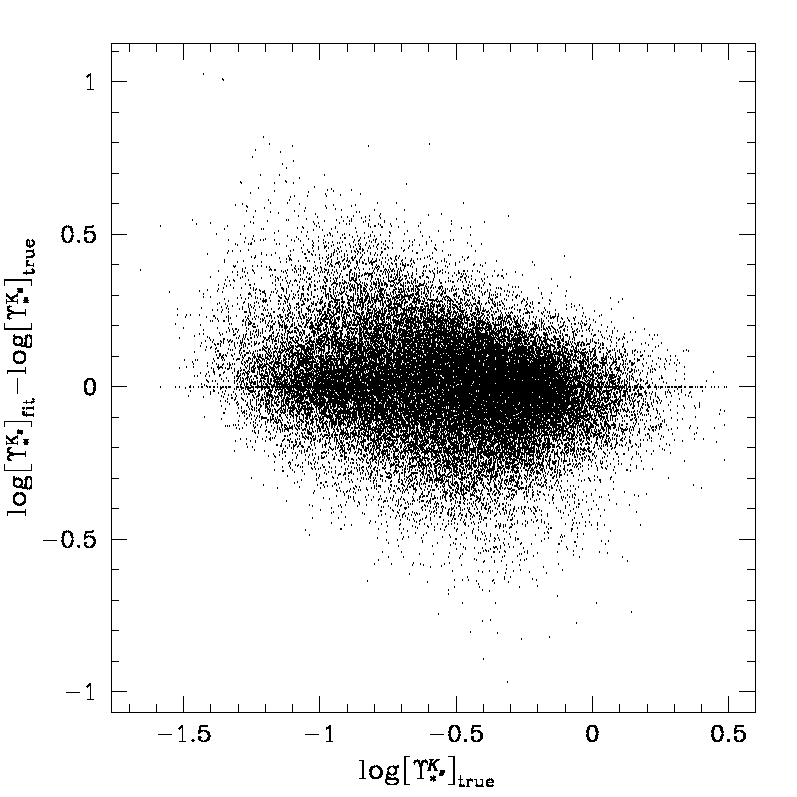}
\caption[f8]{
Fits to colors of 
$\approx 5\times 10^{4}$ mock galaxies (see text).
Noise modeled as a random Gaussian distribution
with $\sigma_{\rm mag} = 0.02$ mag is added to the 
mock objects before fitting them with the noise-free
templates of the MAGPHYS-CB07 library.
The difference is quantified as
$\Delta\log[\Upsilon_{*}^{K_{s}}]=
\log[\Upsilon_{*}^{K_{s}}]_{\rm{fit}}-\log[\Upsilon_{*}^{K_{s}}]_{\rm{true}}$.
The standard deviation of $\Delta\log[\Upsilon_{*}^{K_{s}}]$ is
$\sigma (\Delta\log[\Upsilon_{*}^{K_{s}}])=0.16$ dex.
~\label{fig8}}
\end{figure}

\begin{figure}
\centering
\epsscale{1.0}
\plotone{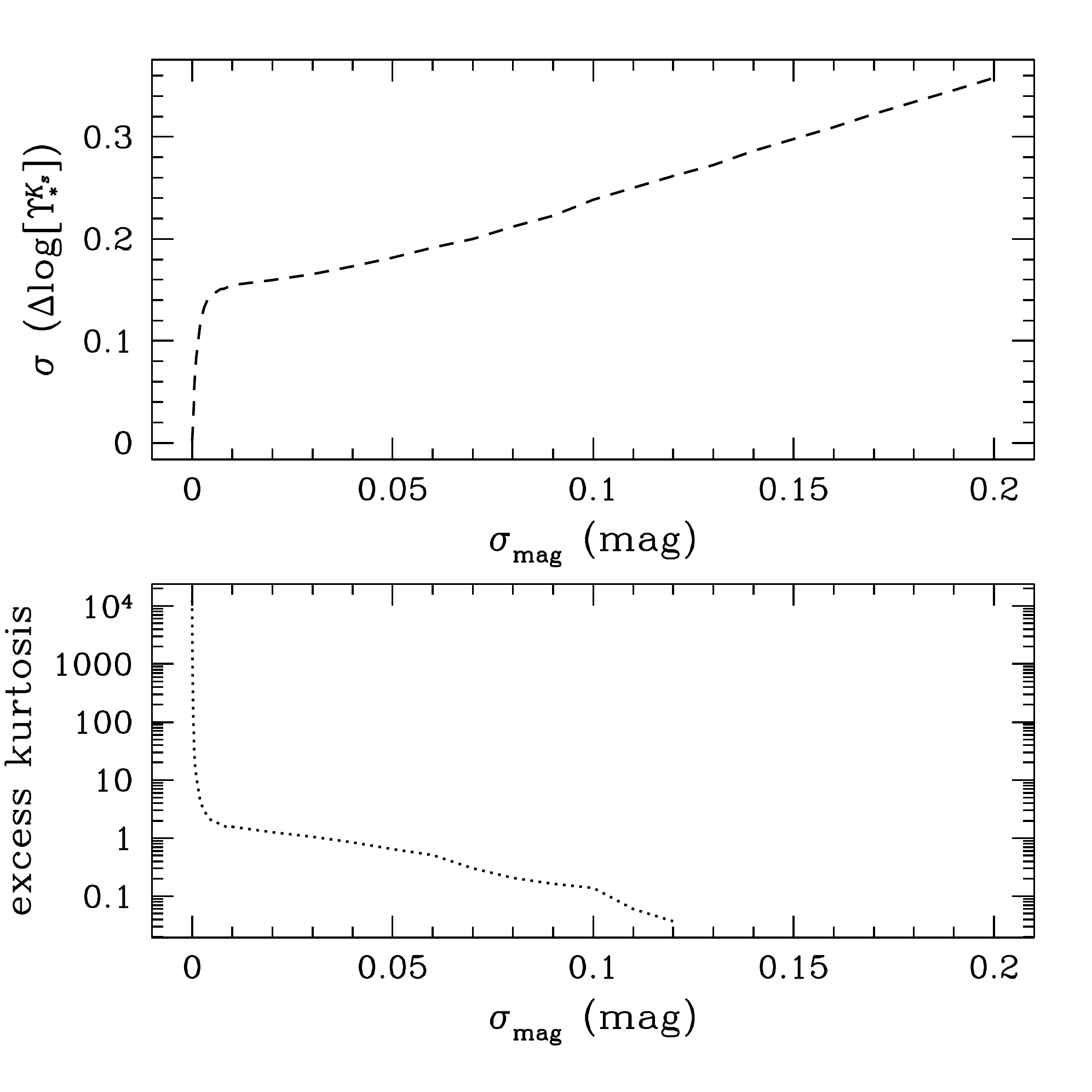}
\caption[f9]{
Statistical parameters of the fits to colors of mock galaxies.
{\it Top:} standard deviation, $\sigma (\Delta\log[\Upsilon_{*}^{K_{s}}])$, vs.\
$\sigma_{\rm mag}$ in the range 0.0-0.2 magnitude 
(see also Figure~\ref{fig8}, where $\sigma_{\rm mag}=0.02$ magnitude).
{\it Bottom:} excess kurtosis (or kurtosis minus 3) of $\Delta\log[\Upsilon_{*}^{K_{s}}]$ vs.\ $\sigma_{\rm mag}$.
~\label{fig9}}
\end{figure}

To better understand the impact of a limited $\Upsilon_{*}$ accuracy
on the resolved mass-maps of galaxies, we build a sample of mock galaxies
drawn from the MAGPHYS-CB07 Monte Carlo SPS library.
Each of the $\approx 5\times 10^{4}$ templates is used as an individual
object in our mock catalog. In order to simulate the photometric error,
we add to each of the $g$, $i$, and $K_{s}$-band magnitudes in our mocks
a random noise component with a Gaussian distribution,
having $\sigma_{\rm mag} = 0.02$ mag ($\sim 2\%$~intensity variation).
We then try to fit the noisy $(g-i)$ and $(i-K_{s})$ values of each simulated object
with the noise-free $(g-i)$ and $(i-K_{s})$ colors,
via $\chi^2$ minimization. Afterwards we compute

\begin{equation}
\Delta\log[\Upsilon_{*}^{K_{s}}]=
\log[\Upsilon_{*}^{K_{s}}]_{\rm{fit}}-\log[\Upsilon_{*}^{K_{s}}]_{\rm{true}},
\end{equation}

\noindent i.e., the ratio between the fitted $\Upsilon_{*}$
and the true value. The results of this test are shown in Figure~\ref{fig8},\footnote{
We notice that~\citet{gal09} obtain a similar plot in spite of neglecting dust corrections,
which indicates that dust is not a decisive factor for $\Upsilon_{*}$ accuracy.}
where we get a dispersion (standard deviation) $\sigma (\Delta\log[\Upsilon_{*}^{K_{s}}])\sim0.16$
dex, as expected.
We carry out the same exercise for different
$\sigma_{\rm mag}$ values and obtain $\sigma (\Delta\log[\Upsilon_{*}^{K_{s}}])$
for each one. The results are shown in Figure~\ref{fig9}, upper panel.
There is a nearly linear decrease of $\sigma (\Delta\log[\Upsilon_{*}^{K_{s}}])$
with diminishing $\sigma_{\rm mag}$ down to $\sigma_{\rm mag}\sim~0.005$.
For lower values of $\sigma_{\rm mag}$,  the shape of the 
$\Delta\log[\Upsilon_{*}^{K_{s}}]$ distribution abruptly
begins to change, from nearly Gaussian with kurtosis $\sim3$, 
going through Laplace distributions,
and finally tending to a Dirac delta function with kurtosis $\rightarrow\infty$.
This effect can be appreciated in the lower panel of Figure~\ref{fig9}, where we plot the
excess kurtosis\footnote{Excess kurtosis is measured with respect to the kurtosis
of any univariate normal distribution, which equals 3.
Therefore, excess kurtosis equals kurtosis minus 3.}
of $\Delta\log[\Upsilon_{*}^{K_{s}}]$ versus $\sigma_{\rm mag}$.

As $\sigma_{\rm mag}$ tends to zero, the dispersion,
$\sigma (\Delta\log[\Upsilon_{*}^{K_{s}}])$, also tends to zero.
A (hypothetical) value of $\sigma (\Delta\log[\Upsilon_{*}^{K_{s}}])=0$ would indicate that 
our adjusted values are equal to the true values (the noise-free models).
We can infer that it is not feasible to get
accurate $\Upsilon_{*}$ values unless the intrinsic errors of the
observations are diminished to zero, i.e., $\sigma_{\rm mag}\rightarrow 0$.
Typical photometric calibration errors are of the
order of $1-2\%$ for the SDSS~\citep{pad08} and other photometric surveys.
Additionally to this, the degeneracies between the different SPS model parameters
(e.g., age-metallicity-reddening) will prevail even when $\sigma_{\rm mag}\rightarrow 0$.

Taking all this into account we can conclude that the features in 
resolved mass-maps, acquired from a simple $\chi^2$ minimization,
will be discrepant from the structures of NIR surface brightness maps,
owing to a limited $\Upsilon_*$ accuracy.
In this manner, the fit we can obtain for some
observed colors will result in a $\Upsilon_{*}$ value near the
statistical mode of similar colors in the SPS
library~\citep[see also the discussion in][]{tay11},
and within $0.1-0.15$ dex of the true value.
Even for the same SPS library, the ``recovered'' $\Upsilon_{*}$ will
depend on the colors used in the fit. 

\section{Bayesian inference aimed at an object}~\label{bay_object}

In this section we introduce the 
Bayesian successive priors (BSP) algorithm, aimed at an individual object,
in order to solve for the mass-map avoiding the bias in the spatial structure.
The idea is to use the previous information regarding the 
stellar surface mass density as deduced from the NIR bands.
The massive older population of a galaxy is mainly traced in the NIR bands,
specially the $K$-band~\citep{rix93}.
Having established this, we can adopt the NIR surface brightness distribution
as a Bayesian prior, in order to infer the ``true'' stellar surface mass density.
In this work, we will use the term ``prior'' in reference to the prior {\it probability distribution function}.
The Bayesian prior is then directed to a particular galaxy, and
not to the entire galaxy population.

\subsection{Bayes' theorem}~\label{bayinf}

Bayesian probability posits that the best outcome of any event 
is found by calculating the probabilities of the various hypotheses involved, using the
rules of probability theory~\citep[e.g.,][]{lor92,lor95}.

The ZCR approach uses a method similar to a Bayesian {\it maximum-likelihood} estimate
by including a uniform (or flat) prior in the fits to the observed colors,
regardless of the SPS library.
In the present work a significant improvement is made in the
calculation of the stellar mass-maps, by introducing a Bayesian method with an informative,
non-uniform, prior.
Applications of Bayesian inference with non-uniform
priors have been used in, e.g.,~\citet{ben00},
for cosmological redshift estimates,~\citet{rov14},
for AGN sources analysis, and~\citet{scho14},
for the determination of stellar parameters.
 
In our case, Bayes' theorem for the most probable
stellar mass-to-light ratio $\Upsilon_{*}$ is given by

\begin{equation}~\label{Bayes}
  P(\Upsilon_{*} \mid C) = \frac{P(C \mid \Upsilon_{*})P(\Upsilon_{*})}{P(C)},
\end{equation}

\noindent
where $P(\Upsilon_{*} \mid C)$ is the {\it{posterior}} probability, 
i.e., the probability of having $\Upsilon_{*}$, for a certain
stellar population, if colors $C$ are observed.

$P(C \mid \Upsilon_{*})$ is the likelihood function
(or the probability of observing colors $C$ given the set of parameters $\Upsilon_{*}$):

\begin{equation}~\label{maxlike}
  P(C \mid \Upsilon_{*}) \propto \frac{1}{\sqrt{2\pi}} \exp \left(-\frac{\chi^{2}}{2}\right),
\end{equation}

\begin{equation}~\label{chi2}
  \chi^{2}=\sum_{n=1}^{N_{\rm colors}} \left(\frac{C_{n}^{\rm obs}-C_{n}^{\rm template}}{\sigma_{\rm col}} \right)^{2},
\end{equation}
where $C_{n}^{\rm obs}$ is the observed $n_{\rm th}$ color with $\sigma_{\rm col}$ photometric error,
and $C_{n}^{\rm template}$ is the color from a certain template in our SPS library.
In our case $N_{\rm colors}=2$, for instance, $(g-i)$ and $(i-K_{s})$, hence $n=1,2$.

$P(\Upsilon_{*})$ represents the previous knowledge we may have about the likely value
of the $\Upsilon_{*}$ parameter, and
\begin{equation}
  P(C) = \sum_{j=1}^{N_{\rm templates}} P(C \mid \Upsilon_{*j}) P(\Upsilon_{*j})
\end{equation}
is a normalization constant, also called the {\it {Bayesian evidence}}~\citep{sav07}.
$N_{\rm templates}$ stands for the number of templates in our SPS library.

\subsection{The Bayesian successive priors (BSP) algorithm}~\label{BSP_algo}

\subsubsection{The {\it{prior}} probability distribution function} 

In order to apply the BSP algorithm,
we have chosen a prior probability distribution function, $P(\Upsilon_{*})$,
of the form

\begin{equation}~\label{prior}
P(\Upsilon_{*}) =\exp \left(-\frac{1}{2}\left[\frac{\Upsilon_{*}^{\rm prior} - \Upsilon_{*}} 
{\sigma_{\Upsilon_{*}}} \right]^{2}\right),
\end{equation}

\noindent where

\begin{equation}
  \sigma_{\Upsilon_{*}} = \left[\frac{\ln(10)}{2.5}\right] \sigma_{\rm mag} \Upsilon_{*}^{\rm prior}.
\end{equation}

\noindent Here, $\sigma_{\rm mag}$ is the photometric error for a certain passband, 
which is related to $\sigma_{\rm col}$ in equation~\ref{chi2} through 
$\sqrt{2}\sigma_{\rm mag} \approx \sigma_{\rm col}$.

Each template in the SPS library corresponds to a single $\Upsilon_{*}$. 
By using equation~\ref{prior} and Bayes' theorem (equation~\ref{Bayes}),
we can effectively marginalize the templates from our SPS library,
as we will demonstrate in the following sections.

\subsubsection{Description of the BSP algorithm}

The BSP algorithm consists of three iterations that are described below. 
The algorithm is intended to work with a SPS library and surface
photometry in several/various bands. In the following we assume that these 
are the optical $g$ and $i$ bands, and the NIR $K_{s}$ filter.
For the library, we use SSAG-BC03 (although the algorithm is designed
to work independently of the choice of
SPS library). The mass-to-light ratio is taken in the $K_{s}$-band,
$\Upsilon_{*}^{K_{s}}$.
Other waveband combinations will be discussed later.
The algorithm is applied on a pixel-by-pixel basis, although
in each iteration all pixels are addressed before moving to the next iteration.

\begin{enumerate}

\item In the first iteration we use a uniform prior,
i.e., $P(\Upsilon_{*})$~=~constant,
and apply equation~\ref{Bayes}.
Then we calculate the absolute maximum (which should be near the median)
of the posterior probability distribution function
$P(\Upsilon_{*}^{K_{s}} \mid C)$, and the 16th and 84th percentiles,
to account for the corresponding error map.
We estimate the percentiles by progressively integrating the area under the posterior
probability curve until we accumulate an area of 0.16
and 0.84 (being the total area equal to 1), for the 16th and 84th percentiles,
respectively.\footnote{ These values are equivalent to $-1\sigma$ and $1\sigma$,
respectively, in a normal distribution.}

Up to this point the method provides a {\it maximum likelihood} estimate and 
is similar to the ZCR algorithm, with the only difference that the templates
are not binned in our case. We call the unbinned
version of the ZCR algorithm ZCR$^\prime$ from now on. 
We then use the results of this step for two purposes.
Firstly, we identify all the pixels for which the difference (absolute value)
between their observed color and the fitted template in the SPS library
is smaller than $3\sigma_{\rm col}$, i.e.,

\begin{equation}
 \left| \Delta C_{n} \right| =
 \left| C_{n}^{\rm obs}-C_{n}^{\rm template} \right| < 3\sigma_{\rm col},
\end{equation}

for $n=1,2$. The pixels that do not fulfill the $3\sigma_{\rm col}$
condition are isolated and flagged.\footnote{These include elements recording emission from AGN activity.}
This step guarantees that we keep only pixels that can be described by 
our SPS library.
Next, we take the resulting $\Upsilon_{*}^{K_{s}}$ values for all the kept pixels
and calculate the statistical median.\footnote{
The number separating the lower and higher value halves of $\Upsilon_{*}^{K_{s}}$.}

\item  In the second iteration this median value of $\Upsilon_{*}^{K_{s}}$, from
iteration number 1, is used as a constant parameter
in equation~\ref{prior}, i.e., 

\begin{equation}
 \Upsilon_{*}^{\rm prior}={\rm constant}
\end{equation}
for all pixels in the disk.\footnote{
A refinement of the method could be achieved by separating the bulge from the
disk of the galaxy, and treating them as objects with different median $\Upsilon_{*}^{K_{s}}$~\citep{por04}.}
The prior, $P(\Upsilon_{*})$, is not uniform in this case,
and adopts the functional form of equation~\ref{prior}.
Now we compute the maximum in $P(\Upsilon_{*}^{K_{s}} \mid C)$,
and the respective 16th and 84th percentiles.
Similarly to iteration number 1, we identify all the pixels where
the difference between the observed colors and the fitted library templates
is smaller than $\alpha\sigma_{\rm P}$, i.e.,

\begin{equation}
  \left| \Delta C_{n} \right| < \alpha\sigma_{\rm P},
\end{equation}

for $n=1,2$.
The value of $\sigma_{\rm P}$ is determined from the resulting $\Delta C_{n}$
(no absolute value) pixel distribution by calculating its 16th and 84th percentiles,
P$_{16}$ and P$_{84}$, respectively, and then using

\begin{equation}~\label{sigma_P}
 \sigma_{\rm P} = ( {\rm P}_{84} - {\rm P}_{16} ) / 2,
\end{equation}

for each color. After some tests (see Appendix~\ref{appA}), we have found that $\alpha=1.0$
is an adequate value that allows us to isolate the pixels that deviate significantly
from $\Delta C_{n}\sim0$.
In a hypothetical case, having $\Delta C_{n}=0$ would indicate that our observed colors
match perfectly the fitted library templates.
The $\left|\Delta C_{n}\right| < \alpha\sigma_{\rm P}$ pixels will be the ``backbone'' of our mass-map,
and represent the locations in the disk where
the $K_{s}$-band is a reliable tracer of the stellar mass surface density,
considering the $\Upsilon_{*}^{\rm prior}={\rm constant}$ condition.
The $\left|\Delta C_{n}\right| > \alpha\sigma_{\rm P}$ pixels
belong mainly to luminous red stars in the asymptotic giant branch, red supergiants,
low surface brightness regions in the outskirts of the disk, and high extinction regions
where $\Upsilon_{*}^{K_{s}}$ does not have the constant (median) value we assumed earlier.
We then need to provide a new $\Upsilon_{*}^{K_{s}}$ value for these
$\left|\Delta C_{n}\right| > \alpha\sigma_{\rm P}$ pixels.
For this purpose we use the information from the ``backbone'' pixels.
We interpolate the stellar mass surface density to fill the places
where we need a new $\Upsilon_{*}^{K_{s}}$ value. The interpolation is
done in the $0\degr$, $45\degr$, $90\degr$, and $135\degr$ directions,
and then an average is taken.
After the interpolation, we visually inspect the resulting
maps to determine whether a minor smoothing is needed.
The smoothing is only applied to the $\left|\Delta C_{n}\right| > \alpha\sigma_{\rm P}$ pixels,
and is performed by replacing each pixel value with
the average of the neighboring pixels.
There are other interpolation techniques that could be used~\citep[see, e.g.,][]{gum13},
but for the present work we will apply the above mentioned procedure to all objects. 
Having established this, the new $\Upsilon_{*}^{K_{s}}$ values
are estimated as the ratio of the interpolated mass-map
and the observed $K_{s}$ photometry.

\item The third and last iteration is intended to deal
only with the $\left|\Delta C_{n}\right| > \alpha\sigma_{\rm P}$ pixels, identified in
iteration number 2. For each pixel, we use the $\Upsilon_{*}^{K_{s}}$ value
also estimated in iteration number 2 to represent
$\Upsilon_{*}^{\rm prior}$ in equation~\ref{prior},
and calculate the absolute maximum of the posterior probability distribution in equation~\ref{Bayes}.
Before this, we may also update the uncertainty in $\Upsilon_{*}$, in equation~\ref{prior};
such uncertainty now reads
\begin{equation}
  \sigma_{\Upsilon_{*}} = \sqrt{\left(\left[\frac{\ln(10)}{2.5}\right] \sigma_{\rm mag} \Upsilon_{*}^{\rm prior}\right)^{2}
                                + \beta^{2}},
\end{equation}

\noindent where $\beta$ accounts for the propagation of uncertainties arising from
the previous iteration (e.g., the mass surface density interpolation from neighboring pixels).
Using bootstrap methods we have estimated that $\beta\approx0.6\%$.

From the resulting $\Upsilon_{*}^{K_{s}}$ map we then obtain 
the stellar mass surface density to complete our mass-map.

\end{enumerate}

As an optional last step, the flagged pixels from iteration number 1
that belong to the inner disk can be interpolated
in mass with the information about the surrounding pixels provided
by all three iterations. For the external disk pixels,
the interpolation is more uncertain.

\begin{figure}
\centering
\epsscale{1.0}
\plotone{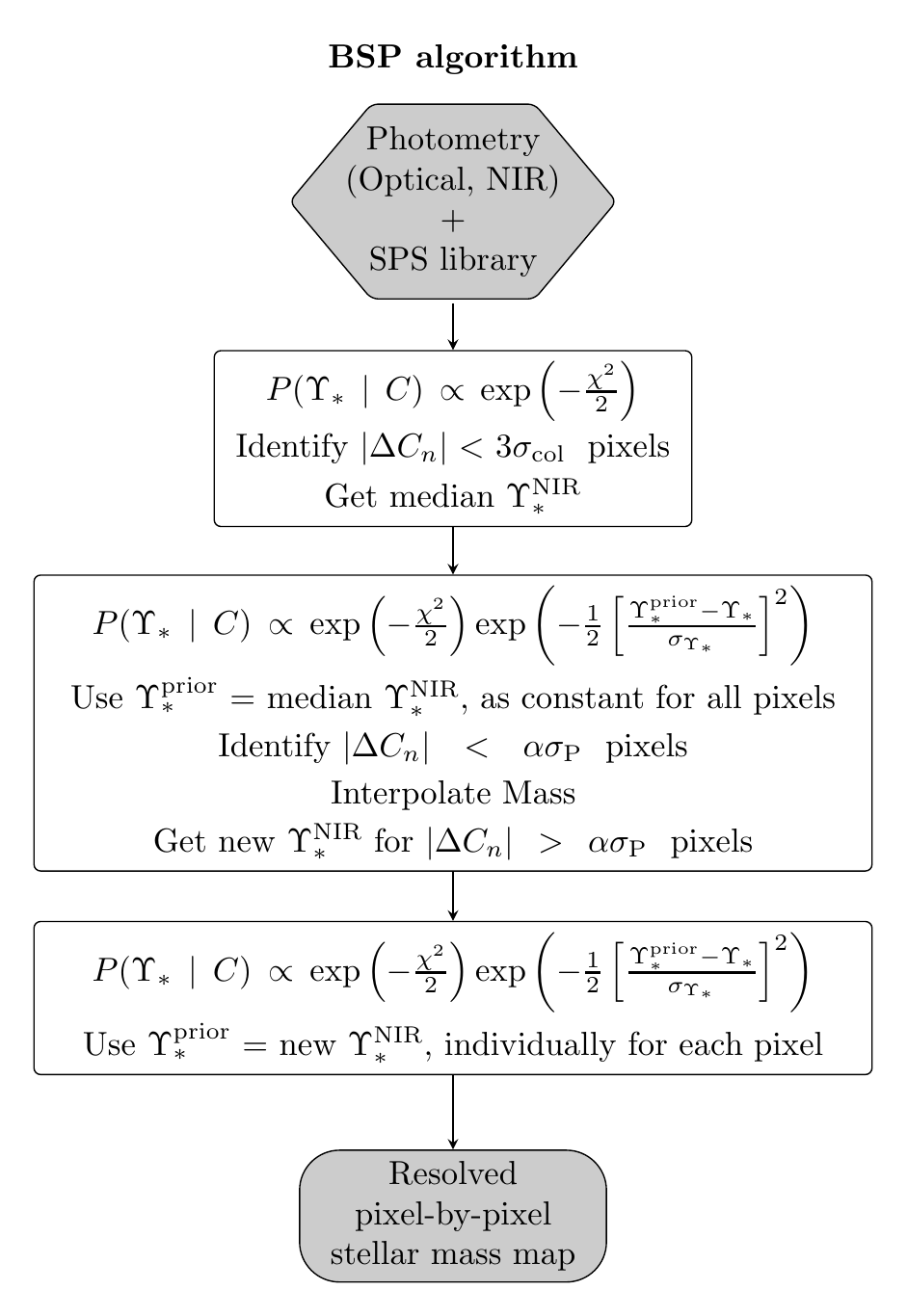}
\caption[f10]{Bayesian successive priors (BSP) flowchart. Photometry in some optical bands
(e.g., SDSS $g$ and $i$) and one NIR band (e.g., $K_{s}$), as well as a SPS library,
are required.
$P(\Upsilon_{*} \mid C)$ denotes the posterior mass-to-light ratio probability distribution;
$\Delta C_{n}$ is the difference between the observed $n_{\rm th}$ color
of a pixel, and the color of the fitted template in the SPS library;
$\sigma_{\rm col}$ is the photometric error in the $n_{\rm th}$ color;
$\sigma_{\rm P}$ is determined from the $\Delta C_{n}$ pixel distribution in iteration number 2
(see equation~\ref{sigma_P}).
Each one of the three rectangular white boxes stands for 
an iteration of the BSP algorithm (see text).
~\label{fig10}}
\end{figure}

We find that adding more iterations does not lead to
any further improvement in the mass-maps.
The flowchart of the BSP algorithm is shown in Figure~\ref{fig10}.

For the BSP algorithm to work properly, the requirement of NIR data
with high S/N ratio is essential; otherwise, any noisy and patchy features
will be transferred to the mass-map. A minimum S/N ratio of $\sim10-20$
in the outskirts of the disk is necessary.
This level can be achieved with techniques as the one
used by the {\tt{Adaptsmooth}} code, or alternatively with
Voronoi two-dimensional binning~\citep{cap03}.

In this investigation we have adopted only two colors,
$(g-i)$ and $(i-K_{s})$, and thus $N_{\rm colors}=2$.
The benefits of using the $g$ and $i$ SDSS data together with one NIR band
are an excellent spatial resolution per element (pixel), and extensive spatial coverage
(of the entire object).
Nevertheless, the BSP algorithm can also be applied by using $N_{\rm colors}>2$, with the
only requirement of the inclusion of one NIR band as described earlier.
In a separate publication we will explore the use of the algorithm to fit optical IFU observations,
for instance, the Calar Alto Legacy Integral Field Area survey~\citep[CALIFA,][]{san12},
and the Mapping Nearby Galaxies at Apache Point Observatory survey~\citep[MaNGA,][]{bun15}.

\section{Application of BSP to M~51}~\label{M51_BSP}

We apply the BSP algorithm to M~51 employing the same data described in section~\ref{m51_maxL}.
We calculate $\sigma_{\rm mag}$ on a pixel-by-pixel basis assuming that

\begin{equation}
  \sigma_{\rm mag} \approx \sqrt{ \sigma^2_{\rm flux} + \sigma^2_{\rm calib} },
\end{equation}

\noindent where $\sigma_{\rm flux}$ is the random error in the flux per pixel,
which we assume to be dominated by the uncertainty in the background~\citep[see also,][]{men12},
and $\sigma_{\rm calib}$ is the calibration uncertainty,
or zero point error, for which we assume $\sigma_{\rm calib}\sim0.01$ mag
for the SDSS images, and $\sigma_{\rm calib}\sim0.03$ mag for the $K_{s}$ image~\citep{jar03}.
We compute $\sigma_{\rm flux}$ in mag by using
$\sigma_{\rm flux} = 1.085736 * \frac{\sigma_{\rm back}}{{\rm flux}}$,
where $\sigma_{\rm back}$ is the standard deviation in the background
(in a sky-subtracted image). We compute $\sigma_{\rm back}$ by sampling
the background statistics in different boxes near the edges of the images.
To account for the use of the {\tt{Adaptsmooth}} procedure we divide
$\sigma_{\rm back}$ by $\sqrt{n_{\rm pix}}$, where $n_{\rm pix}$ is
the number of pixels used to increase the S/N of the
corresponding pixel by {\tt{Adaptsmooth}}.

Without taking into account correlation between
bands, we compute $\sigma_{\rm col}$ by summing in quadrature
the $\sigma_{\rm mag}$ values of each band involved in the color
determination.

\begin{figure*}
\centering
\epsscale{1.0}
\plotone{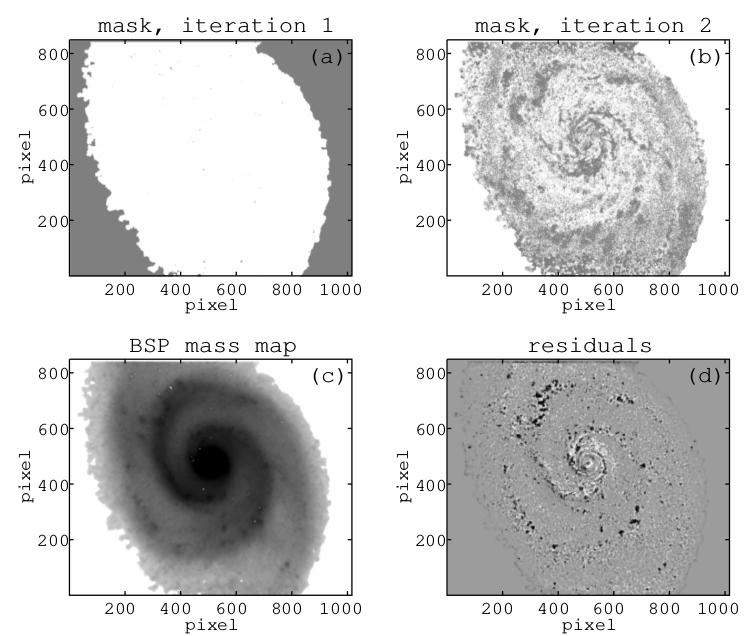}
\caption[f11]{
Application of the BSP algorithm to the spiral galaxy M~51.
The Monte Carlo SPS library used is SSAG-BC03.
{\it{Top left}}: resulting mask after iteration number 1. White regions
have observed colors within 3 $\sigma$ of at least one template in the library.
{\it{Top right}}: resulting mask after iteration number 2. Gray regions
represent pixels where the assumption of a constant $\Upsilon_{*}^{K_{s}}$ for the whole disk
is not fulfilled by the observed colors.
{\it{Bottom left}}: resulting mass-map after iteration number 3.
{\it{Bottom right}}: residuals after subtracting the mass-map obtained at the end of the BSP algorithm (iteration 3),
from a mass-map that assumes a constant $\Upsilon_{*}^{K_{s}}$ (the median after iteration 1).
Dark/white regions represent positive/negative mass differences. 
~\label{fig11}}
\end{figure*}

\begin{figure*}
\centering
\epsscale{1.0}
\plotone{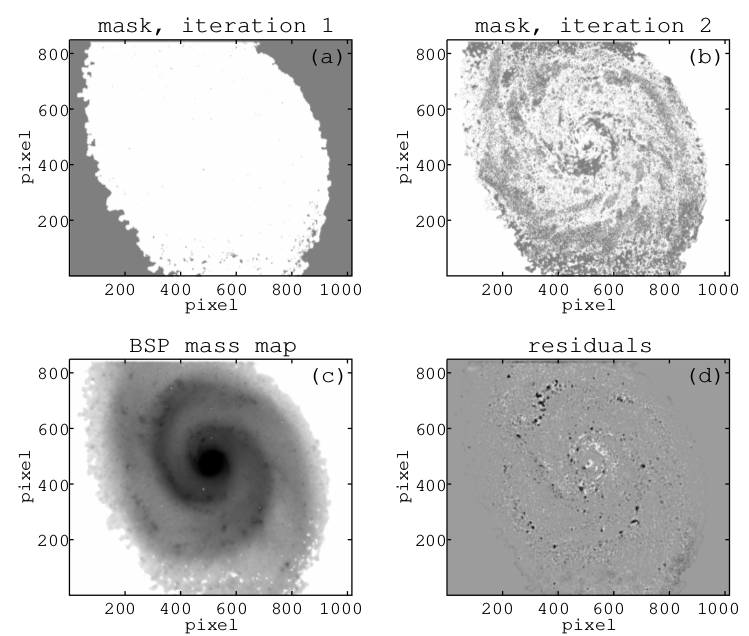}
\caption[f12]{Like Figure~\ref{fig11}, but for the MAGPHYS-CB07 Monte Carlo SPS library.
~\label{fig12}}
\end{figure*}

In Figures~\ref{fig11} and~\ref{fig12}, we show the results
of adopting the SSAG-BC03 and MAGPHYS-CB07 libraries, respectively.
In both figures, the top left panels (a) show the mask obtained after iteration number 1.
White regions represent the pixels where the observed colors
are within $3\sigma$ of at least one SPS-library template (see Figures~\ref{fig4} or~\ref{fig6}).
In the respective top right panels (b), we show the masks obtained after iteration number 2.
For these masks, the gray regions represent the pixels where the color difference (absolute value) 
between the models and the observations, $\left|\Delta C_{n}\right|$, is greater than $\alpha\sigma_{\rm P}$,
with $\alpha=1$ (see section~\ref{BSP_algo}, and Appendix~\ref{appA}),
assuming a constant $\Upsilon_{*}^{K_{s}}$. These regions will
be interpolated in mass with the information of neighboring pixels.
We can also appreciate that the SSAG-BC03 library does a better job at modeling the outskirts
of the disk than the MAGPHYS-CB07 library. 
To investigate the cause of this behavior we obtain a mass map
by using MAGPHYS-BC03.
We obtain very similar masks to those from the SSAG-BC03 library
(Figure~\ref{fig11}, top panels).
With this in mind, most of the differences between BC03
and CB07 mass-maps in our results are mainly due to the distinct treatments
of the TP-AGB stage. To a lesser extent, we also notice an improvement when
SSAG-BC03 is used, instead of MAGPHYS-BC03. We attribute this to the fact
that SSAG covers a wider range of possible star formation histories. 

In the bottom left panels (c) of Figures~\ref{fig11} and~\ref{fig12}, we show the resulting
stellar mass surface density map after
iteration number 3. The filamentary structure is no longer
present, and the maps show greater resemblance to the features in NIR bands,
as expected. Finally, the bottom right panels (d) of both figures
show the ``residuals''; these are the result of subtracting the final output (iteration 3) mass-map using BSP,
from a mass-map that assumes a constant $\Upsilon_{*}^{K_{s}}$ (the median $\Upsilon_{*}$ after iteration number 1).
The dark/white regions represent positive/negative mass differences, i.e.,
where $\Upsilon_{*}$ has been overestimated/underestimated.
For example, the $\Upsilon_{*}$ may be overestimated when young luminous red stars are
mixed with older populations,
and underestimated due to extinction in the NIR bands. 
This is different from the ``outshining bias''~\citep{mara10,sor15},
where the light from young stars eclipses the old population
and the amount of stellar mass is underestimated.
In our case we overestimate the mass (by using a constant $\Upsilon_{*}^{K_{s}}$)
because we are assuming, mistakenly but for convenience, that all the light comes from old stars.

\subsection{Isolating the old massive disk}

\begin{figure}
\centering
\epsscale{1.0}
\plotone{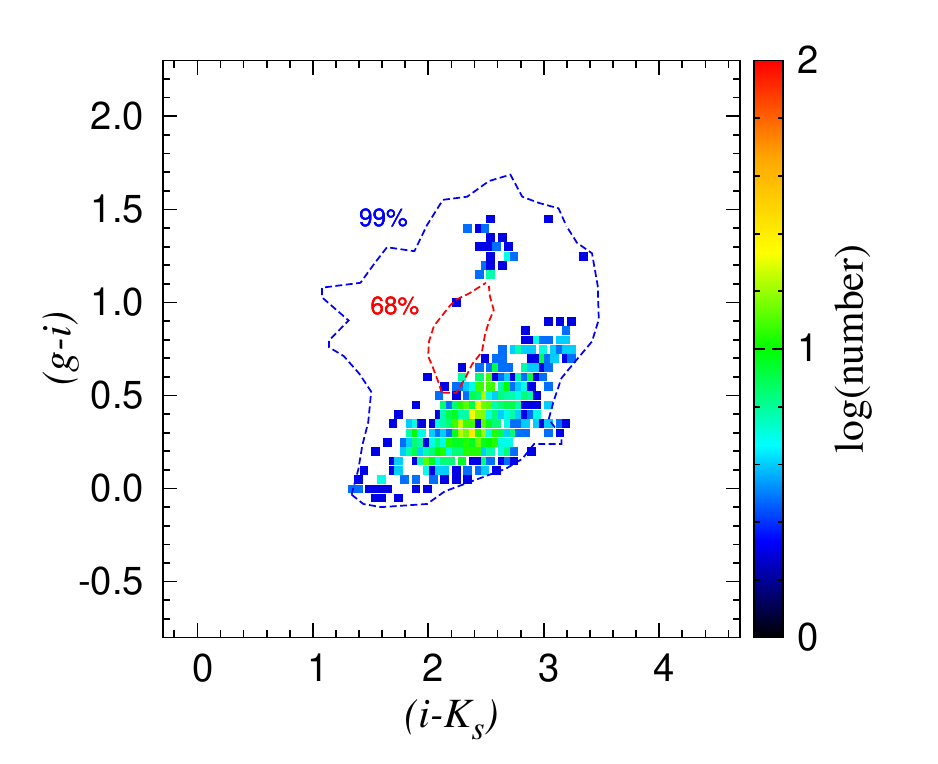}
\caption[f13]{2-D histogram of the observed ``positive mass differences''
in the residuals (see text), after applying BSP to M~51 (dark regions in Figure~\ref{fig12}d, bottom right panel; see text).
The blue/red dashed contour in the plot delimits $99\%$/68\% of all the observed colors (see Figure~\ref{fig4}, left panel).
~\label{fig13}}
\end{figure}

We will now discuss in more detail the positive mass differences in the residuals.
In Figure~\ref{fig13} we plot a 2-D histogram of the colors of the pixels for which the
mass difference is $>2\times 10^4~M_{\sun}$. This cut in the mass was chosen
in order to isolate most of the positive residuals near the spiral arms.
We have excluded the pixels from the bulge region.
We note that most points gather in a group 
with a maximum near $(i-K_{s})\sim2.4$ and $(g-i)\sim0.4$.
Their $(g-i)$ color is relatively blue when compared with all the colors observed
(delimited by the blue dashed contour).
We also note a cluster of points with redder colors, near
$(i-K_{s})\sim2.5$ and $(g-i)\sim1.3$. These pixels mainly correspond
to point sources outside the spiral arms.

In Figure~\ref{fig14} we show the marginalized probability distributions (see Appendix~\ref{appB})
for the $r$-band light-weighted age and for $\Upsilon_{*}^{K_{s}}$,
obtained for M~51 using the MAGPHYS-CB07 library.
The dashed-dotted green line corresponds to the, previously described,
``positive mass differences'' in the residuals, while the blue solid line refers to the
whole disk, both results after BSP.
Interestingly, the excess mass regions are younger (age~$\sim1$~Gyr)
and have a lower $\Upsilon_{*}^{K_{s}}$ (by~$30\%$) than most of the pixels in the disk.
Together with the bluer $(g-i)$ color, the above characteristics
indicate that these regions contain relatively young stars,
that mix with the old stellar population in star forming regions.
These were effectively isolated by BSP!

\begin{figure}
\centering
\includegraphics[angle=-90,scale=0.3]{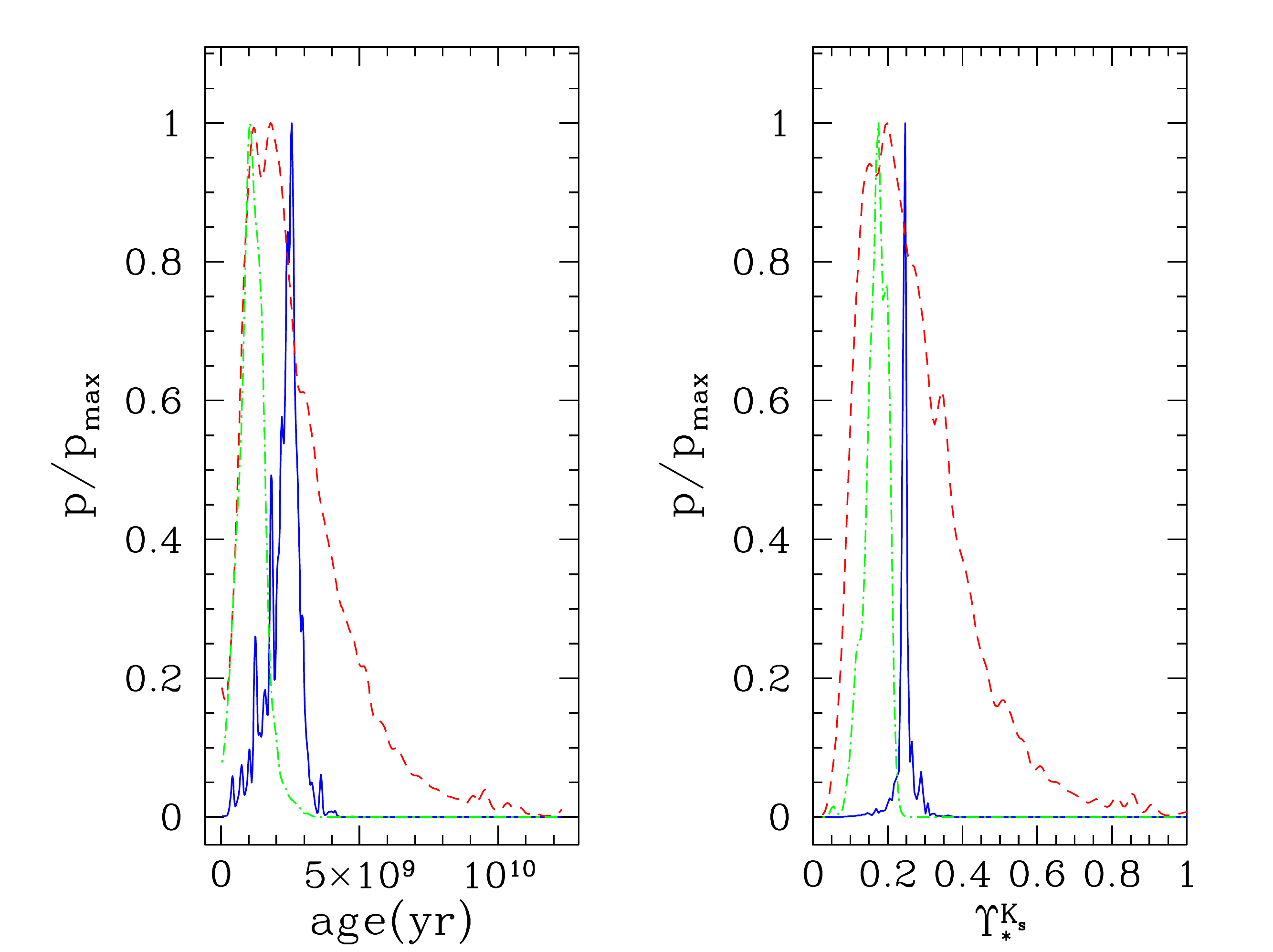}
\caption[f14]{
M 51 probability ($p$) distributions with the
MAGPHYS-CB07 library.
{\it Left:} $r$-band light-weighted age (yr);
{\it right:} mass-to-light ratio $\Upsilon_{*}^{K_{s}}$. 
{\it Green dashed-dotted line:} BSP observed ``positive mass differences'' in the residuals (see text);
{\it blue solid line:} BSP results for the whole disk;
{\it red dashed line:} ZCR$^\prime$ output for the entire disk.
~\label{fig14} }
\end{figure}

The red dashed line in Figure~\ref{fig14} shows the probability
distributions for the whole disk after applying the ZCR$^\prime$ approach.
The light-weighted age yields a larger fraction of younger pixels with ZCR$^\prime$.
As expected from our previous assumptions, the values of $\Upsilon_{*}^{K_{s}}$
are more narrowly confined with BSP, around $\Upsilon_{*}^{K_{s}}=0.2450\pm0.0242$.
This value is dominated by red giant branch stars.

Regarding the output SSAG-BC03 estimates of $\Upsilon_{*}^{K_{s}}$ for the whole disk,
we recover a median $\Upsilon_{*}^{K_{s}}=0.4232$ after BSP iteration number 1.
After iteration number 3 the mean value for the entire disk is $\Upsilon_{*}^{K_{s}}=0.4247\pm0.0386$.
For the $\left|\Delta C_{n}\right| < \alpha\sigma_{\rm P}$ pixels we have $\Upsilon_{*}^{K_{s}}=0.4231\pm0.0034$,
while for the $\left|\Delta C_{n}\right| > \alpha\sigma_{\rm P}$ pixels we obtain $\Upsilon_{*}^{K_{s}}=0.4264\pm0.0556$,
both results after BSP.

Our estimation for $\Upsilon_{*}^{K_{s}}$, derived with MAGPHYS-CB07 and SSAG-BC03, are consistent (within $3.0\sigma$)
with the result derived by~\citet{jus15} for the solar cylinder from star counts ($\Upsilon_{*}^{K_{s}}=0.34$),
and with the average found by~\citet{mts13} for a sample of 30 disk galaxies ($\Upsilon_{*}^{K_{s}}=0.31$).

\subsection{Integrated mass estimates}

\begin{figure}
\centering
\epsscale{1.0}
\plotone{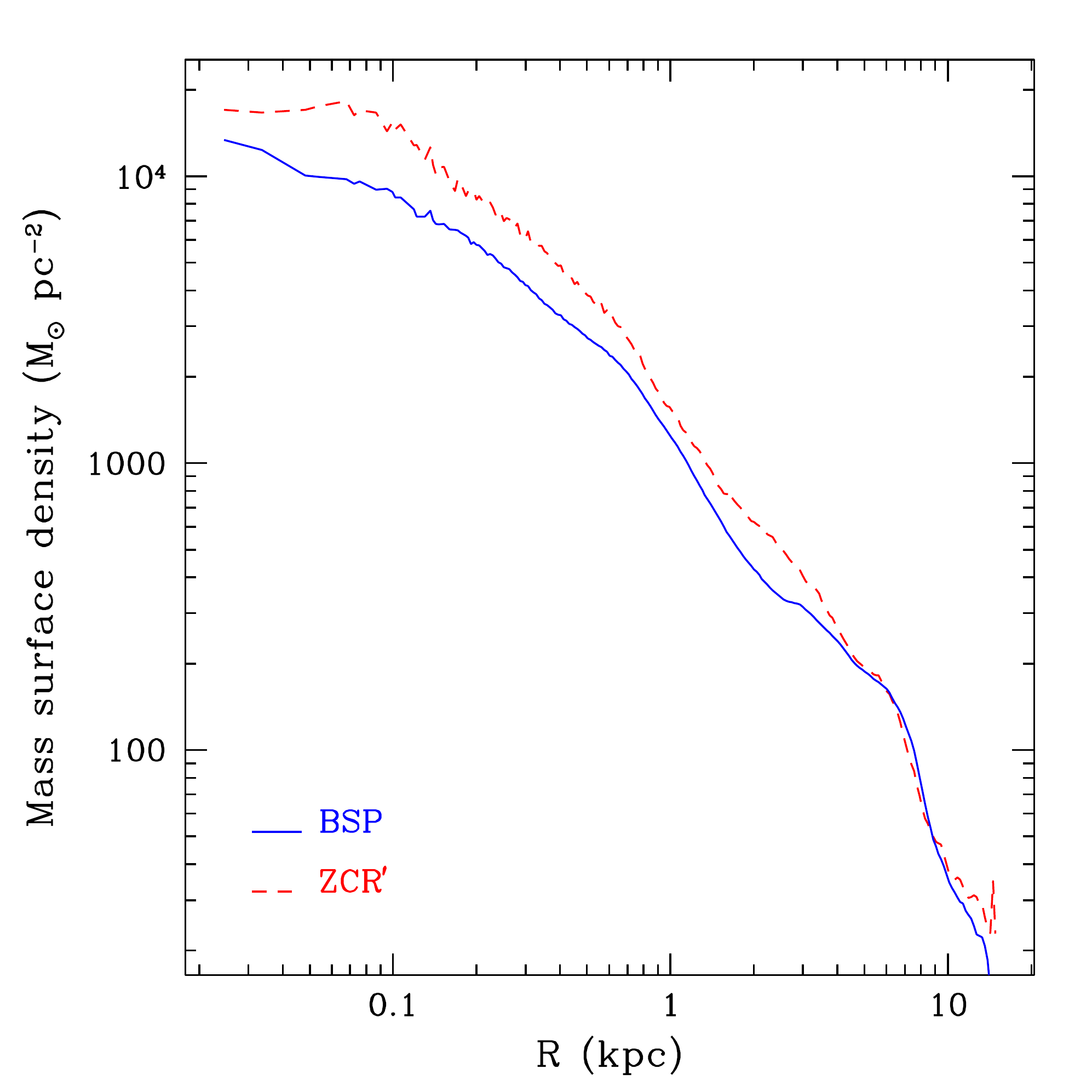}
\caption[f15]{
Azimuthally averaged mass surface density ($M_{\sun}$ pc$^{-2}$) vs.\ radius, $R$ (kpc), 
for M~51 with the SSAG-BC03 library.
Results are for deprojected mass-maps.
{\it Blue solid line:} BSP; {\it red dashed line:} ZCR$^\prime$.
~\label{fig15}}
\end{figure}

\begin{figure*}
\centering
\epsscale{1.0}
\plotone{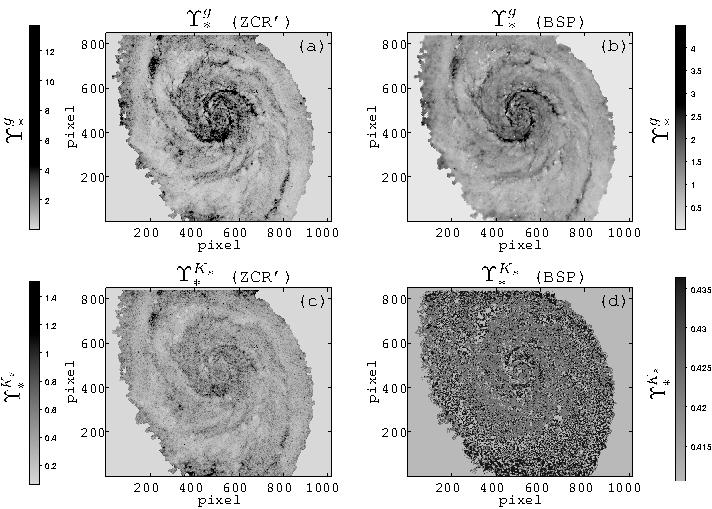}
\caption[f16]{
$\Upsilon_{*}$ maps for M~51 with the SSAG-BC03 library.
{\it{Top left}}: $\Upsilon_{*}^{g}$ obtained with ZCR$^\prime$ method.
{\it{Top right}}: $\Upsilon_{*}^{g}$ with BSP algorithm.
{\it{Bottom left}}: $\Upsilon_{*}^{K_{s}}$, ZCR$^\prime$ method.
{\it{Bottom right}}: $\Upsilon_{*}^{K_{s}}$, BSP algorithm.
Darker pixels indicate higher $\Upsilon_{*}$.
~\label{fig16}}
\end{figure*}

\begin{figure}
\centering
\epsscale{1.0}
\plotone{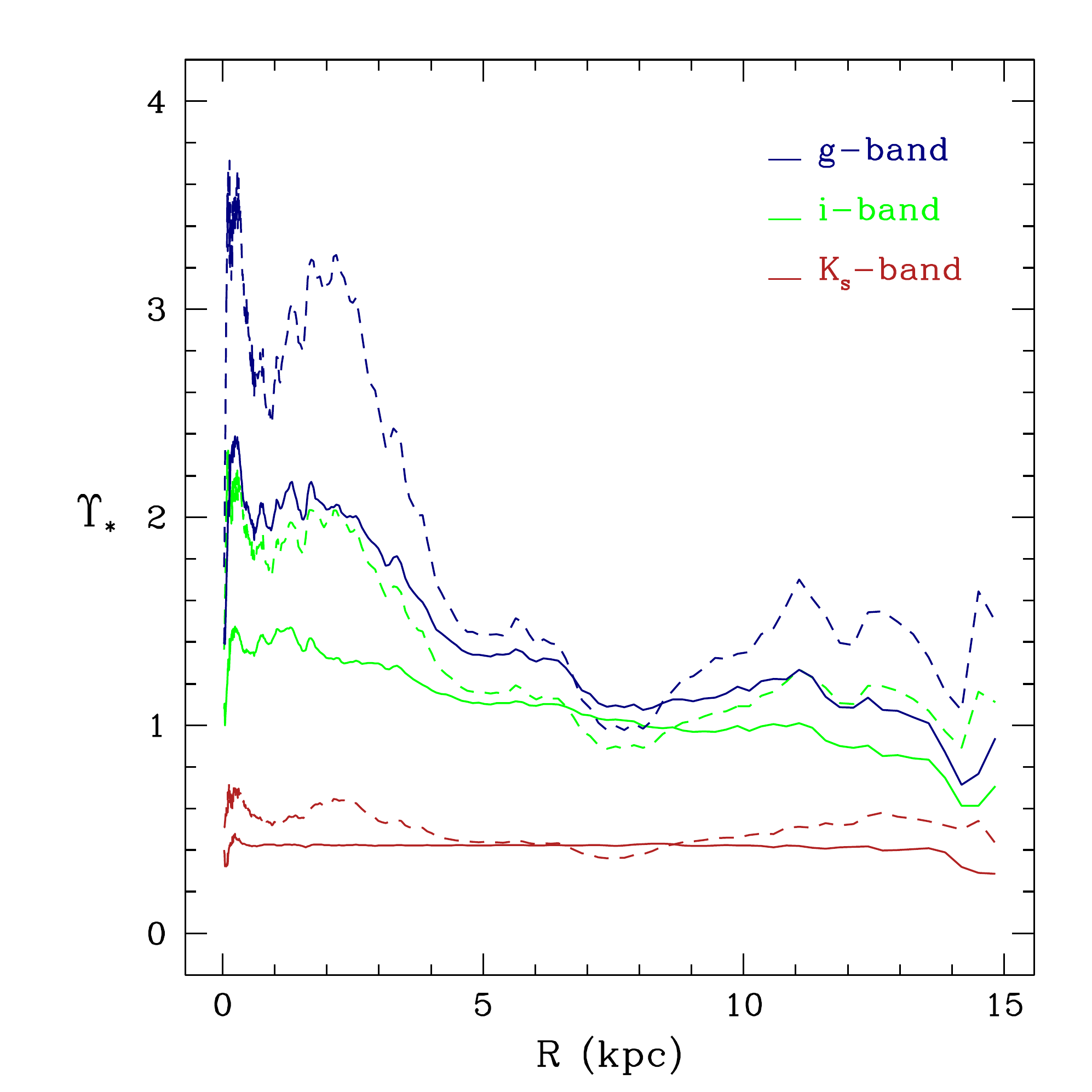}
\caption[f17]{
Azimuthally averaged $\Upsilon_{*}$ as a function of radius, $R$ (kpc).
{\it Solid lines:} BSP; {\it dashed lines:} ZCR$^\prime$.
{\it Dark blue:} $g$-band; {\it green:} $i$-band; {\it dark red:} $K_s$-band. 
Results are for deprojected maps of M~51 with the SSAG-BC03 library.
~\label{fig17}}
\end{figure}

With respect to the total resolved mass, defined as
\begin{equation}~\label{mass_reso}
 M^{\rm resolved}_{*} = \sum\limits_{j}\sum\limits_{i} M_{*ij},
\end{equation}

\noindent where $M_{*ij}$ is the stellar mass 
of the $i^{\rm th},j^{\rm th}$ pixel, we find the following results.
By using the MAGPHYS-CB07 library we obtain for M~51 a total stellar mass
of $M_{*}^{\rm resolved} =3.84\times10^{10}M_{\sun}$ with ZCR$^\prime$,
and $M_{*}^{\rm resolved} =3.22\times10^{10}M_{\sun}$ with BSP.
The SSAG-BC03 library, meanwhile, leads to $M_{*}^{\rm resolved}=6.43\times10^{10}M_{\sun}$ 
with ZCR$^\prime$, and $M_{*}^{\rm resolved}=5.56\times10^{10}M_{\sun}$ with BSP.
The discrepancy between the SSAG-BC03 and MAGPHYS-CB07 mass estimates is mainly due
to the different treatments of the TP-AGB phase~\citep{bru07}.
In Figure~\ref{fig15}, we show the azimuthally averaged surface mass
density vs.\ radius for M~51 obtained with SSAG-BC03. For most of the disk, the
BSP method yields smaller mass estimates than ZCR$^\prime$,
resulting in a $\sim~10\%$ decrease in the total mass.
To complement the analysis, we show in Figures~\ref{fig16}a and ~\ref{fig16}b (top left and top right panels)
the $\Upsilon_{*}^{g}$ maps obtained with the ZCR$^\prime$ method and the BSP algorithm, respectively.
Figures~\ref{fig16}c and~\ref{fig16}d (bottom left and bottom right panels)
present the $\Upsilon_{*}^{K_{s}}$ maps from ZCR$^\prime$ and BSP, respectively.
Figure~\ref{fig17} shows the azimuthally averaged $\Upsilon_{*}$
for the $g$, $i$, and $K_{s}$ bands, as a function of radius.
As expected, the $K_{s}$ profile is virtually constant,
while the $g$ and $i$ profiles show variations with radius,
with lower values at the outskirts of the disk,
as a result of a lower surface brightness and bluer colors~\citep{deJ96,bel01}.

\begin{figure}
\centering
\epsscale{1.0}
\plotone{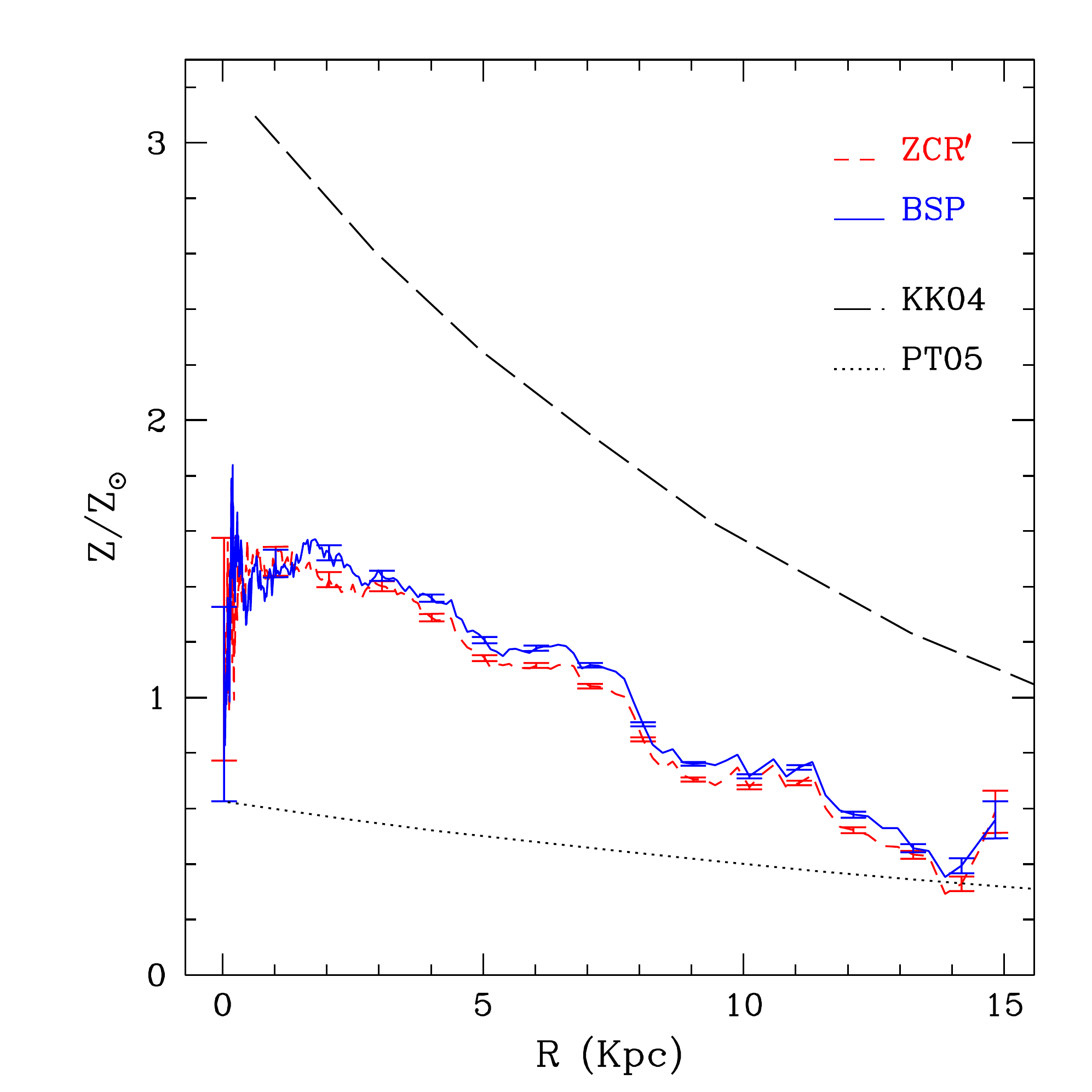}
\caption[f18]{
Azimuthally averaged stellar metallicity, $Z/Z_{\sun}$.
{\it Blue solid line:} BSP; {\it red dashed line:} ZCR$^\prime$.
Results are for deprojected maps of M~51 with the SSAG-BC03 library.
For comparison we show the metallicity abundance gradients of~\citet{mou10},
using the KK04 (black dashed line) and PT05 (black dotted line) calibrations.
~\label{fig18}}
\end{figure}

In Figure~\ref{fig18}, we show the 
azimuthally averaged stellar metallicity, $Z/Z_{\sun}$;
similar results are obtained for both BSP and ZCR$^\prime$.
In this figure we also plot the metallicity abundance gradients for M~51
from~\citet{mou10}. From ancillary data,~\citet{mou10} estimate
radial oxygen abundance gradients for 75 galaxies in the
Spitzer Infrared Nearby Galaxies Survey~\citep[SINGS,][]{ken03},
using both the Kobulnicky \& Kewley (2004; KK04) and the Pilyugin \& Thuan (2005; PT05) calibrations.
We transform~\citet{mou10} oxygen abundance gradients in units of $12+\log(\rm O/H)$,
to units of $Z/Z_{\sun}$, adopting~\citep[e.g.,][]{mart09}
\begin{equation}
   \log(Z/Z_{\sun}) \simeq 3.12 + \log(\rm O/H).
\end{equation}
The stellar metallicity we recover with SSAG-BC03 falls between
the two curves of~\citet{mou10}.~\citet{men12} obtain a similar result for the
Whirlpool galaxy, from optical and infrared photometry.

\subsection{Other filter combinations}

In this section we discuss the application of the BSP algorithm with 
other filter combinations. By using only optical filters, e.g.,
the ($g-i$) color and $\Upsilon_{*}^{i}$ , the method is not able to recover a spatial
structure consistent with the one obtained with optical-NIR combinations. 
This is due to the fact that the information of the prior spatial
structure is missing, as it can only be provided by the NIR bands.
The $\Upsilon_{*}^{i}$ cannot be assumed to be constant through the entire disk
(see Figure~\ref{fig17}); besides, dust lanes can still be noticed near spiral arms,
even at the redder optical wavelengths (see Figure~\ref{fig3}).

For the case when the $u$ filter is included, we were unable
to fit the data satisfactorily.
We have quantified the mean S/N ratio of the imaging data for the entire disk of M~51
(without applying the {\tt{Adaptsmooth}} procedure),
and obtain a value of $2.8$, $23.2$, and $29.6$ for the $u$, $g$, and $i$ bands respectively.
Taking this into account we can deduce that the issues we encounter when trying to
fit the $u$-band SDSS data with our methods are mainly due to their low S/N ratio.
This shortcoming can be remedied with deeper data.
We should also mention that $\Upsilon_{*}$ is more degenerate at shorter wavelengths.

\begin{figure*}
\centering
\epsscale{1.0}
\plotone{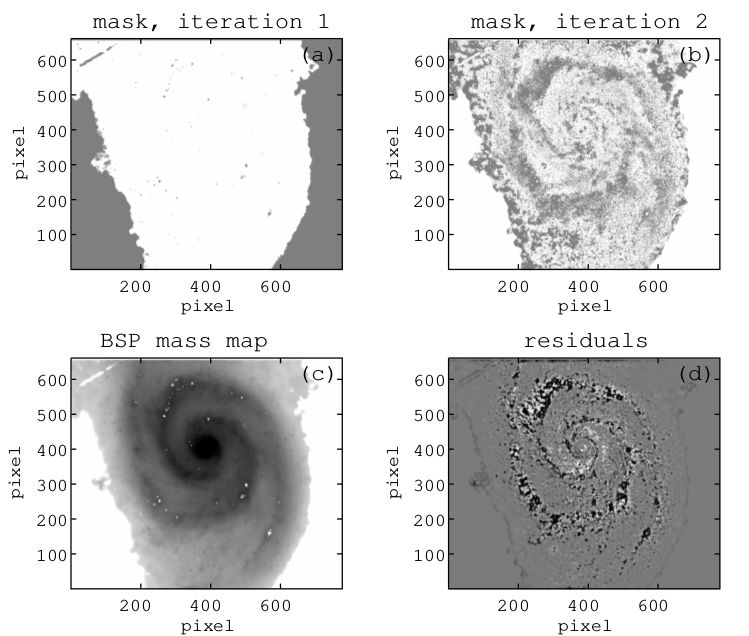}
\caption[f19]{
Application of BSP algorithm to M~51 with ($g-i$), $(i-3.6\micron)$, $\Upsilon_{*}^{3.6\micron}$,
and the MAGPHYS-BC03 SPS library.
Panels organized as in Figures~\ref{fig11} and~\ref{fig12}.
~\label{fig19}}
\end{figure*}

\begin{figure}
\centering
\epsscale{1.0}
\plotone{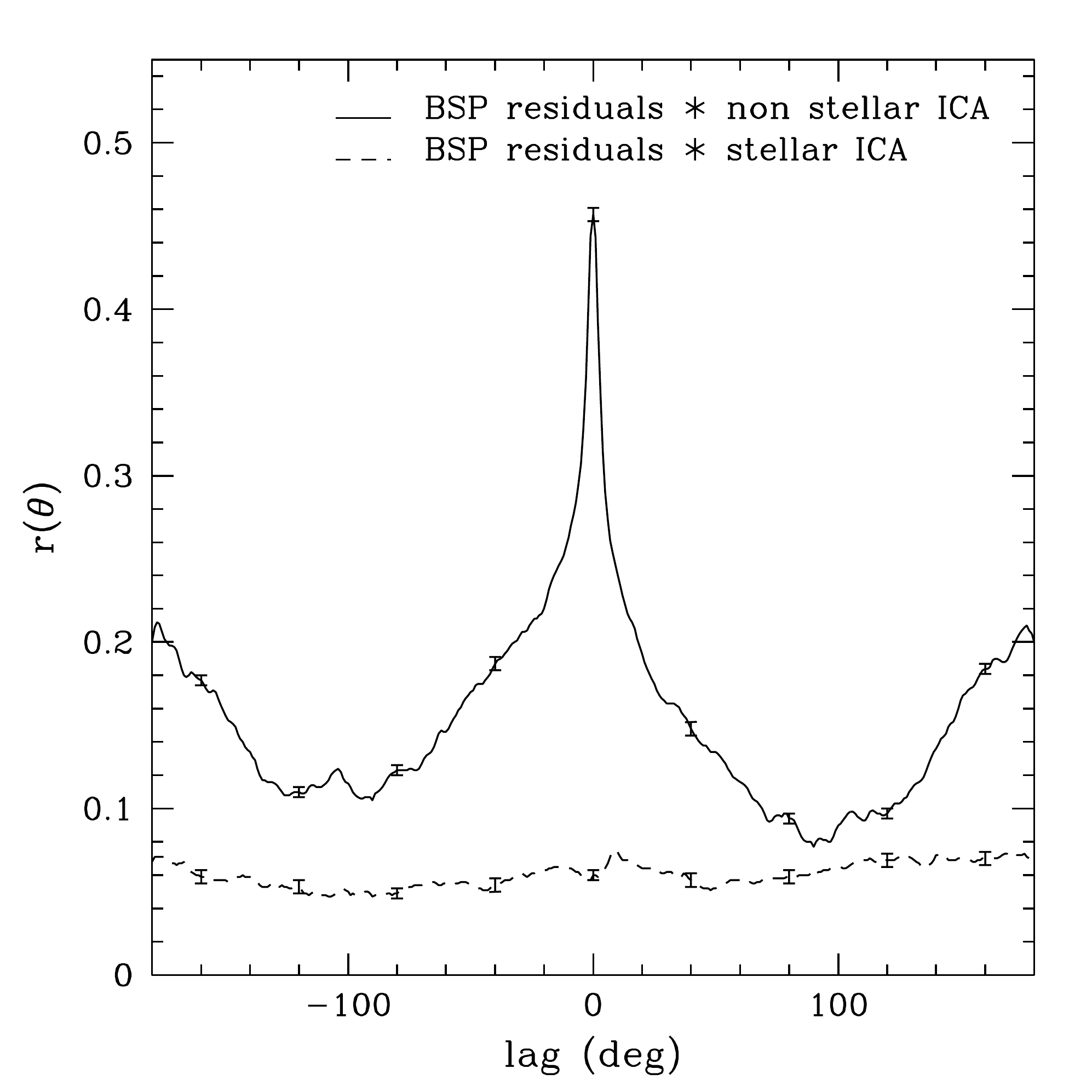}
\caption[f20]{
Cross-correlation functions (see text) for BSP residuals using the 3.6\micron-band.
{\it Solid line:} with the  non-stellar (dust) emission from ICA~\citep{mei12,mei14,que15}; 
the absolute maximum is at $\theta=0\degr$. {\it Dashed line:} 
with the stellar emission from ICA.
Height of error bars is $2\sigma_{cc}$.
~\label{fig20}}
\end{figure}

We also applied the BSP algorithm including the Spitzer-IRAC $3.6\micron$ band.
We used the colors ($g-i$) and $(i-3.6\micron)$, and $\Upsilon_{*}^{3.6\micron}$.
We computed pixel-by-pixel $\sigma_{\rm mag}$ errors as in section~\ref{M51_BSP}, assuming
$\sigma_{\rm calib}\sim0.01$ mag for the SDSS images, and $\sigma_{\rm calib}\sim0.03$ mag
for the $3.6\micron$ band~\citep{rea05}.
We corrected for Galactic extinction as in~\citet{schl11}, and~\citet{chap09}.
The results with the MAGPHYS-BC03 library are shown in Figure~\ref{fig19}. 
It can be noticed that the residuals, i.e., the difference between a mass-map
that assumes a constant $\Upsilon_{*}^{3.6\micron}$ and the output mass-map from BSP 
(Figure~\ref{fig19}d, bottom right panel),
are significantly different from the ones obtained when using the $K_{s}$-band
(see Figures~\ref{fig11}d and~\ref{fig12}d, bottom right panels).
We attribute this to polycyclic aromatic hydrocarbons (PAHs)
and continuum dust emission at $3.6\micron$.
To corroborate this we compare our result to the one derived through
the Independent Component Analysis (ICA) method of~\citet{mei12,mei14}.
This method separates the stellar emission from the dust emission;
\citet{que15} applied it to the Spitzer Survey of Stellar Structure in
Galaxies~\citep[S$^4$G,][]{she10}. We compare quantitatively the
residuals from BSP with the non-stellar (dust) component from ICA for M~51,
by following the same cross-correlation procedure as in section~\ref{m51_maxL}
(equation~\ref{crossco}). The results of this test are shown in Figure~\ref{fig20}.
We find that there is a strong spatial correlation between the ICA dust component
and the BSP residuals, indicated by the sharp peak at~$\theta=0$ in Figure~\ref{fig20}.
We also compare the BSP residuals to the stellar component obtained by ICA,
and find no spatial correlation at~$\theta=0$.
Although our adopted SPS library does not include the emission from dust
in the $3.6\micron$ band\footnote{In principle the emission from dust could be
included because it is predicted by MAGPHYS. Nevertheless,
the number of templates increases from $5\times10^{4}$ to $\sim6.67\times10^{8}$,
and CPU time would be $\sim1\times10^{4}$ times larger.},
the BSP algorithm was able to isolate much of it,
together with that of red luminous young stars.

A discussion of the differences between ICA and BSP would require further analysis
and comparisons using a larger sample of galaxies.
This goes beyond the scope of the present work, and will be investigated in a
separate publication.

\section{Pilot test with other galaxies}~\label{pilot_sample}

\begin{figure}
\centering
\epsscale{1.0}
\plotone{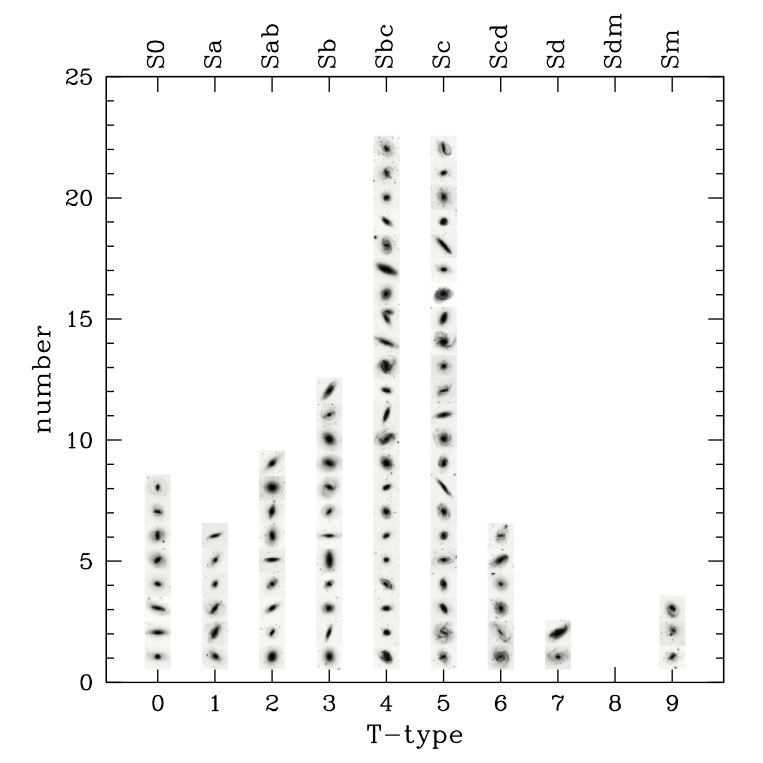}
\caption[f21]{
Bar chart of Hubble types for our galaxy sample (90 objects).
Embedded images from the Digitized Sky Survey, DSS (blue).
~\label{fig21}}
\end{figure}

In order to better understand the differences between
using the BSP algorithm of section~\ref{BSP_algo}, and adopting
the ZCR$^\prime$ method (i.e., a maximum likelihood estimate) to obtain
resolved maps of stellar mass, we analyzed 90 objects with $H$-band imaging
from the Ohio State University Bright Spiral Galaxy Survey~\citep[OSUBSGS,][]{esk02}.
The main statistical results from this sample should hold for other
surveys, such as SINGS and S$^4$G.
Our sample comprises all objects in the OSUBSGS for which SDSS $g$ and $i$ data
are available (see Table~\ref{tbl-1}).
A bar chart of the Hubble types of our OSUBSGS sample is shown in Figure~\ref{fig21}.
We subtracted the $H$-band data ``sky offset''~\citep[see also][]{kas06} with either a constant or a plane,
depending on the object, and then calibrated the resulting frames with 2MASS.
We took optical $g$ and $i$ bands frames from
the eighth release (DR8) of the SDSS~\citep{aih11}, 
and mosaicked them with the SWarp software~\citep{ber10}.
SDSS mosaics were registered and re-sampled to the (lower resolution) 
$H$-band data with the aid of foreground stars.
All foreground stars and background objects
were then removed and replaced with random values from the background.
The {\tt{Adaptsmooth}} code was then used
to increase the S/N ratio at the outskirts of the disk,
while maintaining the relatively higher S/N ratio for the inner disk pixels.
We adopt a minimum S/N ratio per pixel of 10, and a maximum smoothing radius of 10.

Together with the OSUBSGS sample, we also analyzed M~51b (companion of M~51, aka NGC~5195)
using the same data presented in section~\ref{M51_BSP}.

\subsection{Mass-maps results}

We adopt the SSAG-BC03 SPS library for all mass estimates for this sample.
For simplicity we assume that $\sigma_{\rm mag}\sim0.02$ mag for every band and pixel.
The shortcoming of using a constant $\sigma_{\rm mag}$ (and consequently
a constant $\sigma_{\rm col}$) for every band and pixel is that some of
the fitted values could give slightly ($\sim0.3\%$ for individual pixels) different results when compared
to the case where individual errors are computed for every pixel.
The reason for this is the use of equation~\ref{maxlike} together with equation~\ref{chi2}.
In our case we adopt two colors, hence equation~\ref{maxlike} can be seen as the product
of two Gaussian functions (one for each color). In the case where $\sigma_{\rm col}$ differs
for each color, it can be easily demonstrated that this product results in another Gaussian
function with different characteristics, including a distinct maximum,
when compared to the case of two equal Gaussian functions. Also,
the uncertainties in the fitted values will be different.
Despite this, the overall  results for each object will be practically the
same (a $\sim0.1\%$ difference for the resolved total mass estimate).

\begin{figure*}
\centering
\epsscale{1.0}
\plotone{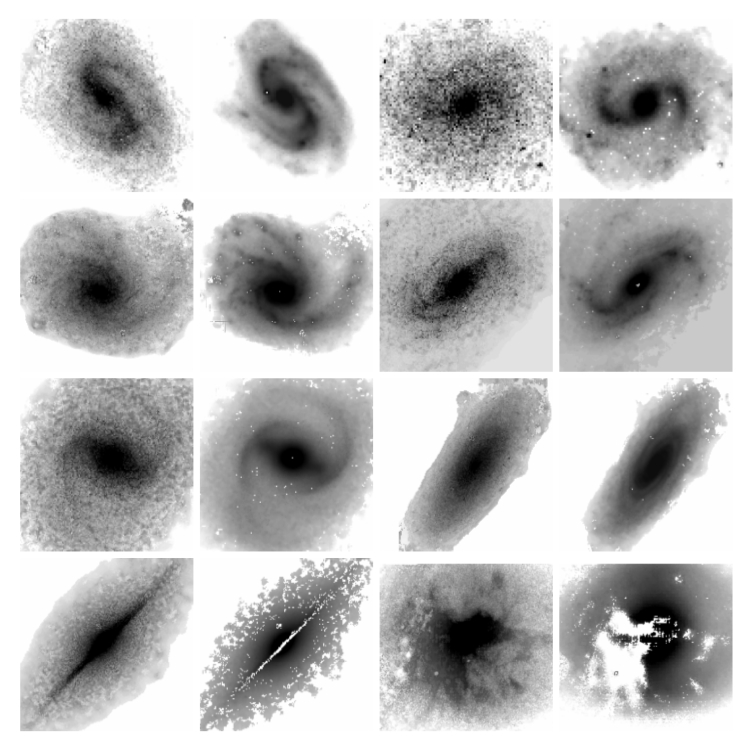}
\caption[f22]{
Resolved maps of stellar mass. {\it Columns 1} and {\it3:} ZCR$^\prime$; 
{\it columns 2} and {\it4:} BSP.
From left to right, in pairs:
NGC~157,  NGC~1042,
NGC~4254, NGC~4051,
NGC~4548, NGC~7606,
NGC~7814, and M~51b.
~\label{fig22}}
\end{figure*}

Some examples of the mass-maps from both the ZCR$^\prime$ approach and BSP are shown
in Figure~\ref{fig22}. The difference in spatial structures is clearly
evident: whereas the ZCR$^\prime$ method gives noisy maps, BSP mass-maps
bear a greater similarity to the structures in the NIR-bands.
Also shown in this figure are two extreme cases, where dust extinction affects
our mass estimates considerably. NGC~7814 is an edge-on spiral with a prominent mid-plane dust lane.
From the first BSP iteration, the colors of the pixels belonging to the dust lanes
are identified (and flagged) as outside of the range available in the SPS library.
A similar phenomenon occurs with M~51b, since 
the dust lanes of one of the arms of M~51 are projected directly on it.
Consequently, a substantial number of pixels are excluded after the first iteration of BSP.
Nevertheless, our recovered stellar mass value for M~51b (see Table~\ref{tbl-1}),
obtained via BSP, is $\sim$ half of the one derived for M~51.
The same result was obtained by~\citet{men12}.

\begin{figure}
\centering
\epsscale{1.0}
\plotone{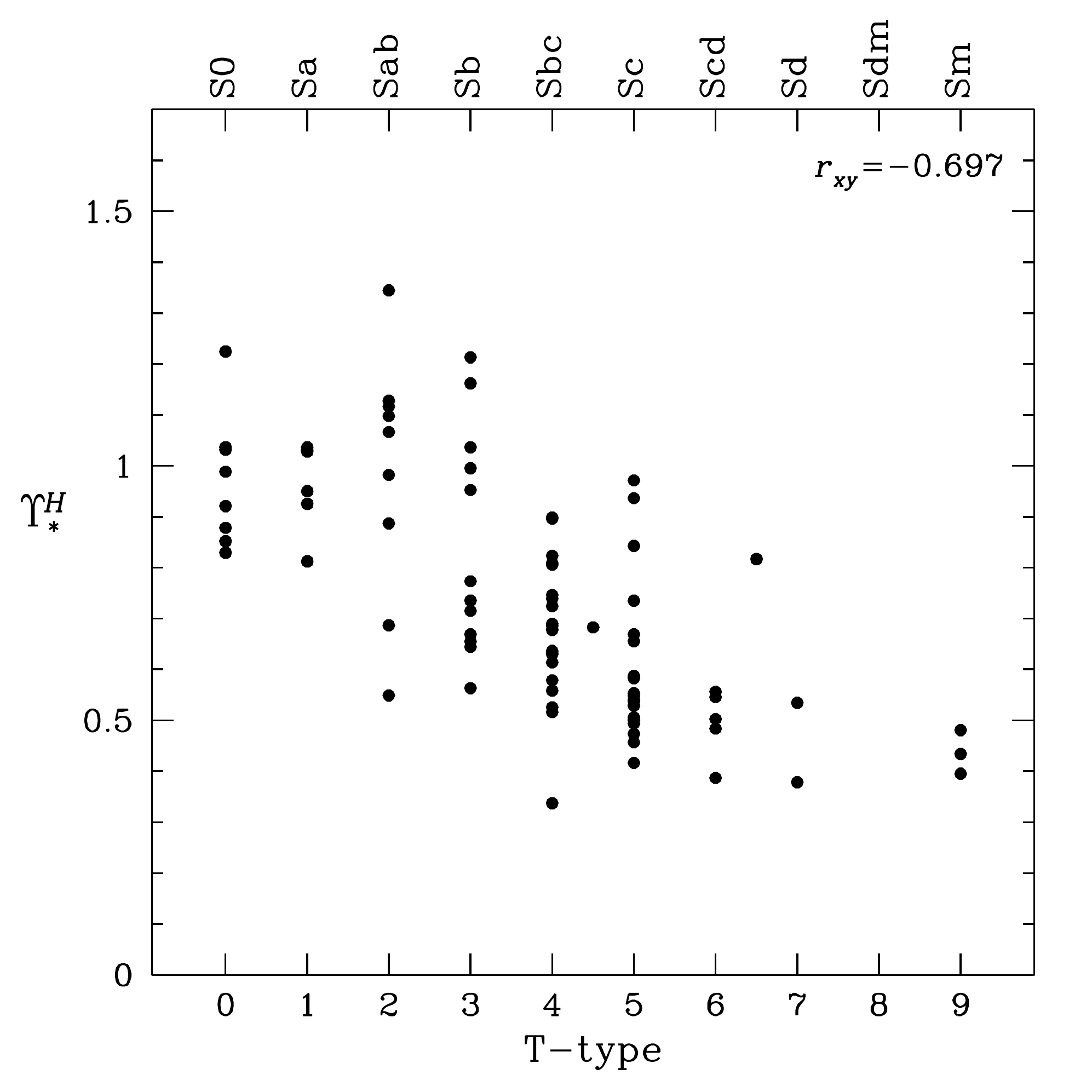}
\caption[f23]{
Median $\Upsilon_{*}^{H}$ vs.\ Hubble T-type after BSP iteration number 1,
applied to the OSUBSGS pilot sample.
Although with some scatter, a strong ($r_{xy}=-0.697$) trend is observed whereby 
$\Upsilon_{*}^{H}$ decreases with increasing T (later Hubble type). 
~\label{fig23}}
\end{figure}

As a result of the application of the BSP algorithm to our pilot sample,
we identify a trend of the median $\Upsilon_{*}^{H}$ (after BSP iteration number 1)
with Hubble type, as predicted by~\citet{por04}
and consistent with more recent star formation/more constant SFHs for later Hubble types.
A strong linear inverse (or negative) correlation with Hubble type is shown in Figure~\ref{fig23},
with a correlation coefficient~\citep{bev69}, $r_{xy}=-0.697$.\footnote{
The value of $r_{xy}$ varies from 0, for no correlation, to $\pm1$, when there is a full correlation.
Generally, $|r_{xy}| \gtrapprox 0.7$ is considered a strong correlation, $|r_{xy}| \approx 0.5$ a moderate correlation,
and $|r_{xy}| \approx 0.3$ a weak correlation.}
Thus, potential biases are introduced when the same $\Upsilon_{*}$ is used
for a sample of galaxies with different Hubble types.

\begin{figure}
\centering
\epsscale{1.0}
\plotone{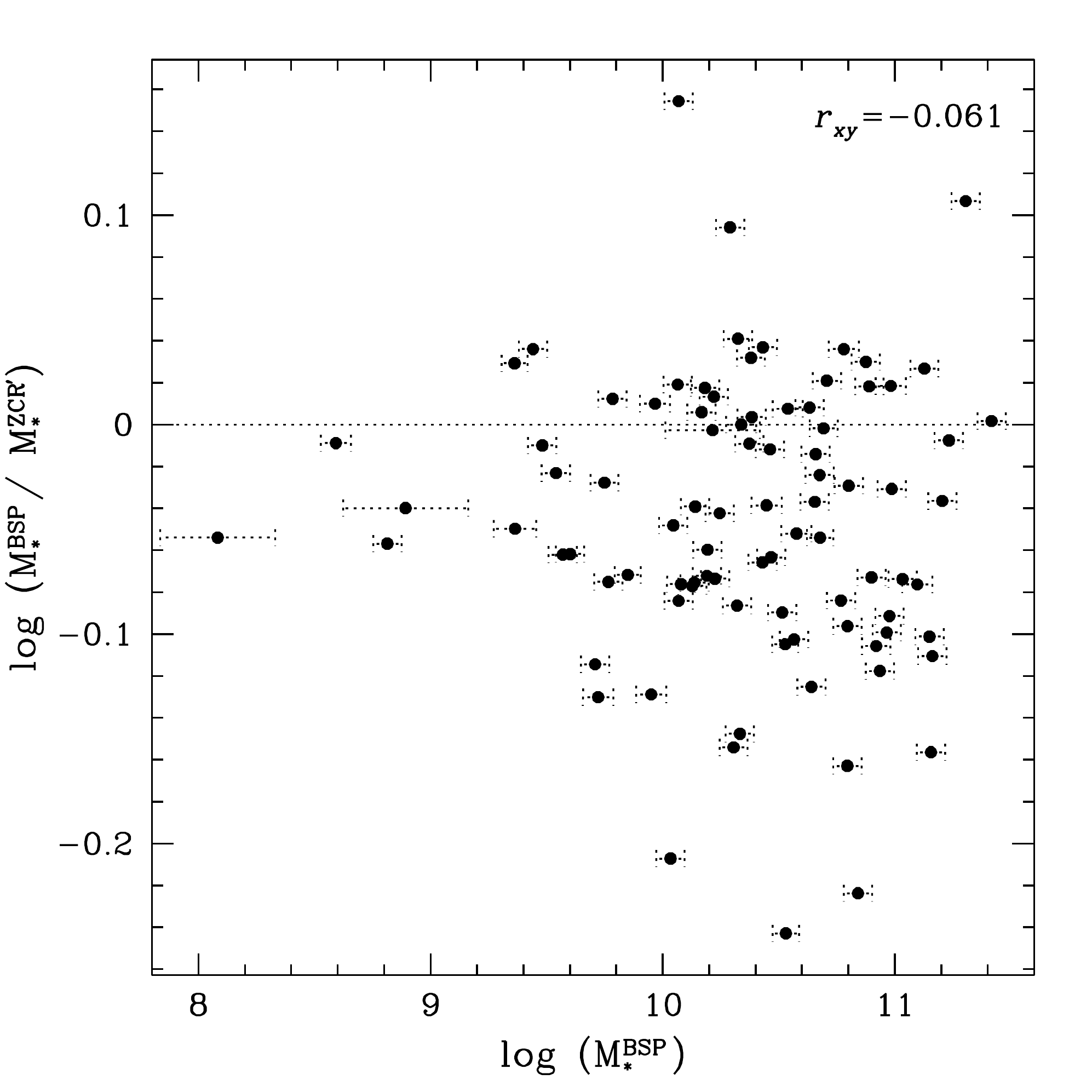}
\caption[f24]{
Comparison of total resolved stellar mass estimates,
$\log(M^{\rm BSP}_{*}/M^{\rm ZCR^\prime}_{*}$) vs.\ $\log(M^{\rm BSP}_{*})$.
Horizontal error bars for $M^{\rm BSP}_{*}$ represent the propagated uncertainty in the distance
to the objects.
~\label{fig24}}
\end{figure}

\begin{figure}
\centering
\epsscale{1.0}
\plotone{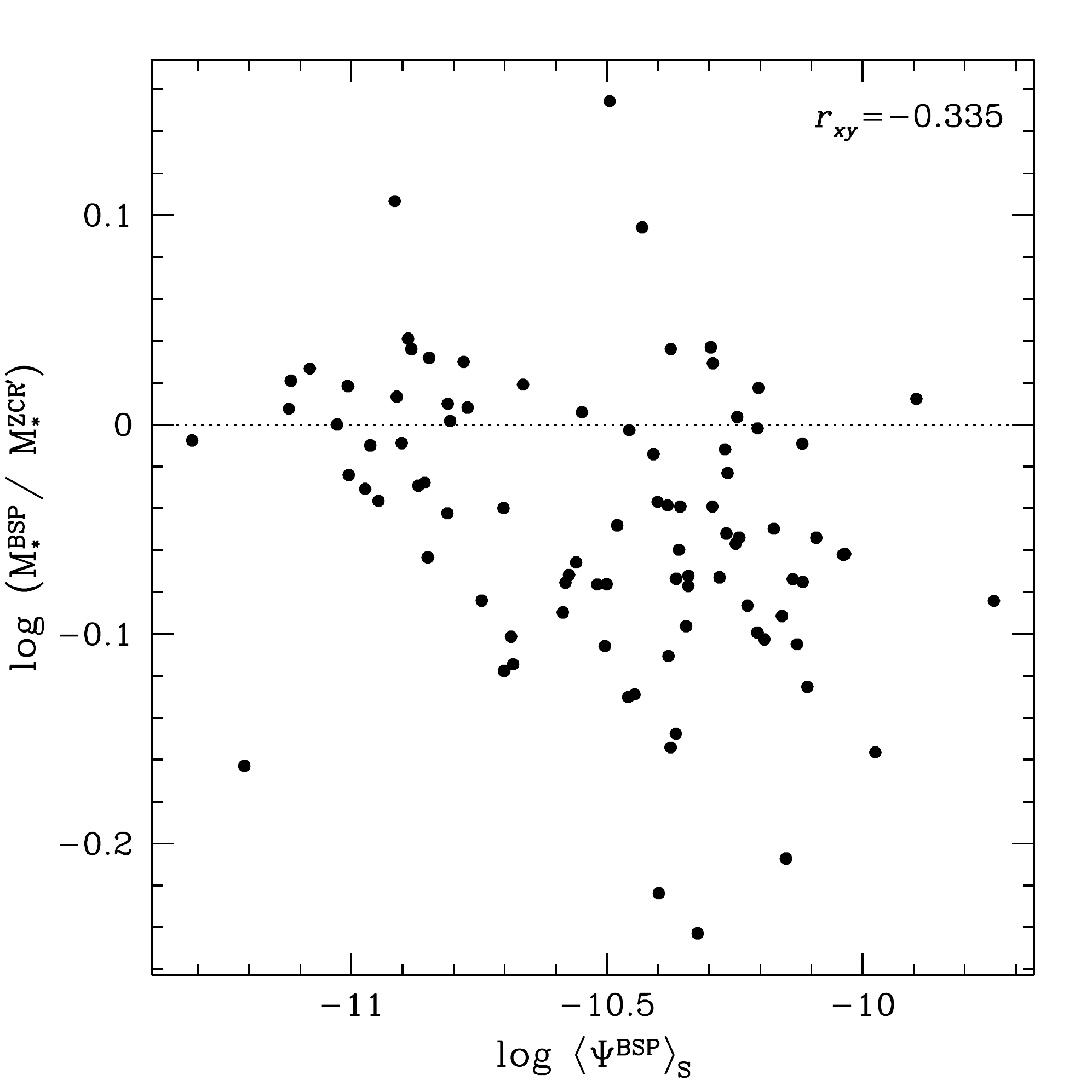}
\caption[f25]{
Decimal logarithm of $(M^{\rm BSP}_{*}/M^{\rm ZCR^\prime}_{*}$) vs.\
$\log\langle\Psi^{\rm BSP}\rangle_{\rm S}$, with $\Psi_{\rm S}$ in yr$^{-1}$.
The resolved specific star formation rate is obtained as the sum of all
pixels in the disk using BSP.
A weak correlation is observed with negative
correlation coefficient $r_{xy}=-0.335$. 
~\label{fig25}}
\end{figure}

The total resolved stellar masses, $M^{\rm resolved}_{*}$ (equation~\ref{mass_reso}),
obtained, respectively, with the BSP algorithm, $M^{\rm BSP}_{*}$,
and with the ZCR$^\prime$ approach, $M^{\rm{ZCR^\prime}}_{*}$, are given in Table~\ref{tbl-1}.
In Figure~\ref{fig24} we display the behavior of the ratio
$M^{\rm BSP}_{*}/M^{\rm{ZCR^\prime}}_{*}$ vs.\ $M^{\rm BSP}_{*}$.
From these data we find that BSP mass estimates are on average $\sim10\%$ lower
than those derived from ZCR$^\prime$, similarly to the M~51 result.
We also investigate possible trends
of the ratio $M^{\rm BSP}_{*}/M^{\rm{ZCR^\prime}}_{*}$ with Hubble Type;
with the ratio of major to minor galaxy axes $a/b$;
with star formation rate, $\Psi$;
and with $V$-band optical depth, $\tau_{V}$.
We find no strong or moderate correlations with these parameters,
except for the star formation rate, having $r_{xy}=-0.061$ for Hubble Type,
$r_{xy}=-0.297$ for galaxy axial ratio (excluding the edge-on object NGC~7814),
and $r_{xy}=0.082$ for the median $\tau_{V}$ for the entire disk, obtained via BSP.
We computed the star formation rate averaged over the last $10^8$ yr
from the parameters of the fitted templates as
\begin{equation}
    \langle\Psi\rangle =
     \frac {\int_{t-t_{\rm last}}^{t} \Psi(t') {\rm d}t'}{t_{\rm last}},
\end{equation}

\noindent where time $t$ corresponds to the current $\Psi$, and $t_{\rm last}=10^8$ yr.
We calculate $\langle\Psi\rangle$ on a pixel-by-pixel basis and then sum over all pixels
(in the same way as the resolved mass estimate).
We also estimate the specific star formation rate averaged over the last $10^8$ yr:
\begin{equation}
    \langle\Psi\rangle_{\rm S} =
    \frac {\int_{t-t_{\rm last}}^{t} \frac{\Psi(t')}{M_{*}(t')} {\rm d}t'}{t_{\rm last}}
    \approx \langle\Psi\rangle {M_{*}}^{-1},
\end{equation}

\noindent where $M_{*}$ is the current stellar mass. 
In this manner we obtain the resolved $\langle\Psi\rangle$, and 
$\langle\Psi\rangle_{\rm S}$, for the corresponding object.
In Figure~\ref{fig25} we show the ratio $M^{\rm BSP}_{*}/M^{\rm{ZCR^\prime}}_{*}$ vs.\
the resolved $\langle\Psi^{\rm BSP}\rangle_{\rm S}$ for the whole disk.
The correlation coefficient is $r_{xy}=-0.335$ indicating a weak inverse correlation.
In the case of the resolved $\langle\Psi\rangle$ we obtain a correlation coefficient of $r_{xy}=-0.240$.
These results suggest that the bias in the resolved mass values $M^{\rm{ZCR^\prime}}_{*}$,
when compared to $M^{\rm BSP}_{*}$, is weakly related to the star formation rate over the disk.

\begin{figure}
\centering
\includegraphics[angle=-90,scale=0.3]{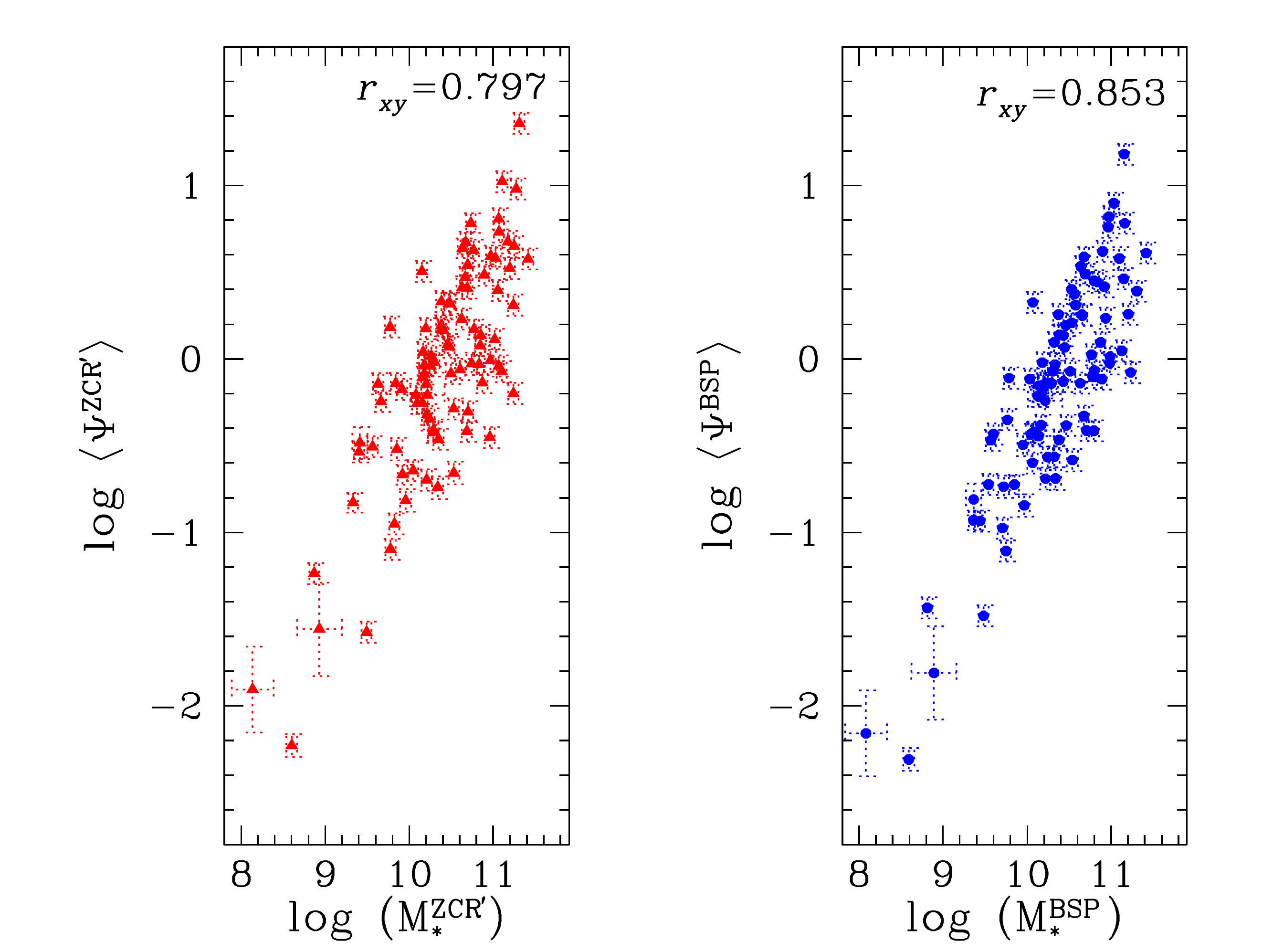}
\caption[f26]{
The resolved ``main sequence'' of star forming galaxies for the OSUBSGS pilot sample
with the SSAG-BC03 library.
{\it Left panel (red triangles):} ZCR$^\prime$; {\it right panel (blue dots):} BSP.
Resolved star formation rate, $\langle\Psi\rangle$, in units of $M_{\sun}$ yr$^{-1}$,
and resolved stellar mass, $M^{\rm resolved}_{*}$, in units of $M_{\sun}$.
~\label{fig26} }
\end{figure}

For completeness, we show in Figure~\ref{fig26} the resolved galaxy ``main sequence'' of star
formation~\citep[see e.g.,][]{noe07,dad07,elb07,sal07}, i.e., the relationship between
resolved $\langle\Psi\rangle$ and\ $M^{\rm resolved}_{*}$.
We find that this correlation is stronger with BSP ($r_{xy}=0.853$) when compared
to ZCR$^\prime$ ($r_{xy}=0.797$).

\subsection{Comparison with unresolved mass estimates}

We also obtain for each object an unresolved mass estimate, $M^{\rm unresolved}_{*}$.
To this end, we fit the global $(g-i)$ and $(i-H)$ colors of the object
to all templates, and get the optimum one via equation~\ref{maxlike}. 
Global magnitudes are calculated by summing the intensities of all the pixels: 

\begin{equation}~\label{unreso}
 {\rm mag}^{\rm global} = -2.5\ {\rm log_{10}}\ \sum\limits_{j}\sum\limits_{i} f_{ij} + zp,
\end{equation}

\noindent where $f_{ij}$ is the intensity of the $i^{\rm th},j^{\rm th}$ pixel
at a certain band, and $zp$ is the appropriate zero point.
The same number of pixels is used in all mass estimates for the same object.

\begin{figure}
\centering
\includegraphics[angle=-90,scale=0.3]{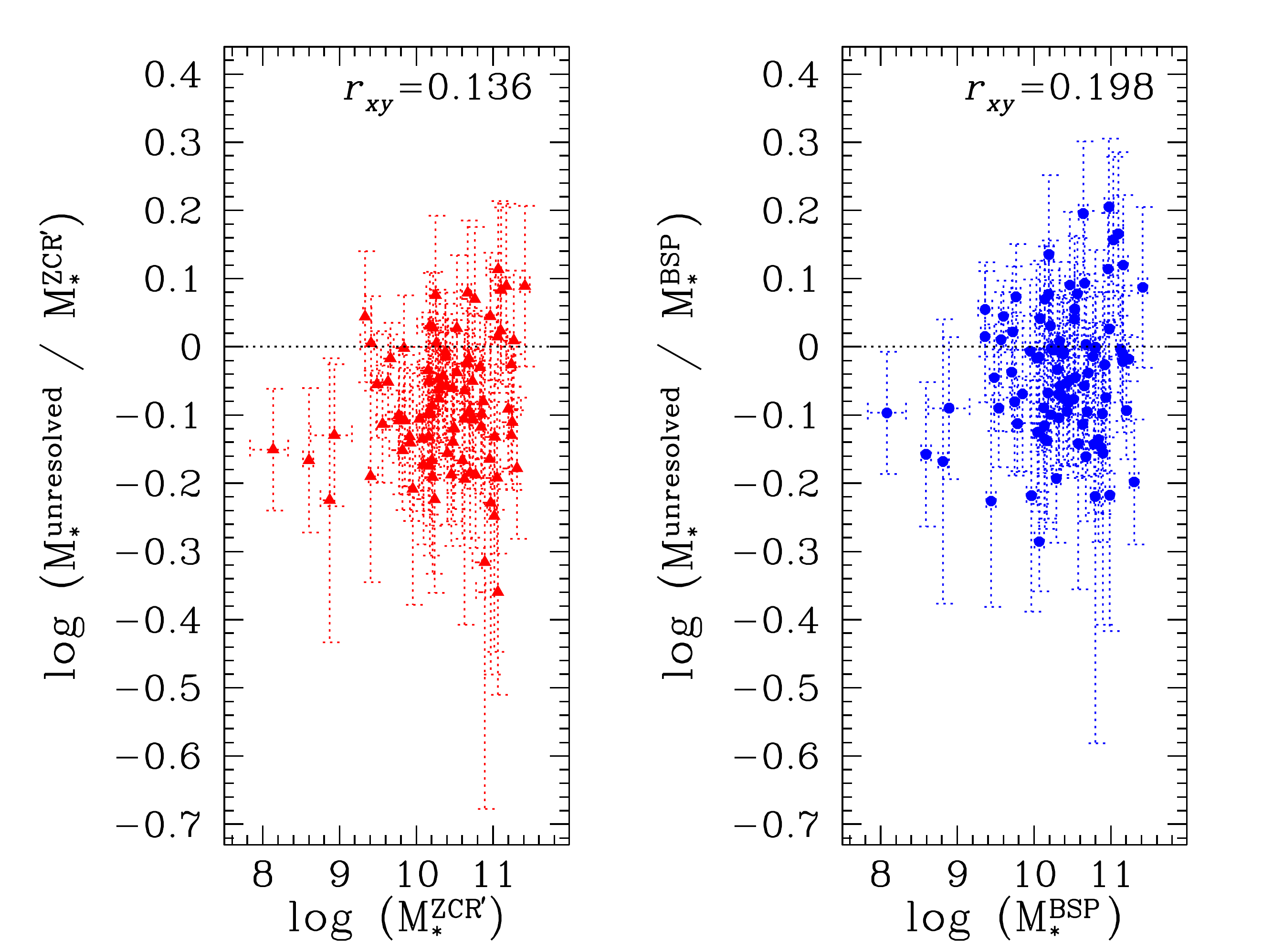}
\caption[f27]{
Comparison of total unresolved and resolved stellar mass estimates.
{\it Left panel (red triangles):} ZCR$^\prime$; {\it right panel (blue dots):} BSP.
~\label{fig27} }
\end{figure}

We compare in Figure~\ref{fig27} $M^{\rm unresolved}_{*}$ with $M^{\rm resolved}_{*}$.
The results for ZCR$^\prime$ are shown in the left panel, and those for BSP are presented on the right.
On average we find that, for our sample of galaxies, unresolved values
underestimate masses by $\sim20\%$ compared to ZCR$^\prime$, but only by $\sim10\%$ relative to BSP.
We also find, however, that for a fraction of the objects (15\% when comparing to ZCR$^\prime$ and 
25\% vis-\`a-vis  BSP) the unresolved mass estimates are actually larger than those determined 
from resolved studies. The estimate we can get for an unresolved mass
depends on how each pixel contributes to the global colors.
Pixels that contain relatively young star forming regions will lead to global bluer colors,
and consequently a lower global $\Upsilon_{*}$ (see Figure~\ref{fig1} or ~\ref{fig6}).
On the other hand, pixels that contain extinction regions, due to dust, will lead to
global redder colors and therefore a higher global $\Upsilon_{*}$.
In spite of these possible effects the error bars for $\log(M^{\rm unresolved}_{*}/M^{\rm resolved}_{*})>0$
(see Figure~\ref{fig27}) are within the $\log(M^{\rm unresolved}_{*}/M^{\rm resolved}_{*})\sim0$
value.

\begin{figure}
\centering
\includegraphics[angle=-90,scale=0.3]{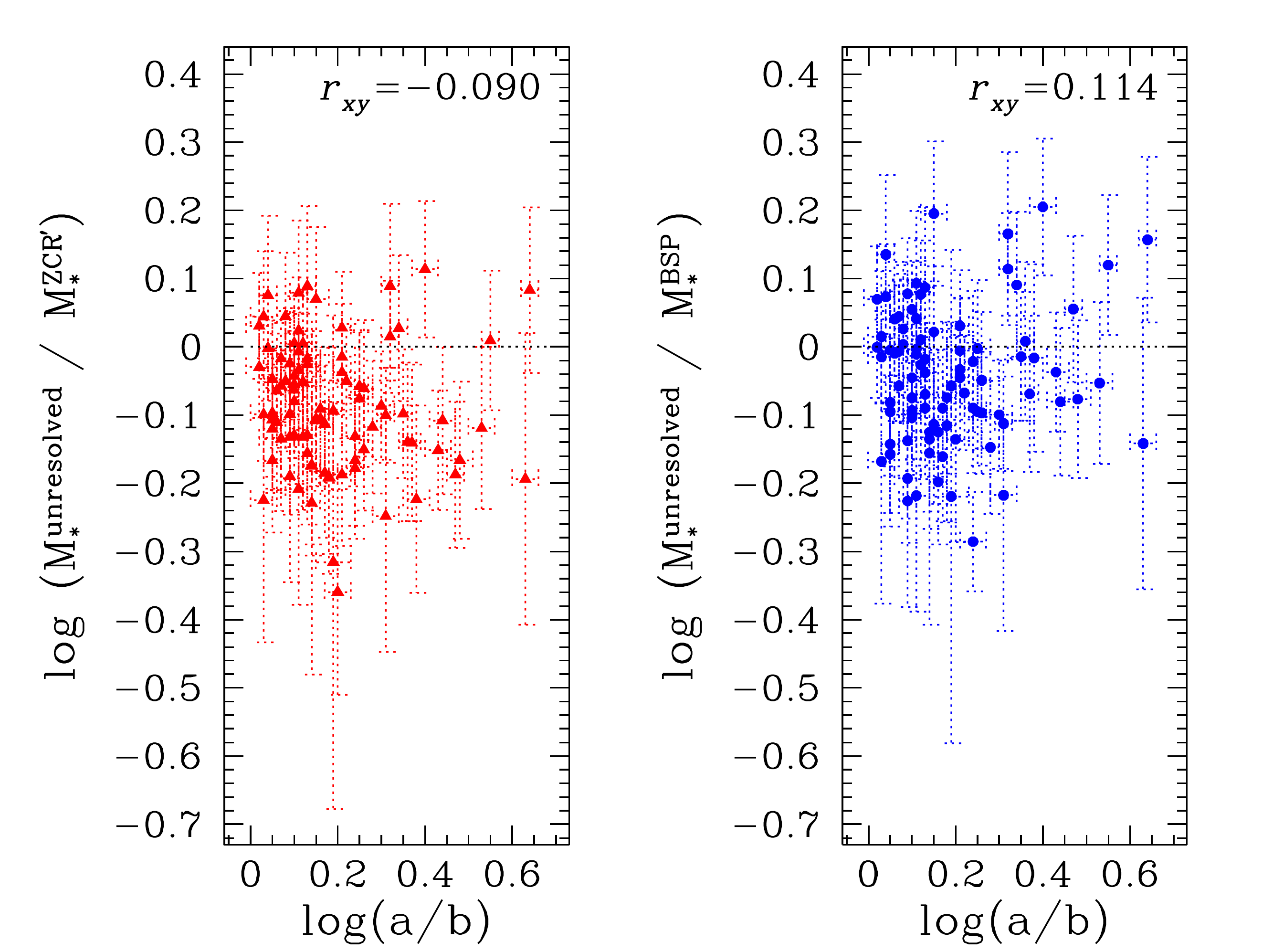}
\caption[f28]{
Ratio of unresolved to resolved stellar mass estimates vs.\
galaxy axial ratio, $a/b$~\citep[from RC3,][]{deV91}.
{\it Left panel (red triangles):} ZCR$^\prime$; {\it right panel (blue dots):} BSP.
~\label{fig28} }
\end{figure}

\begin{figure}
\centering
\includegraphics[angle=-90,scale=0.3]{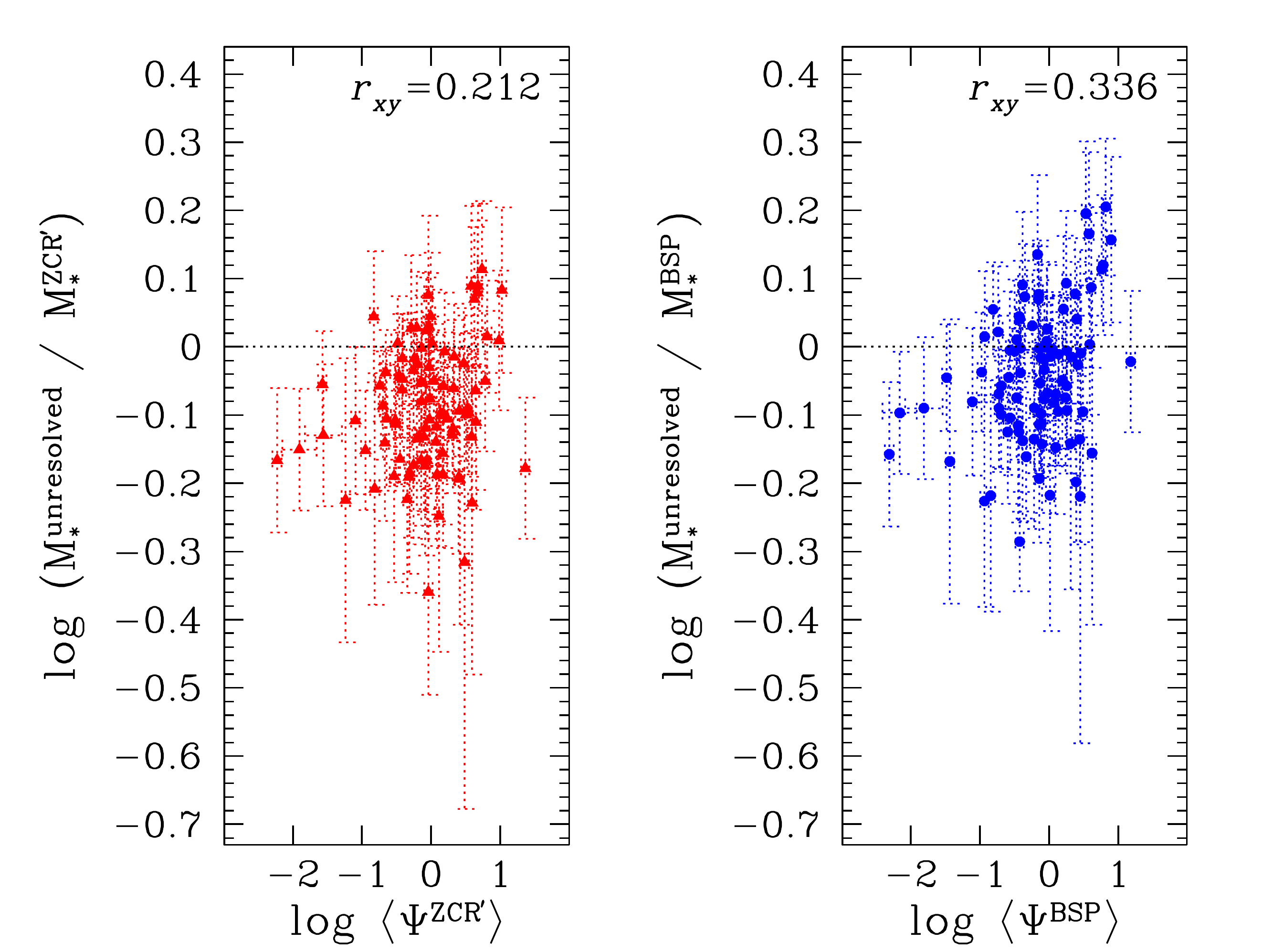}
\caption[f29]{
Ratio of unresolved to resolved stellar mass estimates vs.\
resolved star formation rate $\langle\Psi\rangle$ ($M_{\sun}$ yr$^{-1}$).
{\it Left panel (red triangles):} ZCR$^\prime$; {\it right panel (blue dots):} BSP.
~\label{fig29} }
\end{figure}

We find no correlation of $M^{\rm unresolved}_{*}/M^{\rm resolved}_{*}$
with Hubble type ($r_{xy}= 0.064$ for BSP, and $r_{xy}=0.024$ for ZCR$^\prime$),
global $(g-i)$ color ($r_{xy}=0.057$ for BSP, $r_{xy}=-0.025$ for ZCR$^\prime$),
or median $\tau_{V}$ ($r_{xy}=0.055$ for BSP, $r_{xy}=0.116$ for ZCR$^\prime$).
The correlation test was also negative for galaxy inclination (see Figure~\ref{fig28}), 
with $r_{xy}=0.114$ for BSP, and $r_{xy}=-0.090$ for ZCR$^\prime$.
When comparing the resolved $\langle\Psi\rangle_{\rm S}$ for each object
with the ratio $M^{\rm unresolved}_{*}/M^{\rm resolved}_{*}$,
we find a weak positive correlation ($r_{xy}=0.262$) for BSP, and no correlation ($r_{xy}=0.118$)
for ZCR$^\prime$.
In Figure~\ref{fig29} we show the ratio $M^{\rm unresolved}_{*}/M^{\rm resolved}_{*}$ vs.\
resolved $\langle\Psi\rangle$. The correlation coefficients are $r_{xy}=0.336$
for BSP (right panel), and $r_{xy}=0.212$ for ZCR$^\prime$ (left panel) indicating a weak correlation
in our case test.

\section{Uncertainties in the stellar mass estimates}~\label{mass_errors}

All the stellar mass estimates given in Table~\ref{tbl-1} are for the SSAG-BC03 library;
if, instead, the MAGPHYS-CB07 library is used, the masses will be smaller ($\sim50\%$),
due to the different treatment of the TP-AGB. Hence, the dominant source of error is systematic.

Regarding the uncertainty in the mass per pixel, we obtain a mean value
for the entire disk of $\sim27\%$ with ZCR$^\prime$, and of $\sim3\%$ with BSP.
The reduction of the uncertainty in BSP is due to the inclusion
of equation~\ref{prior} in the calculations.
The random errors in the total resolved mass estimates, on the other hand, are rather small,
given the very large number of pixels involved in the
calculations ($\sim6\times10^{5}$ and $\sim2\times10^{4}$ pixels,
for M~51 and the OSUBSGS objects, respectively).
For an object with $n_{\rm pix}$ pixels the relative uncertainty ($\sigma_{\rm mass}$/mass)
decreases as $\sim\frac{1}{\sqrt{n_{\rm pix}}}$. Hence the random uncertainties in
the total resolved mass estimates tend to be less than 0.1\%.
The random uncertainties in the median $\Upsilon_{*}$ (after iteration number 1),
and $\langle\Psi\rangle$ are also relatively small due to the large number of pixels
involved in the calculations. 
With regard to the systematic uncertainty due to the zero point error $\sigma_{\rm calib}$,
we estimate a $3\%$ relative error in the resolved mass estimates, and the median $\Upsilon_{*}$.
However, this systematic error dominates the relative uncertainties in $M^{\rm unresolved}_{*}$
(see equation~\ref{unreso}), which have a median of $\sim22\%$ (see Table~\ref{tbl-1}).

Another source of systematic error is the uncertainty in the distance to the objects, $\sigma_{\rm dist}$.
Propagating $\sigma_{\rm dist}$ leads to a $\sim14\%$ uncertainty in the mass, and $\langle\Psi\rangle$,
for all galaxies in our pilot sample, with the exception of NGC~3319, NGC~4051, and NGC~4212,
for which the uncertainty in the mass is $\sim55\%$. However, the contribution of this uncertainty is
negligible for the mass ratio of any single galaxy ($M^{\rm BSP}_{*}/M^{\rm{ZCR^\prime}}_{*}$ or
$M^{\rm unresolved}_{*}/M^{\rm resolved}_{*}$), since
all mass estimates are equally affected. Equivalently, $\langle\Psi\rangle_{\rm S}$ is
not affected by $\sigma_{\rm dist}$.

Regarding the choice of the IMF, our default is~\citet{cha03}.
Stellar masses can be $\sim1.7\pm0.3$ times larger with the~\citet{sal55} IMF,
and $\sim1.1\pm0.03$ times larger with the~\citet{kro01} IMF.

We also have quantified that using only a constant $\Upsilon_{*}$ (i.e., skipping iteration number 3)
yields masses per pixel $\sim1\%$ higher on average, and up to $\sim30\%$ larger
in localized regions.

\subsection{Dependence on disk inclination}

Stellar mass is an intrinsic property of galaxies, independent of 
inclination to the line of sight. Stellar mass determinations from
broad-band colors, however, are independent of inclination only as
surface brightness at different wavelengths is independent of it.
~\citet[][]{mal09} study the effects of inclination on mass estimates, by comparing a statistically
significant sample of edge-on ($a/b\geq3.33$) and face-on ($a/b\leq1.18$) SDSS galaxies.
They find no statistical difference for masses derived from $K$-band photometry
by~\citet[][]{bel03} but, on the other hand, point out the very important
corrections with inclination that are necessary for the $B$-band~\citep{dri07}.

We remind the reader that all our calculations are based on the {\it effective} $\Upsilon_{*}$.
Extinction effects may introduce biases with inclination. 
In subsequent publications we will address this issue in more detail.

\section{Conclusions}~\label{conclu}

We have demonstrated quantitatively that resolved maps
of stellar mass obtained by the maximum likelihood
estimate~(as in ZCR) yield biased spatial structures.
The bias consists in a filamentary morphology, and a spatial
coincidence between dust lanes and purported stellar mass surface density.
The bias is due to a limited $\Upsilon_{*}$ accuracy ($\sim0.1-0.15$ dex)
arising from uncertainties inherent to observations, and to
degeneracies between templates of similar colors in the SPS libraries.
Similar observed colors will yield the mode $\Upsilon_{*}$.
Here, we have succeeded in mitigating the bias with the 
BSP algorithm we have developed.
We have applied the new algorithm to M~51 and a pilot sample of 90 spirals.
BSP effectively identifies and isolates the old stellar population, and the
output mass-maps bear more resemblance to NIR structures. 

The results also indicate that total resolved mass estimates obtained by
adding up the pixel-by-pixel contributions 
are on average $\sim 10\%$ lower with BSP than with the ZCR$^\prime$ approach.
Hence, unresolved stellar mass estimates for our pilot sample underestimate the mass
by $\sim 20\%$ when compared to the resolved ZCR$^\prime$ results, but only
by $\sim10\%$ vis-\`a-vis BSP. 

The fact that the same SPS libraries can produce, or not,  
filamentary structures where the mass is supposedly organized
indicates that such structures are merely an artifact of the method,
and not real massive features present in disk galaxies.

An additional advantage of using a {\it spatial structure prior} for
mass estimates is its independence of SPS model parameters
(e.g., SFH, metallicity, dust, age, etc.) or ingredients (e.g., TP-AGB phase, or IMF).
Galaxy masses determined from SPS models should be compared to results of
independent studies, e.g., the Disk Mass Survey~\citep[DMS,][]{bsha10}\footnote{
DMS uses measurements of the vertical velocity dispersion of disk stars
as a dynamical constraint on the mass surface density of spiral disks.}.
Systematic uncertainties may be constrained through these
comparisons~\citep[see also][]{jon07}.

\begin{figure}
\centering
\epsscale{1.0}
\plotone{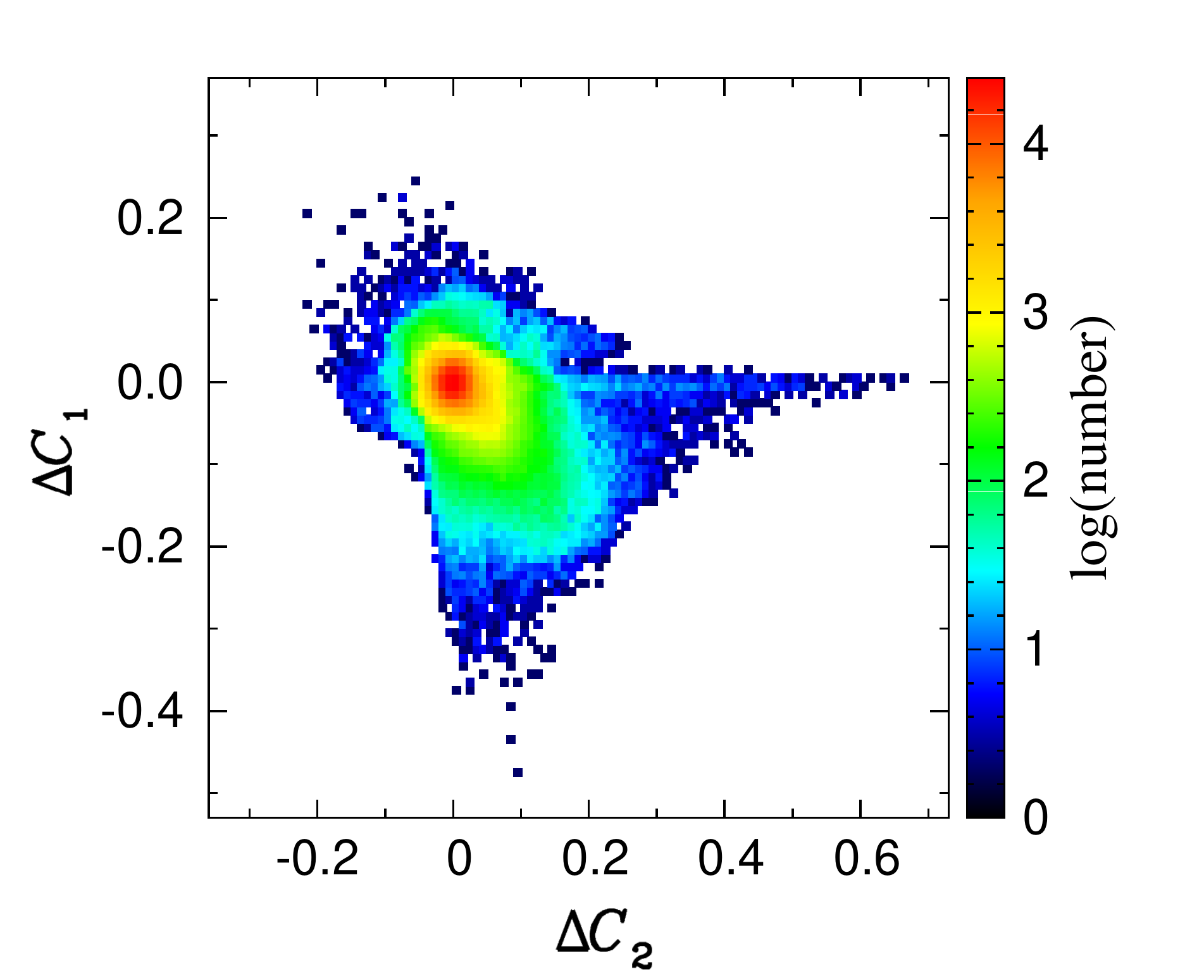}
\caption[f30]{2-D histogram of the 
$\Delta C_{1} = (g-i)^{\rm obs}-(g-i)^{\rm template}$, and
$\Delta C_{2} = (i-K_{s})^{\rm obs}-(i-K_{s})^{\rm template}$
distributions before applying the $\left|\Delta C_{n}\right| < \alpha\sigma_{\rm P}$ condition
at iteration number 2 of BSP. Data pixels for M~51, adopting the MAGPHYS-CB07 library.
~\label{fig30} }
\end{figure}

\begin{figure}
\centering
\epsscale{1.0}
\plotone{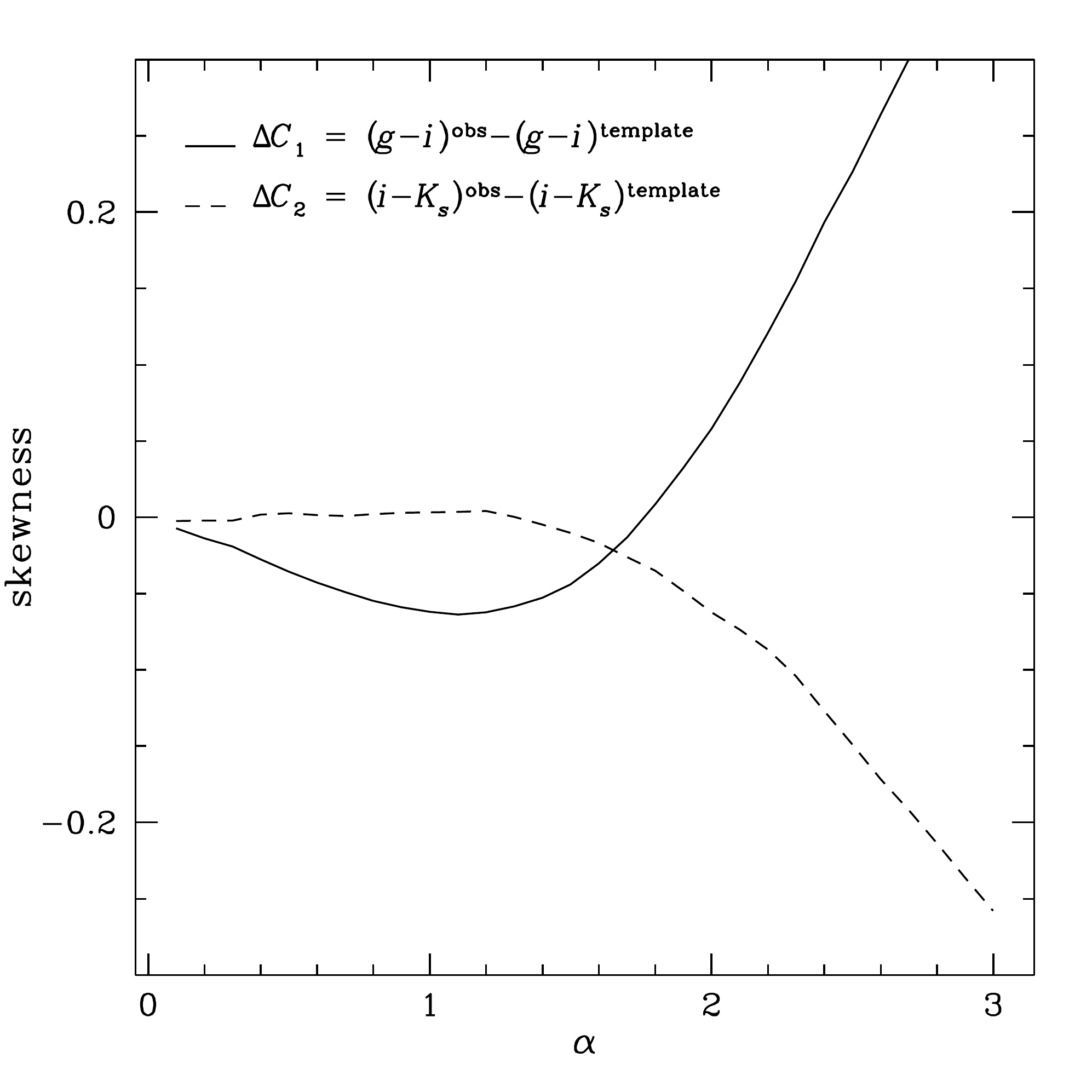}
\caption[f31] {
``Skewness curves'' of the $\Delta C_{1}$ and $\Delta C_{2}$ distributions
after applying the $\left|\Delta C_{n}\right| < \alpha\sigma_{\rm P}$ condition for different $\alpha$ values.
Same data as in Figure~\ref{fig30}.
~\label{fig31} }
\end{figure}

\acknowledgments

We acknowledge the referee for her/his comments and suggestions
that significantly improved the quality of the manuscript.
We appreciate discussions with and comments from Margarita Rosado,
Iv\^anio Puerari, Bernardo Cervantes-Sodi, Sebasti\'an S\'anchez, 
Fabi\'an Rosales-Ortega, Olga Vega, Edgar Ram\'irez, and William Wall.
We thank Alfredo Mej\'ia-Narv\'aez for useful discussions about the SSAG parameters.

EMG acknowledges support from INAOE during the initial stages of this research, and
from IRyA during the development of the project; he gives special thanks to his mother, Gilda Garc\'ia.

RAGL acknowledges the support from DGAPA, UNAM, through project PAPIIT IG100913, and from CONACyT, Mexico,
through project SEP-CONACyT I0017-151671.

GMC acknowledges support from IRyA during the initial stages of this work.

GBA acknowledges support for this work from the National Autonomous University of Mexico (UNAM),
through grant PAPIIT IG100115.

Funding for SDSS-III has been provided by the Alfred P. Sloan Foundation, the Participating Institutions,
the National Science Foundation, and the U.S. Department of Energy Office of Science.
The SDSS-III web site is http://www.sdss3.org/.
SDSS-III is managed by the Astrophysical Research Consortium for the Participating Institutions 
of the SDSS-III Collaboration including the University of Arizona, the Brazilian Participation Group, 
Brookhaven National Laboratory, Carnegie Mellon University, University of Florida, 
the French Participation Group, the German Participation Group, Harvard University, 
the Instituto de Astrof\'isica de Canarias, the Michigan State/Notre Dame/JINA Participation Group, 
Johns Hopkins University, Lawrence Berkeley National Laboratory, Max Planck Institute for Astrophysics, 
Max Planck Institute for Extraterrestrial Physics, New Mexico State University, New York University, 
Ohio State University, Pennsylvania State University, University of Portsmouth, Princeton University, 
the Spanish Participation Group, University of Tokyo, University of Utah, Vanderbilt University, 
University of Virginia, University of Washington, and Yale University. 

This work made use of data from the Ohio State University Bright Spiral Galaxy Survey,
which was funded by grants AST-9217716 and AST-9617006 from the United States National
Science Foundation, with additional support from the Ohio State University.

This research has made use of the NASA/IPAC Extragalactic Database (NED) which is operated
by the Jet Propulsion Laboratory, California Institute of Technology,
under contract with the National Aeronautics and Space Administration. 

The Digitized Sky Surveys were produced at the Space Telescope Science Institute (STScI)
under U.S. Government grant NAG W-2166. DSS images can be found at the
URL~\url{https://archive.stsci.edu/cgi-bin/dss_form}.

This work has made use of the adaptive smoothing code {\tt{Adaptsmooth}},
developed by Stefano Zibetti and available at the
URL~{\scriptsize\url{http://www.arcetri.astro.it/~zibetti/Software/ADAPTSMOOTH.html}}.

\appendix

\section{Determination of the $\alpha$ parameter for BSP.}~\label{appA}

The last step of iteration number 2 is to identify the pixels which
satisfy the condition $\left|\Delta C_{n}\right| < \alpha\sigma_{\rm P}$.
From the definition of $\Delta C_{n} = C_{n}^{\rm obs}-C_{n}^{\rm template}$,
we have:
\begin{equation}
	\Delta C_{1} = (g-i)^{\rm obs}-(g-i)^{\rm template},
\end{equation}
and
\begin{equation}
	\Delta C_{2} = (i-K_{s})^{\rm obs}-(i-K_{s})^{\rm template},
\end{equation}
for the $(g-i)$, and $(i-K_{s})$ colors, respectively. 
The value of $\sigma_{\rm P}$ is computed from equation~\ref{sigma_P}.
In Figure~\ref{fig30} we show a plot of $\Delta C_{1}$ vs.\ $\Delta C_{2}$
for the case of the MAGPHYS-CB07 SPS library, before applying the
$\left|\Delta C_{n}\right| < \alpha\sigma_{\rm P}$ condition to the pixels
of M~51 (see section~\ref{M51_BSP}).
From these $\Delta C_{n}$ distributions we obtain $\sigma_{\rm P}=0.02376$
and $\sigma_{\rm P}=0.02723$, for the $(g-i)$ and $(i-K_{s})$ colors, respectively.
The purpose of applying the $\left|\Delta C_{n}\right| < \alpha\sigma_{\rm P}$
condition is to isolate the pixels that deviate
significantly from the value $\Delta C_{n}\sim0$. In Figure~\ref{fig31}
we show a plot of the skewness (a measure of the degree of asymmetry)
of the $\Delta C_{1}$ and the $\Delta C_{2}$ distributions, after applying 
the $\left|\Delta C_{n}\right| < \alpha\sigma_{\rm P}$ condition for different $\alpha$ values.
The ``skewness curves'' have extrema near $\alpha\sim1$, a minimum for the $\Delta C_{1}$
curve and a maximum for the $\Delta C_{2}$ curve. These extrema values indicate
a transition of the shape of the $\Delta C_{n}$ distributions.
Similar plots are obtained for the MAGPHYS-BC03 and the SSAG-BC03
libraries, which also have extrema
near $\alpha\sim1$. The $\Delta C_{n}$ distributions become extremely asymmetric
for $\alpha>1$. Therefore, in a statistical manner, applying the condition
$\left|\Delta C_{n}\right| < \alpha\sigma_{\rm P}$, with $\alpha\sim1$, fulfills our purposes.

\section{Probability distribution functions for disk parameters.}~\label{appB}

In this section we explain the method we use to obtain the probability distributions for
the disk parameters shown in Figure~\ref{fig14}.
After applying either ZCR$^\prime$ or BSP to a given object, we obtain a set of templates
which were fitted to a group of pixels (e.g., for the whole disk 
we use the pixels shown in Figure~\ref{fig12}a, top left panel).
We then use a Gaussian kernel density method~\citep{kee10} to estimate the probability density function.
In essence, the kernel method produces a smoothed version of a histogram.
First, we build a grid for each parameter (e.g., age) within the range of values given by
the SPS library. The grid contains 512 bins and has a distinct bin width of size $b^{\rm par}_{w}$
for each parameter. A single parameter has bins of equal width, $b^{\rm par}_{w}$, which is estimated
as the difference between the highest value minus the lowest value, divided by 512.
Then we count how many pixels fall into each bin, i.e., we build a histogram of the pixel population
given a certain parameter.
We then calculate, for each histogram, a smoothing parameter called the {\it bandwidth}. 
For a normal distribution with standard deviation $\sigma_{\rm G}^{\rm par}$, the optimal bandwidth, $ \lambda_{G}^{\rm par}$, 
is given by~\citep{kee10}

\begin{equation}
     \lambda_{G}^{\rm par}\approx1.06\sigma_{\rm G}^{\rm par} n_{\rm pix}^{-1/5},
\end{equation}

\noindent where $n_{\rm pix}$ is the number of pixels in our set with
standard deviation $\sigma_{\rm G}^{\rm par}$.
We then convolve the resulting histogram of the pixel population with a Gaussian function
having a standard deviation $\sigma_{\rm conv}= \lambda_{G}^{\rm par}/b^{\rm par}_{w}$.
In the convolution, the Gaussian Kernel extends to $3\sigma_{\rm conv}$.
When building the histogram of the pixel population, all pixels have the same weight.

\clearpage

\LongTables

\begin{deluxetable*}{llccclll}
\tabletypesize{\scriptsize}
\tablecaption{Galaxy parameters~\label{tbl-1}}
\tablewidth{0pt}
\tablehead{
\colhead{Name} &
\colhead{RC3~type} &
\colhead{T-type} &
\colhead{Dist~(Mpc)} &
\colhead{$\Upsilon_{*}^{H}$} &
\colhead{$M^{\rm BSP}_{*}~(M_{\sun})$} &
\colhead{$M^{\rm{ZCR^\prime}}_{*}~(M_{\sun})$} &
\colhead{$M^{\rm unresolved}_{*}~(M_{\sun})$}
}
\startdata

M~51		&	SA(s)bc~pec	&	4.0	&      9.9\tn{a}$\pm$	0.7	&	0.42 ($K_{s}$)	&	5.56$\times10^{10}$	&	6.43$\times10^{10}$	&	(9.02$\pm3.00$)$\times10^{10}$   \\
M~51b		&	I0~pec		&	90.0	&      9.9\tn{a}$\pm$	0.7	&	0.97 ($K_{s}$)	&      2.96$\times10^{10}$\tn{b}&      4.66$\times10^{10}$\tn{b}&       (3.26$\pm0.22$)$\times10^{10}$\tn{b}  \\
NGC~157		&	SAB(rs)bc	&	4.0	&	22.6	$\pm$	1.6	&	0.58	&	4.52$\times10^{10}$	&	4.92$\times10^{10}$	&	(3.96$\pm0.85$)$\times10^{10}$	  \\
NGC~428		&	SAB(s)m		&	9.0	&	15.9	$\pm$	1.1	&	0.40	&	3.71$\times10^{9}$	&	4.28$\times10^{9}$	&	(3.80$\pm0.61$)$\times10^{9}$	  \\
NGC~488		&	SA(r)b		&	3.0	&	30.4	$\pm$	2.1	&	1.04	&	2.61$\times10^{11}$	&	2.60$\times10^{11}$	&	(3.19$\pm0.87$)$\times10^{11}$	  \\
NGC~779		&	SAB(r)b		&	3.0	&	18.5	$\pm$	1.3	&	0.72	&	2.69$\times10^{10}$	&	3.13$\times10^{10}$	&	(2.38$\pm0.65$)$\times10^{10}$	  \\
NGC~864		&	SAB(rs)c	&	5.0	&	20.9	$\pm$	1.5	&	0.53	&	1.55$\times10^{10}$	&	1.83$\times10^{10}$	&	(1.85$\pm0.34$)$\times10^{10}$	  \\
NGC~1042	&	SAB(rs)cd	&	6.0	&	18.1	$\pm$	1.3	&	0.56	&	1.20$\times10^{10}$	&	1.43$\times10^{10}$	&	(1.32$\pm0.26$)$\times10^{10}$	  \\
NGC~1073	&	SB(rs)c		&	5.0	&	16.1	$\pm$	1.1	&	0.51	&	5.83$\times10^{9}$	&	6.93$\times10^{9}$	&	(6.90$\pm1.23$)$\times10^{9}$	  \\
NGC~1084	&	SA(s)c		&	5.0	&	18.6	$\pm$	1.3	&	0.66	&	2.70$\times10^{10}$	&	2.48$\times10^{10}$	&	(2.17$\pm0.40$)$\times10^{10}$	  \\
NGC~1087	&	SAB(rs)c	&	5.0	&	20.1	$\pm$	1.4	&	0.54	&	1.52$\times10^{10}$	&	1.46$\times10^{10}$	&	(1.30$\pm0.25$)$\times10^{10}$	  \\
NGC~1309	&	SA(s)bc:	&	4.0	&	28.3	$\pm$	2.0	&	0.34	&	1.17$\times10^{10}$	&	1.42$\times10^{10}$	&	(1.13$\pm0.18$)$\times10^{10}$	  \\
NGC~2775	&	SA(r)ab		&	2.0	&	21.4	$\pm$	1.5	&	1.12	&	1.34$\times10^{11}$	&	1.26$\times10^{11}$	&	(1.33$\pm0.24$)$\times10^{11}$	  \\
NGC~2964	&	SAB(r)bc:	&	4.0	&	23.2	$\pm$	1.6	&	0.69	&	2.90$\times10^{10}$	&	2.98$\times10^{10}$	&	(2.59$\pm0.60$)$\times10^{10}$	  \\
NGC~3166	&	SAB(rs)0/a	&	0.0	&	22.0	$\pm$	1.5	&	0.99	&	9.69$\times10^{10}$	&	1.04$\times10^{11}$	&	(5.87$\pm2.69$)$\times10^{10}$	  \\
NGC~3169	&	SA(s)a~pec	&	1.0	&	19.9	$\pm$	1.4	&	0.95	&	6.93$\times10^{10}$	&	1.16$\times10^{11}$	&	(5.07$\pm1.76$)$\times10^{10}$	  \\
NGC~3227	&	SAB(s)a~pec	&	1.0	&	20.3	$\pm$	1.4	&	1.03	&	4.74$\times10^{10}$	&	5.01$\times10^{10}$	&	(3.27$\pm0.71$)$\times10^{10}$	  \\
NGC~3319	&	SB(rs)cd	&	6.0	&	3.3	$\pm$	0.9	&	0.48	&	1.21$\times10^{8}$	&	1.37$\times10^{8}$	&	(9.68$\pm1.99$)$\times10^{7}$	  \\
NGC~3338	&	SA(s)c		&	5.0	&	23.2	$\pm$	1.6	&	0.47	&	2.02$\times10^{10}$	&	2.88$\times10^{10}$	&	(1.87$\pm0.45$)$\times10^{10}$	  \\
NGC~3423	&	SA(s)cd		&	6.0	&	14.1	$\pm$	1.0	&	0.39	&	3.99$\times10^{9}$	&	4.60$\times10^{9}$	&	(4.42$\pm0.53$)$\times10^{9}$	  \\
NGC~3504	&	(R)SAB(s)ab	&	2.0	&	27.8	$\pm$	1.9	&	0.69	&	4.56$\times10^{10}$	&	4.71$\times10^{10}$	&	(5.65$\pm1.38$)$\times10^{10}$	  \\
NGC~3507	&	SB(s)b		&	3.0	&	15.0	$\pm$	1.1	&	0.56	&	8.92$\times10^{9}$	&	1.20$\times10^{10}$	&	(8.79$\pm2.13$)$\times10^{9}$	  \\
NGC~3583	&	SB(s)b		&	3.0	&	35.7	$\pm$	2.5	&	0.67	&	6.25$\times10^{10}$	&	7.80$\times10^{10}$	&	(3.77$\pm3.14$)$\times10^{10}$	  \\
NGC~3593	&	SA(s)0/a	&	0.0	&	5.6	$\pm$	0.4	&	1.04	&	5.11$\times10^{9}$	&	6.65$\times10^{9}$	&	(4.69$\pm0.95$)$\times10^{9}$	  \\
NGC~3596	&	SAB(rs)c	&	5.0	&	22.5	$\pm$	1.6	&	0.54	&	1.38$\times10^{10}$	&	1.51$\times10^{10}$	&	(1.62$\pm0.29$)$\times10^{10}$	  \\
NGC~3646	&	RING		&	4.0	&	65.2	$\pm$	4.6	&	0.52	&	1.43$\times10^{11}$	&	2.05$\times10^{11}$	&	(1.36$\pm0.32$)$\times10^{11}$	  \\
NGC~3675	&	SA(s)b		&	3.0	&	14.3	$\pm$	1.0	&	1.21	&	7.50$\times10^{10}$	&	7.00$\times10^{10}$	&	(5.34$\pm1.20$)$\times10^{10}$	  \\
NGC~3681	&	SAB(r)bc	&	4.0	&	24.9	$\pm$	1.7	&	0.90	&	2.40$\times10^{10}$	&	2.23$\times10^{10}$	&	(2.02$\pm0.42$)$\times10^{10}$	  \\
NGC~3684	&	SA(rs)bc	&	4.0	&	22.8	$\pm$	1.6	&	0.81	&	1.16$\times10^{10}$	&	1.11$\times10^{10}$	&	(8.70$\pm2.11$)$\times10^{9}$	  \\
NGC~3686	&	SB(s)bc		&	4.0	&	22.6	$\pm$	1.6	&	0.63	&	2.42$\times10^{10}$	&	2.40$\times10^{10}$	&	(2.36$\pm0.47$)$\times10^{10}$	  \\
NGC~3705	&	SAB(r)ab	&	2.0	&	13.2	$\pm$	0.9	&	0.55	&	1.08$\times10^{10}$	&	1.74$\times10^{10}$	&	(1.04$\pm0.33$)$\times10^{10}$	  \\
NGC~3810	&	SA(rs)c		&	5.0	&	10.7	$\pm$	0.8	&	0.49	&	5.27$\times10^{9}$	&	7.11$\times10^{9}$	&	(5.54$\pm1.23$)$\times10^{9}$	  \\
NGC~3877	&	SA(s)c:		&	5.0	&	17.8	$\pm$	1.3	&	0.84	&	3.77$\times10^{10}$	&	4.25$\times10^{10}$	&	(2.72$\pm1.34$)$\times10^{10}$	  \\
NGC~3893	&	SAB(rs)c:	&	5.0	&	19.4	$\pm$	1.4	&	0.53	&	2.36$\times10^{10}$	&	2.41$\times10^{10}$	&	(2.33$\pm0.42$)$\times10^{10}$	  \\
NGC~3938	&	SA(s)c		&	5.0	&	15.5	$\pm$	1.1	&	0.50	&	1.56$\times10^{10}$	&	1.79$\times10^{10}$	&	(2.13$\pm0.57$)$\times10^{10}$	  \\
NGC~3949	&	SA(s)bc:	&	4.0	&	15.8	$\pm$	1.1	&	0.69	&	1.17$\times10^{10}$	&	8.20$\times10^{9}$	&	(6.06$\pm1.02$)$\times10^{9}$	  \\
NGC~4030	&	SA(s)bc		&	4.0	&	26.4	$\pm$	1.8	&	0.61	&	7.93$\times10^{10}$	&	9.38$\times10^{10}$	&	(5.54$\pm3.21$)$\times10^{10}$	  \\
NGC~4051	&	SAB(rs)bc	&	4.0	&	2.9	$\pm$	0.9	&	0.68	&	7.80$\times10^{8}$	&	8.55$\times10^{8}$	&	(6.34$\pm1.52$)$\times10^{8}$	  \\
NGC~4062	&	SA(s)c		&	5.0	&	10.4	$\pm$	0.7	&	0.67	&	7.07$\times10^{9}$	&	8.34$\times10^{9}$	&	(6.03$\pm1.59$)$\times10^{9}$	  \\
NGC~4100	&	(R')SA(rs)bc	&	4.0	&	21.5	$\pm$	1.5	&	0.72	&	3.27$\times10^{10}$	&	4.02$\times10^{10}$	&	(2.74$\pm0.73$)$\times10^{10}$	  \\
NGC~4123	&	SB(r)c		&	5.0	&	27.3	$\pm$	1.9	&	0.55	&	2.09$\times10^{10}$	&	2.55$\times10^{10}$	&	(1.78$\pm0.43$)$\times10^{10}$	  \\
NGC~4136	&	SAB(r)c		&	5.0	&	6.7	$\pm$	0.5	&	0.42	&	6.51$\times10^{8}$	&	7.42$\times10^{8}$	&	(4.42$\pm2.12$)$\times10^{8}$	  \\
NGC~4145	&	SAB(rs)d	&	7.0	&	20.3	$\pm$	1.4	&	0.53	&	1.38$\times10^{10}$	&	1.51$\times10^{10}$	&	(1.01$\pm0.31$)$\times10^{10}$	  \\
NGC~4151	&	(R')SAB(rs)ab:	&	2.0	&	20.0	$\pm$	1.4	&	0.98	&	4.29$\times10^{10}$	&	4.21$\times10^{10}$	&	(3.30$\pm0.58$)$\times10^{10}$	  \\
NGC~4212	&	SAc:		&	4.5	&     16.3\tn{c}$\pm$	3.8	&	0.68	&	1.64$\times10^{10}$	&	1.65$\times10^{10}$	&	(1.76$\pm0.33$)$\times10^{10}$	  \\
NGC~4254	&	SA(s)c		&	5.0	&     16.5\tn{d}$\pm$	1.1	&	0.46	&	3.37$\times10^{10}$	&	4.29$\times10^{10}$	&	(3.70$\pm0.71$)$\times10^{10}$	  \\
NGC~4293	&	(R)SB(s)0/a	&	0.0	&	14.1	$\pm$	1.0	&	0.85	&	2.93$\times10^{10}$	&	3.39$\times10^{10}$	&	(3.61$\pm0.89$)$\times10^{10}$	  \\
NGC~4303	&	SAB(rs)bc	&	4.0	&	13.6	$\pm$	1.0	&	0.56	&	2.80$\times10^{10}$	&	3.06$\times10^{10}$	&	(2.32$\pm0.50$)$\times10^{10}$	  \\
NGC~4314	&	SB(rs)a		&	1.0	&	17.8	$\pm$	1.3	&	1.03	&	6.03$\times10^{10}$	&	5.55$\times10^{10}$	&	(4.34$\pm1.01$)$\times10^{10}$	  \\
NGC~4388	&	SA(s)b:~sp	&	3.0	&	41.4	$\pm$	2.9	&	0.64	&	1.08$\times10^{11}$	&	1.28$\times10^{11}$	&	(1.55$\pm0.43$)$\times10^{11}$	  \\
NGC~4394	&	(R)SB(r)b	&	3.0	&	14.1	$\pm$	1.0	&	0.77	&	1.76$\times10^{10}$	&	1.94$\times10^{10}$	&	(1.74$\pm0.35$)$\times10^{10}$	  \\
NGC~4414	&	SA(rs)c?	&	5.0	&	9.0	$\pm$	0.6	&	0.74	&	1.68$\times10^{10}$	&	1.99$\times10^{10}$	&	(1.67$\pm0.39$)$\times10^{10}$	  \\
NGC~4448	&	SB(r)ab		&	2.0	&	7.0	$\pm$	0.5	&	0.89	&	5.61$\times10^{9}$	&	5.98$\times10^{9}$	&	(4.66$\pm1.16$)$\times10^{9}$	  \\
NGC~4450	&	SA(s)ab		&	2.0	&	14.1	$\pm$	1.0	&	1.07	&	5.09$\times10^{10}$	&	4.85$\times10^{10}$	&	(4.66$\pm0.71$)$\times10^{10}$	  \\
NGC~4457	&	(R)SAB(s)0/a	&	0.0	&	13.6	$\pm$	1.0	&	1.03	&	2.18$\times10^{10}$	&	2.18$\times10^{10}$	&	(1.91$\pm0.23$)$\times10^{10}$	  \\
NGC~4490	&	SB(s)d~pec	&	7.0	&	9.2	$\pm$	0.7	&	0.38	&	6.08$\times10^{9}$	&	5.91$\times10^{9}$	&	(4.69$\pm0.75$)$\times10^{9}$	  \\
NGC~4496A	&	SB(rs)m		&	9.0	&	13.6	$\pm$	1.4	&	0.43	&	2.31$\times10^{9}$	&	2.59$\times10^{9}$	&	(2.62$\pm0.42$)$\times10^{9}$	  \\
NGC~4527	&	SAB(s)bc	&	4.0	&	13.5	$\pm$	0.9	&	0.81	&	3.39$\times10^{10}$	&	5.93$\times10^{10}$	&	(3.85$\pm0.95$)$\times10^{10}$	  \\
NGC~4548	&	SB(rs)b		&	3.0	&	3.7	$\pm$	0.3	&	0.95	&	3.03$\times10^{9}$	&	3.10$\times10^{9}$	&	(2.73$\pm0.49$)$\times10^{9}$	  \\
NGC~4568	&	SA(rs)bc	&	4.0	&	13.9	$\pm$	1.0	&	0.75	&	2.15$\times10^{10}$	&	3.02$\times10^{10}$	&	(2.19$\pm0.59$)$\times10^{10}$	  \\
NGC~4571	&	SA(r)d		&	6.5	&	2.6	$\pm$	0.2	&	0.82	&	3.91$\times10^{8}$	&	3.99$\times10^{8}$	&	(2.72$\pm0.66$)$\times10^{8}$	  \\
NGC~4579	&	SAB(rs)b	&	3.0	&	13.9	$\pm$	1.0	&	1.00	&	7.76$\times10^{10}$	&	7.44$\times10^{10}$	&	(6.19$\pm1.19$)$\times10^{10}$	  \\
NGC~4580	&	SAB(rs)a~pec	&	1.0	&	13.6	$\pm$	1.0	&	0.93	&	9.26$\times10^{9}$	&	9.05$\times10^{9}$	&	(5.60$\pm2.19$)$\times10^{9}$	  \\
NGC~4618	&	SB(rs)m		&	9.0	&	8.8	$\pm$	0.6	&	0.48	&	2.76$\times10^{9}$	&	2.54$\times10^{9}$	&	(1.64$\pm0.59$)$\times10^{9}$	  \\
NGC~4643	&	SB(rs)0/a	&	0.0	&	27.3	$\pm$	1.9	&	1.22	&	1.71$\times10^{11}$	&	1.74$\times10^{11}$	&	(1.64$\pm0.12$)$\times10^{11}$	  \\
NGC~4647	&	SAB(rs)c	&	5.0	&	13.9	$\pm$	1.0	&	0.97	&	2.11$\times10^{10}$	&	1.92$\times10^{10}$	&	(1.66$\pm0.32$)$\times10^{10}$	  \\
NGC~4651	&	SA(rs)c		&	5.0	&	14.0	$\pm$	1.0	&	0.59	&	1.37$\times10^{10}$	&	1.63$\times10^{10}$	&	(1.05$\pm0.34$)$\times10^{10}$	  \\
NGC~4654	&	SAB(rs)cd	&	6.0	&	13.9	$\pm$	1.0	&	0.55	&	1.34$\times10^{10}$	&	1.60$\times10^{10}$	&	(1.09$\pm0.24$)$\times10^{10}$	  \\
NGC~4665	&	SB(s)0/a	&	0.0	&	13.5	$\pm$	0.9	&	0.88	&	4.76$\times10^{10}$	&	5.39$\times10^{10}$	&	(4.80$\pm1.14$)$\times10^{10}$	  \\
NGC~4666	&	SABc:		&	5.0	&	27.5	$\pm$	1.9	&	0.94	&	1.45$\times10^{11}$	&	1.87$\times10^{11}$	&	(1.91$\pm0.45$)$\times10^{11}$	  \\
NGC~4689	&	SA(rs)bc	&	4.0	&	14.0	$\pm$	1.0	&	0.74	&	1.47$\times10^{10}$	&	1.45$\times10^{10}$	&	(1.07$\pm0.28$)$\times10^{10}$	  \\
NGC~4691	&	(R)SB(s)0/a~pec	&	0.0	&	17.0	$\pm$	1.2	&	0.83	&	1.95$\times10^{10}$	&	1.57$\times10^{10}$	&	(1.25$\pm0.27$)$\times10^{10}$	  \\
NGC~4698	&	SA(s)ab		&	2.0	&	13.7	$\pm$	1.0	&	1.13	&	3.46$\times10^{10}$	&	3.40$\times10^{10}$	&	(3.12$\pm0.45$)$\times10^{10}$	  \\
NGC~4699	&	SAB(rs)b	&	3.0	&	22.9	$\pm$	1.6	&	1.16	&	2.02$\times10^{11}$	&	1.58$\times10^{11}$	&	(1.28$\pm0.27$)$\times10^{11}$	  \\
NGC~4772	&	SA(s)a		&	1.0	&	13.3	$\pm$	0.9	&	1.04	&	1.66$\times10^{10}$	&	1.61$\times10^{10}$	&	(1.32$\pm0.24$)$\times10^{10}$	  \\
NGC~4900	&	SB(rs)c		&	5.0	&	9.1	$\pm$	0.6	&	0.55	&	2.30$\times10^{9}$	&	2.15$\times10^{9}$	&	(2.38$\pm0.53$)$\times10^{9}$	  \\
NGC~5005	&	SAB(rs)bc	&	4.0	&	19.3	$\pm$	1.4	&	0.82	&	1.25$\times10^{11}$	&	1.49$\times10^{11}$	&	(1.83$\pm0.51$)$\times10^{11}$	  \\
NGC~5334	&	SB(rs)c		&	5.0	&	24.2	$\pm$	1.7	&	0.59	&	1.11$\times10^{10}$	&	1.24$\times10^{10}$	&	(8.32$\pm2.23$)$\times10^{9}$	  \\
NGC~5371	&	SAB(rs)bc	&	4.0	&	42.8	$\pm$	3.0	&	0.90	&	1.60$\times10^{11}$	&	1.74$\times10^{11}$	&	(1.29$\pm0.21$)$\times10^{11}$	  \\
NGC~5448	&	(R)SAB(r)a	&	1.0	&	35.2	$\pm$	2.5	&	0.81	&	5.86$\times10^{10}$	&	7.11$\times10^{10}$	&	(5.67$\pm1.23$)$\times10^{10}$	  \\
NGC~5676	&	SA(rs)bc	&	4.0	&	36.5	$\pm$	2.6	&	0.64	&	9.23$\times10^{10}$	&	1.16$\times10^{11}$	&	(1.20$\pm0.23$)$\times10^{11}$	  \\
NGC~5701	&	(R)SB(rs)0/a	&	0.0	&	26.7	$\pm$	1.9	&	0.92	&	6.33$\times10^{10}$	&	6.77$\times10^{10}$	&	(6.32$\pm1.08$)$\times10^{10}$	  \\
NGC~5713	&	SAB(rs)bc~pec	&	4.0	&	31.3	$\pm$	2.2	&	0.63	&	4.93$\times10^{10}$	&	4.95$\times10^{10}$	&	(3.96$\pm0.86$)$\times10^{10}$	  \\
NGC~5850	&	SB(r)b		&	3.0	&	41.6	$\pm$	2.9	&	0.74	&	1.41$\times10^{11}$	&	1.78$\times10^{11}$	&	(1.38$\pm0.32$)$\times10^{11}$	  \\
NGC~5921	&	SB(r)bc		&	4.0	&	26.2	$\pm$	1.8	&	0.53	&	3.68$\times10^{10}$	&	4.66$\times10^{10}$	&	(4.40$\pm0.83$)$\times10^{10}$	  \\
NGC~5962	&	SA(r)c		&	5.0	&	34.2	$\pm$	2.4	&	0.50	&	4.37$\times10^{10}$	&	5.83$\times10^{10}$	&	(6.85$\pm1.67$)$\times10^{10}$	  \\
NGC~6384	&	SAB(r)bc	&	4.0	&	29.2	$\pm$	2.0	&	0.68	&	8.62$\times10^{10}$	&	1.13$\times10^{11}$	&	(7.26$\pm1.86$)$\times10^{10}$	  \\
NGC~7217	&	(R)SA(r)ab	&	2.0	&	16.5	$\pm$	1.2	&	1.10	&	9.60$\times10^{10}$	&	9.20$\times10^{10}$	&	(1.02$\pm0.22$)$\times10^{11}$	  \\
NGC~7479	&	SB(s)c		&	5.0	&	33.7	$\pm$	2.4	&	0.58	&	8.31$\times10^{10}$	&	1.06$\times10^{11}$	&	(7.82$\pm1.92$)$\times10^{10}$	  \\
NGC~7606	&	SA(s)b		&	3.0	&	31.3	$\pm$	2.2	&	0.66	&	9.48$\times10^{10}$	&	1.17$\times10^{11}$	&	(1.52$\pm0.35$)$\times10^{11}$	  \\
NGC~7741	&	SB(s)cd		&	6.0	&	12.5	$\pm$	0.9	&	0.50	&	3.47$\times10^{9}$	&	3.66$\times10^{9}$	&	(2.82$\pm0.56$)$\times10^{9}$	  \\
NGC~7814	&	SA(s)ab:~sp	&	2.0	&	15.7	$\pm$	1.1	&	1.34	&      6.24$\times10^{10}$\tn{b}&      9.08$\times10^{10}$\tn{b}&     (6.21$\pm0.52$)$\times10^{10}$\tn{b}\\

\enddata
\tablenotetext{a}{~\citet{tik09}}
\tablenotetext{b}{~Lower limit.}
\tablenotetext{c}{~\citet{sor14}}
\tablenotetext{d}{~\citet{mei07}}
\tablecomments{Col. 1: galaxy name. 
Col. 2: RC3 type~\citep{deV91}.
Col. 3: T Hubble type~\citep{deV91}.
Col. 4: distance to object in Mpc, from NED (Virgo + GA + Shapley), unless otherwise indicated.
Col. 5: median $\Upsilon_{*}^{H}$ after BSP iteration number 1.
For M~51 and M~51b the median $\Upsilon_{*}^{K_{s}}$ is tabulated instead of $\Upsilon_{*}^{H}$.
Col. 6: total resolved stellar mass obtained from the BSP algorithm, $M^{\rm BSP}_{*}$, in solar units.
Col. 7: total resolved stellar mass obtained from ZCR$^\prime$, $M^{\rm{ZCR^\prime}}_{*}$, in solar units.
Col. 8: unresolved stellar mass, $M^{\rm unresolved}_{*}$, in solar units.
All the masses given in this table have been calculated using the SSAG-BC03 SPS library.
The uncertainties in $M^{\rm unresolved}_{*}$ correspond to the propagation of the systematic error
due to the zero point calibration, which affects the values of $M^{\rm BSP}_{*}$ and $M^{\rm{ZCR^\prime}}_{*}$
by only $\sim3\%$. The systematic uncertainty in the distance to the objects is not
quoted in this table (see section~\ref{mass_errors}).
}
\end{deluxetable*}




\begin{thebibliography}{}

\bibitem[Abraham et al.(1999)]{abr99} Abraham, R.~G., Ellis, R.~S., Fabian, A.~C., Tanvir, N.~R., \& Glazebrook, K.\ 1999, \mnras, 303, 641 
\bibitem[Aihara et al.(2011)]{aih11} Aihara, H., Allende Prieto, C., An, D., et al.\ 2011, \apjs, 193, 29 
\bibitem[Alam et al.(2015)]{ala15} Alam, S., Albareti, F.~D., Allende Prieto, C., et al.\ 2015, \apjs, 219, 12
\bibitem[Ben{\'{\i}}tez(2000)]{ben00} Ben{\'{\i}}tez, N.\ 2000, \apj, 536, 571 
\bibitem[Bell \& de Jong(2001)]{bel01} Bell, E.~F., \& de Jong, R.~S.\ 2001, \apj, 550, 212 
\bibitem[Bell et al.(2003)]{bel03} Bell, E.~F., McIntosh, D.~H., Katz, N., \& Weinberg, M.~D.\ 2003, \apjs, 149, 289 
\bibitem[Bershady et al.(2010)]{bsha10} Bershady, M.~A., Verheijen, M.~A.~W., Swaters, R.~A., et al.\ 2010, \apj, 716, 198 
\bibitem[Bertin(2010)]{ber10} Bertin, E.\ 2010, Astrophysics Source Code Library, ascl:1010.068
\bibitem[Bevington(1969)]{bev69} Bevington, P.~R.\ 1969, Data reduction and error analysis for the physical sciences, New York: McGraw-Hill  
\bibitem[Bhavsar(1990)]{bha90} Bhavsar, S.~P.\ 1990, Errors, Bias and Uncertainties in Astronomy, 107 
\bibitem[Blanton \& Roweis(2007)]{bla07} Blanton, M.~R., \& Roweis, S.\ 2007, \aj, 133, 734 
\bibitem[Block \& Wainscoat(1991)]{blo91} Block, D.~L., \& Wainscoat, R.~J.\ 1991, \nat, 353, 48 
\bibitem[Block et al.(1994)]{blo94} Block, D.~L., Bertin, G., Stockton, A., et al.\ 1994, \aap, 288, 365
\bibitem[Bothun(1986)]{bot86} Bothun, G.~D.\ 1986, \aj, 91, 507 
\bibitem[Bruzual \& Charlot(2003)]{bru03} Bruzual, G., \& Charlot, S.\ 2003, \mnras, 344, 1000
\bibitem[Bruzual(2007)]{bru07} Bruzual, A.~G.\ 2007, in IAU Symp. 241, Stellar Populations as Building Blocks of Galaxies, ed. A. Vazdekis \& R. F. Peletier (Cambridge: Cambridge Univ. Press), 241, 125 
\bibitem[Bundy et al.(2015)]{bun15} Bundy, K., Bershady, M.~A., Law, D.~R., et al.\ 2015, \apj, 798, 7 
\bibitem[Cappellari \& Copin(2003)]{cap03} Cappellari, M., \& Copin, Y.\ 2003, \mnras, 342, 345
\bibitem[Chapman et al.(2009)]{chap09} Chapman, N.~L., Mundy, L.~G., Lai, S.-P., \& Evans, N.~J., II 2009, \apj, 690, 49
\bibitem[Chabrier(2003)]{cha03} Chabrier, G.\ 2003, \pasp, 115, 763 
\bibitem[Charlot \& Fall(2000)]{cha00} Charlot, S., \& Fall, S.~M.\ 2000, \apj, 539, 718
\bibitem[Chen et al.(2012)]{che12} Chen, Y.-M., Kauffmann, G., Tremonti, C.~A., et al.\ 2012, \mnras, 421, 314 
\bibitem[Conroy et al.(2009)]{conr09} Conroy, C., Gunn, J.~E., \& White, M.\ 2009, \apj, 699, 486 
\bibitem[Conti et al.(2003)]{cont03} Conti, A., Connolly, A.~J., Hopkins, A.~M., et al.\ 2003, \aj, 126, 2330 
\bibitem[Courteau et al.(2014)]{cou14} Courteau, S., Cappellari, M., de Jong, R.~S., et al.\ 2014, Reviews of Modern Physics, 86, 47 
\bibitem[da Cunha et al.(2008)]{daC08} da Cunha, E., Charlot, S., \& Elbaz, D.\ 2008, \mnras, 388, 1595
\bibitem[Daddi et al.(2007)]{dad07} Daddi, E., Dickinson, M., Morrison, G., et al.\ 2007, \apj, 670, 156 
\bibitem[de Jong(1996)]{deJ96} de Jong, R.~S.\ 1996, \aap, 313, 377 
\bibitem[de Jong \& Bell(2007)]{jon07} de Jong, R.~S., \& Bell, E.~F.\ 2007, Astrophysics and Space Science Proceedings, 3, 107 
\bibitem[de Vaucouleurs et al.(1991)]{deV91} de Vaucouleurs, G., de Vaucouleurs, A., Corwin, H.~G., Jr., et al.\ 1991, Third Reference Catalogue of Bright Galaxies (RC3)
\bibitem[Driver et al.(2007)]{dri07} Driver, S.~P., Popescu, C.~C., Tuffs, R.~J., et al.\ 2007, \mnras, 379, 1022
\bibitem[Egusa et al.(2016)]{egu16} Egusa, F., Mentuch Cooper, E., Koda, J., \& Baba, J.\ 2016, arXiv:1610.06642 
\bibitem[Elbaz et al.(2007)]{elb07} Elbaz, D., Daddi, E., Le Borgne, D., et al.\ 2007, \aap, 468, 33 
\bibitem[Eskridge et al.(2002)]{esk02} Eskridge, P.~B., Frogel, J.~A., Pogge, R.~W., et al.\ 2002, \apjs, 143, 73 
\bibitem[Eskridge et al.(2003)]{esk03} Eskridge, P.~B., Frogel, J.~A., Taylor, V.~A., et al.\ 2003, \apj, 586, 923 
\bibitem[Foyle et al.(2010)]{foy10} Foyle, K., Rix, H.-W., \& Zibetti, S.\ 2010, \mnras, 407, 163 
\bibitem[Gallazzi \& Bell(2009)]{gal09} Gallazzi, A., \& Bell, E.~F.\ 2009, \apjs, 185, 253 
\bibitem[Gonzalez \& Graham(1996)]{gon96} Gonzalez, R.~A., \& Graham, J.~R.\ 1996, \apj, 460, 651
\bibitem[Grosb{\o}l \& Dottori(2008)]{gros08} Grosb{\o}l, P., \& Dottori, H.\ 2008, \aap, 490, 87
\bibitem[Grosb{\o}l et al.(2006)]{gros06} Grosb{\o}l, P., Dottori, H., \& Gredel, R.\ 2006, \aap, 453, L25 
\bibitem[Gumus \& Sen(2013)]{gum13} Gumus, K., \& Sen, A.\ 2013, Geodetski Vestnik, 57-3, 523
\bibitem[Into \& Portinari(2013)]{int13} Into, T., \& Portinari, L.\ 2013, \mnras, 430, 2715 
\bibitem[James \& Seigar(1999)]{jam99} James, P.~A., \& Seigar, M.~S.\ 1999, \aap, 350, 791 
\bibitem[Jarrett et al.(2003)]{jar03} Jarrett, T.~H., Chester, T., Cutri, R., Schneider, S.~E., \& Huchra, J.~P.\ 2003, \aj, 125, 525 
\bibitem[Just et al.(2015)]{jus15} Just, A., Fuchs, B., Jahrei{\ss}, H., et al.\ 2015, \mnras, 451, 149 
\bibitem[Kassin et al.(2003)]{kas03} Kassin, S.~A., Frogel, J.~A., Pogge, R.~W., Tiede, G.~P., \& Sellgren, K.\ 2003, \aj, 126, 1276 
\bibitem[Kassin et al.(2006)]{kas06} Kassin, S.~A., de Jong, R.~S., \& Pogge, R.~W.\ 2006, \apjs, 162, 80 
\bibitem[Keen(2010)]{kee10} Keen, K.~J., Graphics for Statistics and Data Analysis with R, Chapman~\& Hall/CRC, 2010
\bibitem[Kennicutt et al.(2003)]{ken03} Kennicutt, R.~C., Jr., Armus, L., Bendo, G., et al.\ 2003, \pasp, 115, 928 
\bibitem[Kobulnicky \& Kewley(2004)]{kob04} Kobulnicky, H.~A., \& Kewley, L.~J.\ 2004, \apj, 617, 240
\bibitem[Kroupa(2001)]{kro01} Kroupa, P.\ 2001, \mnras, 322, 231  
\bibitem[Lanyon-Foster et al.(2007)]{lan07} Lanyon-Foster, M.~M., Conselice, C.~J., \& Merrifield, M.~R.\ 2007, \mnras, 380, 571 
\bibitem[Leauthaud et al.(2012)]{lea12} Leauthaud, A., George, M.~R., Behroozi, P.~S., et al.\ 2012, \apj, 746, 95 
\bibitem[Lepage \& Billard(1992)]{lep92} Lepage, R., \& Billard, L.\ 1992, Wiley Series in Probability and Mathematical Statistics, New York: Wiley, 1992
\bibitem[Leroy et al.(2008)]{ler08} Leroy, A.~K., Walter, F., Brinks, E., et al.\ 2008, \aj, 136, 2782
\bibitem[Loredo(1992)]{lor92} Loredo, T.~J.\ 1992, Statistical Challenges in Modern Astronomy, 275
\bibitem[Loredo(1995)]{lor95} Loredo, T.~J.\ 1995, Ph.D.~Thesis, Univ. Chicago
\bibitem[Magris et al.(2015)]{mag15} Magris C., G., Mateu P., J., Mateu, C., et al.\ 2015, \pasp, 127, 16 
\bibitem[Maller et al.(2009)]{mal09} Maller, A.~H., Berlind, A.~A., Blanton, M.~R., \& Hogg, D.~W.\ 2009, \apj, 691, 394 
\bibitem[Maraston et al.(2006)]{mara06} Maraston, C., Daddi, E., Renzini, A., et al.\ 2006, \apj, 652, 85 
\bibitem[Maraston et al.(2010)]{mara10} Maraston, C., Pforr, J., Renzini, A., et al.\ 2010, \mnras, 407, 830 
\bibitem[Mart{\'{\i}}nez-Garc{\'{\i}}a et al.(2009)]{mart09} Mart{\'{\i}}nez-Garc{\'{\i}}a, E.~E., Gonz{\'a}lez-L{\'o}pezlira, R.~A., \& Bruzual-A, G.\ 2009, \apj, 694, 512
\bibitem[Mart{\'{\i}}nez-Garc{\'{\i}}a \& Gonz{\'a}lez-L{\'o}pezlira(2013)]{mart13} Mart{\'{\i}}nez-Garc{\'{\i}}a, E.~E., \& Gonz{\'a}lez-L{\'o}pezlira, R.~A.\ 2013, \apj, 765, 105 
\bibitem[Martinsson et al.(2013)]{mts13} Martinsson, T.~P.~K., Verheijen, M.~A.~W., Westfall, K.~B., et al.\ 2013, \aap, 557, A131 
\bibitem[McGaugh \& Schombert(2014)]{mcg14} McGaugh, S.~S., \& Schombert, J.~M.\ 2014, \aj, 148, 77 
\bibitem[McGaugh et al.(2016)]{mcg16} McGaugh, S., Lelli, F., \& Schombert, J.\ 2016, \prl, 117, 201101
\bibitem[Mei et al.(2007)]{mei07} Mei, S., Blakeslee, J.~P., C{\^o}t{\'e}, P., et al.\ 2007, \apj, 655, 144 
\bibitem[Meidt et al.(2012)]{mei12} Meidt, S.~E., Schinnerer, E., Knapen, J.~H., et al.\ 2012, \apj, 744, 17 
\bibitem[Meidt et al.(2014)]{mei14} Meidt, S.~E., Schinnerer, E., van de Ven, G., et al.\ 2014, \apj, 788, 144 
\bibitem[Mentuch Cooper et al.(2012)]{men12} Mentuch Cooper, E., Wilson, C.~D., Foyle, K., et al.\ 2012, \apj, 755, 165 
\bibitem[Mitchell et al.(2013)]{mit13} Mitchell, P.~D., Lacey, C.~G., Baugh, C.~M., \& Cole, S.\ 2013, \mnras, 435, 87
\bibitem[Moustakas et al.(2010)]{mou10} Moustakas, J., Kennicutt, R.~C., Jr., Tremonti, C.~A., et al.\ 2010, \apjs, 190, 233-266  
\bibitem[Noeske et al.(2007)]{noe07} Noeske, K.~G., Weiner, B.~J., Faber, S.~M., et al.\ 2007, \apjl, 660, L43 
\bibitem[Padmanabhan et al.(2008)]{pad08} Padmanabhan, N., Schlegel, D.~J., Finkbeiner, D.~P., et al.\ 2008, \apj, 674, 1217-1233 
\bibitem[Patsis et al.(2001)]{pat01} Patsis, P.~A., H{\'e}raudeau, P., \& Grosb{\o}l, P.\ 2001, \aap, 370, 875
\bibitem[Pilyugin \& Thuan(2005)]{pil05} Pilyugin, L.~S., \& Thuan, T.~X.\ 2005, \apj, 631, 231  
\bibitem[Portinari et al.(2004)]{por04} Portinari, L., Sommer-Larsen, J., \& Tantalo, R.\ 2004, \mnras, 347, 691 
\bibitem[Querejeta et al.(2015)]{que15} Querejeta, M., Meidt, S.~E., Schinnerer, E., et al.\ 2015, \apjs, 219, 5
\bibitem[Reach et al.(2005)]{rea05} Reach, W.~T., Megeath, S.~T., Cohen, M., et al.\ 2005, \pasp, 117, 978 
\bibitem[Repetto et al.(2013)]{rep13} Repetto, P., Mart{\'{\i}}nez-Garc{\'{\i}}a, E.~E., Rosado, M., \& Gabbasov, R.\ 2013, \apj, 765, 7 
\bibitem[Repetto et al.(2015)]{rep15} \samename 2015, \mnras, 451, 353 
\bibitem[Rhoads(1998)]{rho98} Rhoads, J.~E.\ 1998, \aj, 115, 472 
\bibitem[Rix \& Rieke(1993)]{rix93} Rix, H.-W., \& Rieke, M.~J.\ 1993, \apj, 418, 123 
\bibitem[Roediger \& Courteau(2015)]{rod15} Roediger, J.~C., \& Courteau, S.\ 2015, \mnras, 452, 3209 
\bibitem[Rovilos et al.(2014)]{rov14} Rovilos, E., Georgantopoulos, I., Akylas, A., et al.\ 2014, \mnras, 438, 494
\bibitem[Salpeter(1955)]{sal55} Salpeter, E.~E.\ 1955, \apj, 121, 161 
\bibitem[Salim et al.(2007)]{sal07} Salim, S., Rich, R.~M., Charlot, S., et al.\ 2007, \apjs, 173, 267 
\bibitem[S{\'a}nchez et al.(2012)]{san12} S{\'a}nchez, S.~F., Kennicutt, R.~C., Gil de Paz, A., et al.\ 2012, \aap, 538, A8 
\bibitem[Savage \& Oliver(2007)]{sav07} Savage, R.~S., \& Oliver, S.\ 2007, \apj, 661, 1339 
\bibitem[Schlafly \& Finkbeiner(2011)]{schl11} Schlafly, E.~F., \& Finkbeiner, D.~P.\ 2011, \apj, 737, 103
\bibitem[Sch{\"o}nrich \& Bergemann(2014)]{scho14} Sch{\"o}nrich, R., \& Bergemann, M.\ 2014, \mnras, 443, 698
\bibitem[Sheth et al.(2010)]{she10} Sheth, K., Regan, M., Hinz, J.~L., et al.\ 2010, \pasp, 122, 1397
\bibitem[Skrutskie et al.(2006)]{skr06} Skrutskie, M.~F., Cutri, R.~M., Stiening, R., et al.\ 2006, \aj, 131, 1163
\bibitem[Sorba \& Sawicki(2015)]{sor15} Sorba, R., \& Sawicki, M.\ 2015, \mnras, 452, 235 
\bibitem[Sorce et al.(2014)]{sor14} Sorce, J.~G., Tully, R.~B., Courtois, H.~M., et al.\ 2014, \mnras, 444, 527 
\bibitem[Taylor et al.(2011)]{tay11} Taylor, E.~N., Hopkins, A.~M., Baldry, I.~K., et al.\ 2011, \mnras, 418, 1587
\bibitem[Tody(1993)]{tod93} Tody, D.\ 1993, in ASP Conf. Ser. 52, Astronomical Data Analysis Software and Systems II, ed. R. J. Hanisch, R. J. V. Brissenden, \& J. Barnes (San Francisco, CA: ASP), 173
\bibitem[Tikhonov et al.(2009)]{tik09} Tikhonov, N.~A., Galazutdinova, O.~A., \& Tikhonov, E.~N.\ 2009, Astronomy Letters, 35, 599
\bibitem[Welikala et al.(2008)]{wel08} Welikala, N., Connolly, A.~J., Hopkins, A.~M., Scranton, R., \& Conti, A.\ 2008, \apj, 677, 970
\bibitem[Wuyts et al.(2012)]{wuy12} Wuyts, S., F{\"o}rster Schreiber, N.~M., Genzel, R., et al.\ 2012, \apj, 753, 114 
\bibitem[Zibetti(2009)]{zbt09} Zibetti, S.\ 2009, arXiv:0911.4956 
\bibitem[Zibetti, Charlot,~\& Rix(2009)]{zcr09} Zibetti, S., Charlot, S., \& Rix, H.-W.\ 2009, \mnras, 400, 1181, ZCR
 
\end{thebibliography}
\end{document}